\documentclass[english]{achemso}
\usepackage[T1]{fontenc}
\usepackage[utf8]{inputenc}
\usepackage{color}
\usepackage{babel}
\usepackage{array}
\usepackage{booktabs}
\usepackage{textcomp}
\usepackage{multirow}
\usepackage{amsmath}
\usepackage{amssymb}
\usepackage{cancel}
\usepackage{graphicx}
\usepackage[numbers]{natbib}
\usepackage[unicode=true,pdfusetitle,
 bookmarks=true,bookmarksnumbered=false,bookmarksopen=false,
 breaklinks=false,pdfborder={0 0 0},pdfborderstyle={},backref=false,colorlinks=false]
 {hyperref}

\makeatletter

\title{First-Principles Study of Penta-CN$_{2}$ Quantum Dots for Efficient
Hydrogen Evolution Reaction}

\author{Rupali Jindal, Rachana Yogi, and Alok Shukla}

\affiliation{Department of Physics, Indian Institute of Technology Bombay, Powai,
Mumbai 400076, India}

\email{rupalijindalpu@gmail.com, yogirachana04@gmail.com, shukla@iitb.ac.in}

\newcommand*{\LyXFourPerEmSpace}{\hskip0.25em\relax}
\providecommand{\tabularnewline}{\\}

\makeatother

\begin{document}
\begin{abstract}
\textcolor{black}{The objective of our research is to investigate
the electrocatalytic properties of novel metal-free quantum dots (QDs)
composed of the recently discovered 2D material penta-CN$_{2}$, with
the aim of replacing costly and scarce catalysts such as Pt and Pd.
Employing a first-principles density functional theory (DFT) based
approach, the geometries of the three penta-CN$_{2}$ quantum dots
(QDs) of increasing sizes, $3\times3$, $3\times4$, $4\times4$ are
optimized. Through comprehensive analysis, our research extensively
explored the structural stability of penta-CN$_{2}$ QDs, delved into
their electronic properties, and assessed their catalytic performance
concerning the Hydrogen Evolution Reaction (HER). Notably, the H-adsorbed
penta-CN$_{2}$ QDs exhibit a significant reduction in the HOMO-LUMO
gap (E$_{g}$) ranging from 35\% to 49\% compared to the pristine
QD. This observation underscores the crucial impact of H-adsorption
on Penta-CN$_{2}$ QD and is further supported by the appearance of
mid-gap states in total and partial density of states plots. Next,
we investigated their catalytic performance relevant to HER, using
well-known descriptors: (i) adsorption energy, (ii) over-potential,
(iii) Gibbs free energy and (iv) exchange current density along with
the volcano curve. As far as size dependence of the catalytic performance
is concerned, the value of $\Delta G^{(av)}$ is minimum for $3\times3$
penta--CN$_{2}$ QD, with those of $3\times4$ and $4\times4$ QDs
being slightly larger. Our calculations predict a high value of exchange
current density $2.24\times10^{-3}$ A-cm$^{-2}$ for one of the sites
(N11 for $3\times3$ QD), which we believe will lead to significantly
enhanced HER properties. The minimum value of $\Delta G=0.158\:\mathrm{eV}$
for a 3$\times3$ $\mathrm{\mathrm{penta-CN_{2}}}$ QD implies that
its catalytic performance is at least as effective or perhaps better
than most of the metal-free hybrid and non-hybrid structures. Our
research outcomes hold great promise in advancing the discovery of
abundant, non-toxic, and cost-effective catalysts for HER, playing
a vital role in facilitating large-scale hydrogen production.}\\
\textcolor{black}{{} }\textbf{Keywords}: Penta-CN$_{2}$ quantum dots,
Metal-free electrocatalyst, Density functional theory, Water splitting
\end{abstract}

\section{Introduction}

With the increasing population, the biggest challenge confronting
the humanity is to fulfill global energy requirements without severely
impacting the climate. The power consumption of the entire world is
expected to increase to $\sim27$ TW in 2050, which is double of the
value reported in 2001, i.e., $\sim13.5$ TW\citep{lewis2006powering}.
However, the adverse impact of power production on the climate can
be judged from the facts that it accounts for three-quarters of greenhouse
gas emissions, 66\% of $\mathrm{NO_{x}}$, and most of the particulate
matter emissions \citep{wang2019air}. However, in order to prevent
drastic weather and climate changes in the future, we must shift to
clean and green fuels in the energy sector with a high conversion
rate, and low $\mathrm{CO_{2}}$ emissions \citep{levin2010challenges,hasani2019two}.
The $\mathrm{H_{2}}$ economy emerges as a potential solution to satisfy
the worldwide energy demands and to resolve this future environmental
crisis\citep{dincer2012green}. The $\mathrm{H_{2}}$ fuel besides
its key properties such as clean, non-toxic, and sustainability, also
has the highest gravimetric energy density $\mathrm{(142\:MJ/Kg)}$,
which make it one of the most attractive fuels as compared to other
energy sources \citep{zhu2020two}. Hydrogen covers $90\%$ of the
universe's total mass, but it is not available in its molecular form
on earth\citep{olabi2021large}. $\mathrm{H_{2}}$ can be produced
via thermal \citep{turn1998experimental,cilogullari2017investigation},
electrolytic \citep{hughes2021polymer,mahmood2017efficient}, or photocatalytic
\citep{fujishima1972electrochemical} methods using conventional or
non-conventional resources. However, $\mathrm{H_{2}}$ production
through electrolysis of water using decarbonized sources is an efficient
way to confront the sustainable future energy challenges\citep{kalamaras2013hydrogen}.
The electrolysis of water ($\mathrm{2H^{+}+2e^{-}\rightarrow H_{2})}$
is capable of generating ultra-pure hydrogen ($>99.999\%$) after
water vapor removal\citep{turner2008renewable}. But, at present,
only $\sim4\%$ contribution of electrolysis has been seen towards
$\mathrm{H_{2}}$ production as water splitting requires a large thermodynamic
potential of 1.23 V at ambient conditions due to kinetic barriers
\citep{dubouis2019hydrogen}. Therefore, robust electrocatalysts of
low cost, high efficiency, and excellent durability are required to
overcome this high thermodynamic potential, for sustainable H$_{2}$
production. The electrocatalysts can be categorized into: a) Pt-based
electrocatalysts\citep{yu2020pt,shao2018advanced}, b) Transition
metal (TM) based electrocatalysts\citep{tsai2014tuning,er2018prediction},
and c) Metal-free electrocatalysts\citep{govindaraju2021graphitic,liu2016carbon}.
Pt catalysts and their derivatives exhibit the highest exchange current
density and negligible overpotential values, but their scarcity and
high cost limit their use to a large extent\citep{conway2002interfacial}.
The second class of TM electrocatalysts such as Co, Ni, Al, etc. undergo
in-built corrosion that restricts their catalytic activities and needs
alloying for practical implementations\citep{frankel2013evolution,frankel2015introductory}.
Because of these limitations of metal-based catalysts, it is of utmost
importance to explore the metal-free catalyst for HER.

Recent progress in synthesizing materials at the nanoscale allows
us to explore the low-dimensional materials so as to exploit their
unique properties from the perspective of catalysis \citep{jun2013three}.
Numerous studies of 2D nanomaterials have been devoted to overcome
the low catalytic stability of materials towards HER \citep{hasani2019two,voiry2016recent}.
\textcolor{black}{Carbon-based materials are effective alternatives
for applications in diverse areas such as electrocatalysis, microwave
absorption materials, impedance matching, etc.\citep{zheng2023core,wang2023ion,wang2023heterostructure,lim2015carbon}.}
The well-developed carbon-based nanomaterials encapsulated or decorated
with non-metals or metal-rich nanoparticles, show excellent catalytic
activity \citep{paunovic2007effect}. Combined carbon and nitrogen
based electrocatalysts are of great interest due to their non-corrosive,
low cost, and unique properties and have been explored widely to achieve
better HER performance\citep{pachaiappan2021recent,zheng2014hydrogen}.
Recently, graphitic carbon nitride (g-$\mathrm{C_{3}N_{4})}$ structures
have been shown to have attractive electrocatalytic properties because
of their electron-rich nature, H-bonding capabilities, and enhanced
surface characteristics \citep{chen2020review,hong2017rational,han2017graphene}.
Besides 1D and 2D nanomaterials, 0D structures, i.e., quantum dots
(QDs) have also attracted significant interest because of their size-dependent
excited-state properties\citep{brus1984electron,jindal2023first},
charge transfer\citep{wu2016quantum} and surface characteristics
\citep{fan2019quantum,liu2021graphene}. Because of these desirable
characteristics, QDs are potential competitors in the fields of energy
conversion\citep{mohanty2021mxene,molaei2020optical}, optoelectronics\citep{arakawa2002progress,litvin2017colloidal},
sensors\citep{frasco2009semiconductor,raeyani2020optical,lesiak2019optical},
photocatalysis\citep{li2012carbon,lim2015carbon}, \emph{etc}. In
particular, they have been extensively researched for their HER electrocatalytic
and photocatalytic activities\citep{jindal2022density,mohanty2020mos2}.
\textcolor{black}{Various other QDs have been reported for their catalytic
performance which includes inorganic QDs such as TM dichalcogenides
QDs, TiO$_{2}$ QDs, CdS QDs, etc. and carbon QDs (CQDs) of different
morphologies for high performance HER\citep{sargin2019green,gogoi2020enhanced,wang2019novel,meng2020carbon}.
Inorganic QDs are known for high efficiency and widely used for photoelectrochemical
devices, however, their use is limited due to the toxicity and scarcity
of elements such as Pb, Cd, Ag, In, etc.\citep{wang2021colloidal}.
Besides having comparable optical characteristics to inorganic QDs,
CQDs possess favorable attributes such as low toxicity, eco-friendliness,
affordability, and easily attainable synthetic pathways\citep{wang2023doped,lim2015carbon}.
The properties of these QDs can be easily modified by substitutional
doping or surface functionalization\citep{meng2023self}. These characteristics
render them suitable for diverse applications such as bioimaging,
drug delivery, sensing, and catalysis\citep{meng2023self}.}\textcolor{blue}{{}
}Despite the intense research interest in 2D carbon nitride nanomaterials
for boosting the HER efficiency, so far its QDs have not been studied
for HER.

The novel properties of penta-CN$_{2}$ 2D-nanosheet first proposed
by Zhang \citep{zhang2016beyond}, and its 1D nanoribbons studied
by Cheng and He\citep{cheng2018electronic}, have inspired us to study
0D structures, i.e., QDs to investigate their possible applications
as HER catalysts\citep{zhang2016beyond}. Reducing the dimensionality
of $\mathrm{penta-CN_{2}}$ nanosheet to QDs offers edge active and
highly exposed surface sites, which is likely to improve its HER catalytic
performance \citep{smith2010semiconductor}. In this work, we study
rationally designed penta-$\mathrm{\mathrm{CN_{2}}}$ QDs using a
formalism based on first-principles density-functional theory (DFT),
and show that they have unique structural, electronic, and catalytic
properties. The edge dangling bonds of the considered QDs are passivated
by H atoms to prevent the edge reconstruction\citep{smith2010semiconductor}.
The C and N both exhibit $\mathrm{sp^{3}}$ hybridization having a
highly puckered structure, with $50\%$ higher N weight as compared
to C. For a QD of the size 3$\times$3, we obtained an overpotential
of 158 mV which is lower than most of the CN based systems such as
CN nanoribbons (207 mV)\citep{zhao2014graphitic}, BCN sheet decorated
by graphene capsules (333 mV)\citep{liu2021graphene}, a hybrid composed
of graphene oxide and graphitic carbon nitride (180 mV)\citep{shabnam2020doping},
and mesoporous graphitic carbon nitride electrocatalyst (272 mV)\citep{idris2019mesoporous}.
Thus, our calculations indicate that penta-CN$_{2}$ QDs can be excellent
metal-free catalysts for HER.

\section{Computational Methods}

The present work employed the computational chemistry software Gaussian16\citep{frisch2016gaussian}
to calculate the electronic, structural, and HER catalytic properties
of $\mathrm{\mathrm{penta-CN_{2}}}$ QDs within the framework of \emph{ab-initio}
density functional theory (DFT). The hybrid exchange-correlation functional
B3LYP integrated with Becke’s three parameters (B3) exchange functional\citep{becke1988density}
and Lee, Yang, and Parr (LYP) non-local correlation functional\citep{lee1988development}
was chosen in all the calculations. The split valence basis set 6-31G(d,p)
with B3LYP has been found to be adequate for obtaining reliable results
for H-adsorption \citep{jindal2022density}. For the structure visualization
and $\mathrm{\mathrm{penta-CN_{2}}}$ QD input file generation Gaussview06\citep{dennington2016gaussview}
was used. The vibrational frequency calculations were performed on
all the geometry-optimized QDs, and no imaginary frequencies were
found indicating that all the considered structures are dynamically
stable. Gaussian16 formatted checkpoint file is used to generate the
total density of states (TDOS), partial density of states (PDOS),
and Mulliken charge analysis plots using Multiwfn software\citep{lu2012multiwfn}.
The results of the DFT calculations also yield quantities like HOMO-LUMO
gap $\mathrm{(E_{g}})$, chemical potential $(\mu$), chemical softness
$(S)$, etc.

We performed calculations on three different-sized QD structures constructed
from penta-CN$_{2}$ supercells of sizes $3\times3$, $3\times4$,
and $4\times4$ to understand the effect of size on their structural,
electronic and the HER related properties. In the main paper we have
presented figures corresponding only to the final optimized geometries
of the $3\times3$ QD, with and without the adsorbed hydrogen atom.
The figures showing the initial configurations for starting the geometry
optimization process for all the QDs, along with the optimized geometries
of the $3\times4$ and $4\times4$ QDs, are presented in Figs. S1--S7
of supporting information (SI). An H adatom configurations is defined
by the label of the atom on which the hydrogen atom is adsorbed. Thus,
N2 configuration implies that the adsorbed H (circled red) atom is
located on top of the nitrogen atom labeled as 2. For the $3\times3$
QD, all possible H adatom configurations, i.e., on the top of N, C,
and center of pentagon are labeled as NX, CY, and PZ (where X=1-12,
Y=13-18, Z=1-6) (see Fig. S3) were studied to rigorously analyze its
behavior. According to our calculations, for all the QDs considered
in this work, the initial and the final geometries have C$_{1}$ point
group, i.e. no symmetry element is present.

Further, the formation energy per atom $E_{f}$ of a $\mathrm{\mathrm{penta-CN_{2}}}$
QD is calculated to check its structural stability using the formula
\begin{equation}
E_{f}=\frac{1}{N_{atom}}(E_{penta-CN_{2}}-n_{H}E_{H}-n_{C}E_{C}-n_{N}E_{N}),\label{eq:formation}
\end{equation}

where $n_{H},\:n_{C},\:n_{N},\:N_{atom}$, respectively, represent
the numbers of hydrogen, carbon, nitrogen, and all the atoms present
in the given QD. Furthermore, $E_{penta-CN_{2}}$ denotes the total
energy of the considered QD, while $E_{H}$, $E_{C}$, and $E_{N}$
represent the energies of the isolated hydrogen, carbon, and nitrogen
atom, respectively.

Various HER parameters are studied to check the catalytic activity
of designed $\mathrm{\mathrm{penta-CN_{2}}}$ QD. The initial parameter
for HER performance is the adsorption energy $\Delta E_{ads}$ calculated
as
\begin{equation}
\Delta E_{\mathrm{ads}}=E_{penta-CN_{2}+H}-(E_{penta-CN_{2}}+E_{H}),\label{eq:adsorption_energy}
\end{equation}

where $E_{\mathrm{penta-CN_{2}}}$ is the optimized energy of the
$\mathrm{penta-CN_{2}}$ QD, $E_{\mathrm{H}}$ is the energy of isolated
hydrogen atom, and $E_{penta-CN_{2}\,+\,H}$ is the energy of hydrogen
adsorbed $\mathrm{penta-CN_{2}}$ QD. The Gibb's free energy difference
$\Delta G$ is the conventional descriptor to analyze the HER catalytic
activity can be obtained as\citep{norskov2005trends}
\begin{equation}
\Delta G=\Delta E_{ads}+\Delta E_{ZPE}-T\Delta S_{H},\label{eq:Gibbs_Free_energy}
\end{equation}

where, $\Delta E_{ZPE}$ represents the zero point energy corrections
can be estimated as 
\begin{equation}
\Delta E_{ZPE}=E_{ZPE}^{nH}-\mathit{E_{ZPE}^{(n-1)H}-\frac{1}{2}E_{\mathrm{ZPE}}^{H_{2}}.}\label{eq:zero_point_energy}
\end{equation}

Here, $\mathrm{E_{\mathrm{ZPE}}^{\mathrm{nH}}}$ and $\mathrm{E_{\mathrm{ZPE}}^{(\mathrm{n-1)H}}}$
are the zero point energies of $n$ and $n-1$ adsorbed H on top of
$\mathrm{\mathrm{penta-CN_{2}}}$ QD, whereas $E_{ZPE}^{H_{2}}$ is
the zero point energy of gas phase $\mathrm{H_{2}}$ molecule. The
$\mathrm{\Delta E_{\mathrm{ZPE}}}$ lies between 0.01 to 0.04 eV for
$\mathrm{H_{2}}$\citep{tsai2014tuning}. The last $T\Delta S_{H}$
term can be calculated as $\approx\frac{1}{2}T\Delta S_{H_{2}}$,
and has value 0.205 eV at the standard conditions\citep{norskov2005trends},
where, $\Delta S$ is the entropy change. As per the $\mathrm{N\cancel{o}rskov}$
approximation\citep{norskov2005trends}, the $\Delta E_{ZPE}-T\Delta S$
can be replaced by a constant, and $\Delta G$ can be approximated
as (in eV)
\begin{equation}
\Delta G=\Delta E_{ads}+0.24.\label{eq:net_deltaG}
\end{equation}

The free energy change determines the rate of HER and should be close
to zero for the best catalytic performance\citep{sabatier1911hydrogenations}.
To theoretically estimate the potential barrier encountered by HER,
the overpotential is calculated as:

\begin{equation}
\varphi=-\frac{|\Delta G|}{e}.\label{eq:overpotential}
\end{equation}

In addition to these parameters, exchange current density $i_{0}$
is another parameter to measure the catalytic activity of a penta-CN$_{2}$
QD towards HER. A high value of exchange current density indicates
better catalytic activity, and the following relation can be used
to calculate its value\citep{norskov2005trends}

\begin{equation}
\mathit{i_{0}}=-e\mathit{k_{0}}\frac{1}{1+\exp(|\Delta G|/k_{B}T)},\label{eq:exchange current}
\end{equation}

where, $k_{B}$ is the Boltzmann constant, and $k_{0}$ represents
the unknown rate constant set to one for the calculation purposes.

\section{Results and Discussion}

In this section we present and discuss the results of our calculations
performed on $3\times3$, $3\times4$, and $4\times4$ penta-CN$_{2}$
QDs. We have discussed results on the $3\times3$ QD in greater detail,
while, to avoid repetition, those on $3\times4$, and $4\times4$
QDs are presented in the SI.

\subsection{Optimized Structures}

\subsubsection{Pristine QDs}

\begin{figure}[H]
\centering{}\includegraphics[scale=0.5]{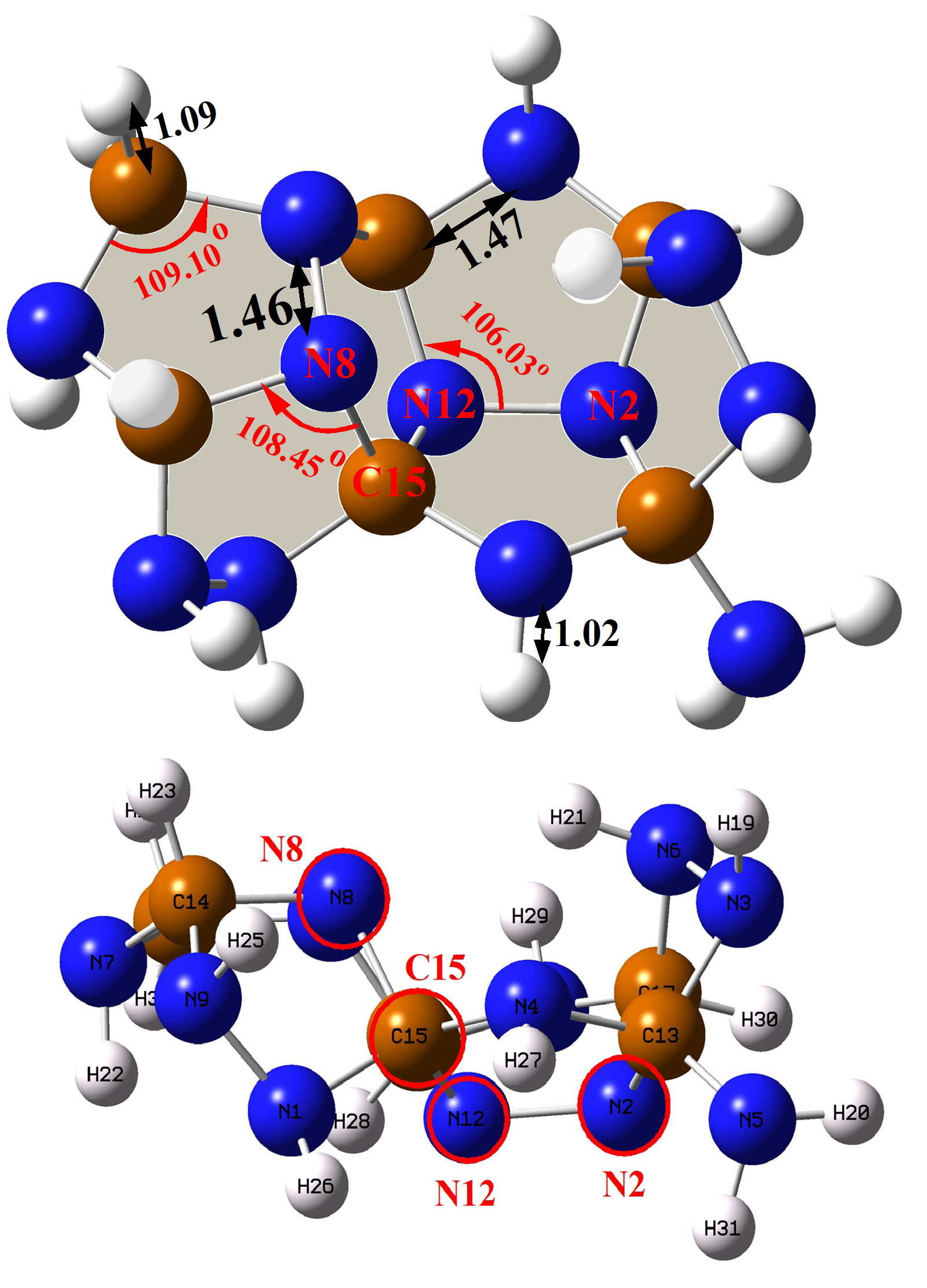}\caption{\label{fig:optimized CN2-penta}The top and side views of the optimized
structure of $3\times3$ penta-CN$_{2}$ QD, along with the atom labels.
Brown, blue, and gray spheres denote carbon, nitrogen, and hydrogen
atoms, respectively. In the top view, the average bond lengths and
bond angles are also indicated.}
\end{figure}

The buckled structure of $3\times3$ penta-CN$_{2}$ QD consists of
six pentagons with three N and two C atoms in each pentagon, while
the edge atoms are passivated with hydrogens to saturate the dangling
bonds. In this structure, the first 12 labels correspond to nitrogen
(N), 13-18 labels are assigned to carbon (C), and 19-32 labels are
used for hydrogen (H). For each QD, the initial structures, using
which the geometry optimization was started, are presented in Fig.
S1 of SI. The final optimized structure of the $3\times3$ QD is presented
in Fig. \ref{fig:optimized CN2-penta}, while those of the bigger
QDs are given in Fig. S2 of the SI. From the side view of the optimized
structures of the QDs it is obvious that, as compared to the initial
configurations, after optimization there are distortions on the H-passivated
edges. The average bond lengths and bond angles of the optimized structure
are also indicated in the top view of Fig. \ref{fig:optimized CN2-penta},
from which it is obvious that there are no C-C bonds in the system,
whereas the QD has four N-N bonds. All the individual bond lengths
and bond angles of the $3\times3$ QD are presented, respectively,
in Tables S1 and S2 of the SI.

In the optimized structure of the $3\times3$ QD, the N-N bond lengths
are found to be in the range 1.44--1.48 $\text{Å}$, with the average
of 1.46 $\text{Å}$. Similarly, C-N, N-H, and C-H bond lengths are
in the ranges 1.44--1.52 $\text{Å}$, 1.01--1.03 $\text{Å}$,
and 1.09--1.10 $\text{Å}$, with the corresponding average values
of 1.47 $\text{Å}$, 1.02 $\text{Å}$, and 1.09 $\text{Å}$, respectively.
Thus, all the average bond lengths indicate $\sigma$-type single
bond characteristics, with the values in good agreement with the existing
literature\citep{cheng2018electronic,zhang2016beyond}. In the pentagons,
three main bond angles present are $\mathrm{\angle NNC}$, $\mathrm{\angle CNC}$,
and $\mathrm{\angle NCN}$, and their average values are $106^{\mathrm{o}}$,
$\mathrm{108.4^{o}}$, and $\mathrm{109.1^{o}}$, respectively. Clearly,
these values are fairly close to the angle 108$^{\text{°}}$ of a
perfect pentagon. From Table S2 of the SI we can conclude that the
bond angles minimum and maximum (min., max.) values for the $\angle\mathrm{NNC}$,
$\mathrm{\angle CNC}$, and $\mathrm{\angle NCN}$ are ($\mathrm{98.3^{o}}$,
$\mathrm{109.6^{o}}$), ($\mathrm{103.6^{o}}$, $\mathrm{118.1^{o}}$),
and ($\mathrm{102.2^{o}}$, $\mathrm{117.3^{o}}$), respectively.
This large variation in bond angles demonstrates the non-planar nature
of the system. 

\subsubsection{H-adsorbed QDs}

\begin{figure}[H]
\centering{}\includegraphics[scale=0.5]{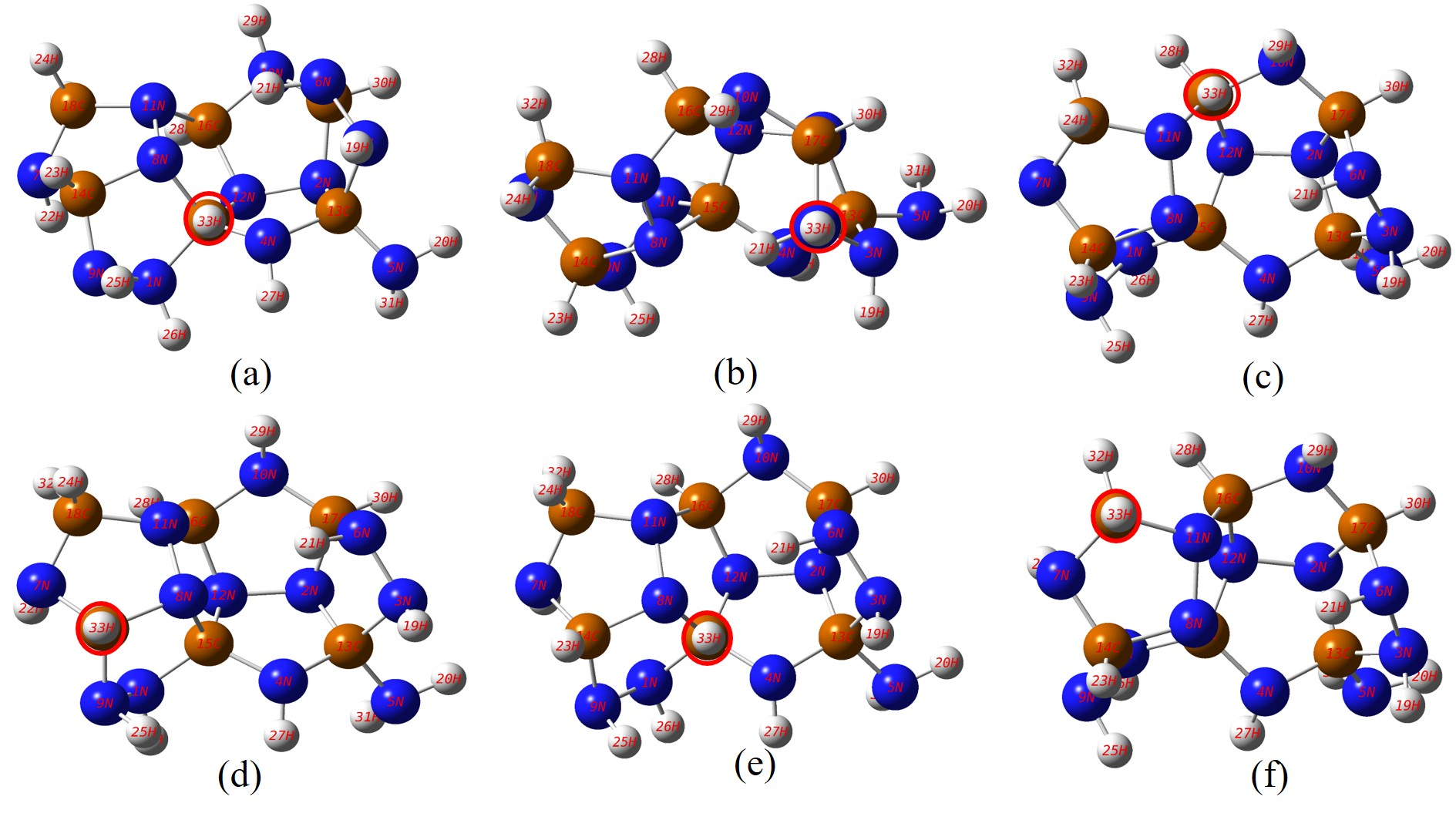}\caption{\label{fig:Final-structure-H-adsorbed}Final structures of $3\times3$
penta-CN$_{2}$ QD in which initially H atoms was placed on the top
of \textcolor{black}{N4, N6, N11 C14, C15, and C18 sites, }where carbon,
nitrogen and hydrogen atoms are shown in brown, blue, and gray colors,
respectively.}
\end{figure}

In order to assess the HER performance of the penta-CN$_{2}$ QDs,
we considered their hydrogen-adsorbed configurations. The geometries
of the H-adsorbed QDs were optimized after placing the H atom $\sim2\text{Å}$
above the chosen site of a QD, using its optimized geometry discussed
in the previous section. The optimized geometries of the pristine
discussed in the previous section were used for optimizing their hydrogen-adsorbed
structures. The initial configurations of the $3\times3$, $3\times4$,
and $4\times4$ H-adsorbed QDs are presented, respectively, in Figs.
S3, S4, and S5, of the SI. \textcolor{black}{For the $3\times3$ QD,
all distinct atomic positions were considered for H adsorption to
comprehensively analyze the localized active sites of penta-CN$_{2}$
QDs in the context of HER. For a concise and clear presentation, only
four configurations were provided in the SI (Fig. S3).} The top view
of the optimized H-adsorbed structures of the $3\times3$ QD are presented
in Fig. \ref{fig:Final-structure-H-adsorbed}, in which the adsorbed
H atoms are eclipsing the atoms on the sites N4, N6, N11, C14, C15
and C18. In Table \ref{tab:Vertical_distances-1} we present the vertical
distance of the H atom corresponding to each adsorption site in the
relaxed configuration of the $3\times3$ QD and note that it lies
between 2.17-2.46 Å for N4, N6 and N11 configurations. While in case
of H-adsorbed on the top of carbon atom, post relaxation H atom moves
at least 1 Å further away from the QD as compared to the initial configurations.
Final, optimized configurations for various H adsorption sites for
the $3\times4$, and $4\times4$ penta-CN$_{2}$ QDs are presented,
respectively, in Figs. S6 and S7 of the SI.

\begin{table}[H]
\centering{}%
\begin{tabular}{cccc}
\toprule 
\begin{tabular}{c}
Initial Adsorption \tabularnewline
Site\tabularnewline
\end{tabular} & %
\begin{tabular}{c}
Final Adsorption\tabularnewline
Site\tabularnewline
\end{tabular} & %
\begin{tabular}{c}
Initial Distance \tabularnewline
($\text{Å}$)\tabularnewline
\end{tabular} & %
\begin{tabular}{c}
Final Distance\tabularnewline
($\text{Å}$)\tabularnewline
\end{tabular}\tabularnewline
\midrule
\midrule 
\begin{tabular}{c}
$\mathrm{N1}$\tabularnewline
\end{tabular} & %
\begin{tabular}{c}
Unstable\tabularnewline
\end{tabular} & %
\begin{tabular}{c}
2\tabularnewline
\end{tabular} & %
\begin{tabular}{c}
-\tabularnewline
\end{tabular}\tabularnewline
\begin{tabular}{c}
$\mathrm{N2}$\tabularnewline
\end{tabular} & %
\begin{tabular}{c}
$\mathrm{N4}$\tabularnewline
\end{tabular} & %
\begin{tabular}{c}
2\tabularnewline
\end{tabular} & %
\begin{tabular}{c}
2.46\tabularnewline
\end{tabular}\tabularnewline
\begin{tabular}{c}
$\mathrm{N3}$\tabularnewline
\end{tabular} & %
\begin{tabular}{c}
$\mathrm{C15}$\tabularnewline
\end{tabular} & %
\begin{tabular}{c}
2\tabularnewline
\end{tabular} & %
\begin{tabular}{c}
3.15\tabularnewline
\end{tabular}\tabularnewline
\begin{tabular}{c}
$\mathrm{N4}$\tabularnewline
\end{tabular} & %
\begin{tabular}{c}
$\mathrm{N4}$\tabularnewline
\end{tabular} & %
\begin{tabular}{c}
2\tabularnewline
\end{tabular} & %
\begin{tabular}{c}
2.46\tabularnewline
\end{tabular}\tabularnewline
\begin{tabular}{c}
$\mathrm{N5}$\tabularnewline
\end{tabular} & %
\begin{tabular}{c}
$\mathrm{N4}$\tabularnewline
\end{tabular} & %
\begin{tabular}{c}
2\tabularnewline
\end{tabular} & %
\begin{tabular}{c}
2.46\tabularnewline
\end{tabular}\tabularnewline
\begin{tabular}{c}
$\mathrm{N6}$\tabularnewline
\end{tabular} & %
\begin{tabular}{c}
$\mathrm{N6}$\tabularnewline
\end{tabular} & %
\begin{tabular}{c}
2\tabularnewline
\end{tabular} & %
\begin{tabular}{c}
2.3\tabularnewline
\end{tabular}\tabularnewline
\begin{tabular}{c}
$\mathrm{N7}$\tabularnewline
\end{tabular} & %
\begin{tabular}{c}
$\mathrm{C14}$\tabularnewline
\end{tabular} & %
\begin{tabular}{c}
2\tabularnewline
\end{tabular} & %
\begin{tabular}{c}
3.59\tabularnewline
\end{tabular}\tabularnewline
\begin{tabular}{c}
$\mathrm{N8}$\tabularnewline
\end{tabular} & %
\begin{tabular}{c}
$\mathrm{C14}$\tabularnewline
\end{tabular} & %
\begin{tabular}{c}
2\tabularnewline
\end{tabular} & %
\begin{tabular}{c}
3.68\tabularnewline
\end{tabular}\tabularnewline
\begin{tabular}{c}
$\mathrm{N9}$\tabularnewline
\end{tabular} & %
\begin{tabular}{c}
$\mathrm{N4}$\tabularnewline
\end{tabular} & %
\begin{tabular}{c}
2\tabularnewline
\end{tabular} & %
\begin{tabular}{c}
2.46\tabularnewline
\end{tabular}\tabularnewline
\begin{tabular}{c}
$\mathrm{N10}$\tabularnewline
\end{tabular} & %
\begin{tabular}{c}
$\mathrm{N11}$\tabularnewline
\end{tabular} & %
\begin{tabular}{c}
2\tabularnewline
\end{tabular} & %
\begin{tabular}{c}
2.17\tabularnewline
\end{tabular}\tabularnewline
\begin{tabular}{c}
$\mathrm{N11}$\tabularnewline
\end{tabular} & %
\begin{tabular}{c}
$\mathrm{N11}$\tabularnewline
\end{tabular} & %
\begin{tabular}{c}
2\tabularnewline
\end{tabular} & %
\begin{tabular}{c}
2.17\tabularnewline
\end{tabular}\tabularnewline
\begin{tabular}{c}
$\mathrm{N12}$\tabularnewline
\end{tabular} & %
\begin{tabular}{c}
$\mathrm{C15}$\tabularnewline
\end{tabular} & %
\begin{tabular}{c}
2\tabularnewline
\end{tabular} & %
\begin{tabular}{c}
3.15\tabularnewline
\end{tabular}\tabularnewline
\begin{tabular}{c}
$\mathrm{C13}$\tabularnewline
\end{tabular} & %
\begin{tabular}{c}
Unstable\tabularnewline
\end{tabular} & %
\begin{tabular}{c}
2\tabularnewline
\end{tabular} & %
\begin{tabular}{c}
-\tabularnewline
\end{tabular}\tabularnewline
\begin{tabular}{c}
$\mathrm{C14}$\tabularnewline
\end{tabular} & %
\begin{tabular}{c}
$\mathrm{C15}$\tabularnewline
\end{tabular} & %
\begin{tabular}{c}
2\tabularnewline
\end{tabular} & %
\begin{tabular}{c}
3.15\tabularnewline
\end{tabular}\tabularnewline
\begin{tabular}{c}
$\mathrm{C15}$\tabularnewline
\end{tabular} & %
\begin{tabular}{c}
$\mathrm{C15}$\tabularnewline
\end{tabular} & %
\begin{tabular}{c}
2\tabularnewline
\end{tabular} & %
\begin{tabular}{c}
3.15\tabularnewline
\end{tabular}\tabularnewline
\begin{tabular}{c}
$\mathrm{C16}$\tabularnewline
\end{tabular} & %
\begin{tabular}{c}
$\mathrm{N11}$\tabularnewline
\end{tabular} & %
\begin{tabular}{c}
2\tabularnewline
\end{tabular} & %
\begin{tabular}{c}
2.17\tabularnewline
\end{tabular}\tabularnewline
\begin{tabular}{c}
$\mathrm{C17}$\tabularnewline
\end{tabular} & %
\begin{tabular}{c}
$\mathrm{N6}$\tabularnewline
\end{tabular} & %
\begin{tabular}{c}
2\tabularnewline
\end{tabular} & %
\begin{tabular}{c}
2.3\tabularnewline
\end{tabular}\tabularnewline
\begin{tabular}{c}
$\mathrm{C18}$\tabularnewline
\end{tabular} & %
\begin{tabular}{c}
$\mathrm{C18}$\tabularnewline
\end{tabular} & %
\begin{tabular}{c}
2\tabularnewline
\end{tabular} & %
\begin{tabular}{c}
3.55\tabularnewline
\end{tabular}\tabularnewline
\begin{tabular}{c}
$\mathrm{P1}$\tabularnewline
\end{tabular} & %
\begin{tabular}{c}
$\mathrm{C14}$\tabularnewline
\end{tabular} & %
\begin{tabular}{c}
2\tabularnewline
\end{tabular} & %
\begin{tabular}{c}
3.59\tabularnewline
\end{tabular}\tabularnewline
\begin{tabular}{c}
$\mathrm{P2}$\tabularnewline
\end{tabular} & %
\begin{tabular}{c}
$\mathrm{C15}$\tabularnewline
\end{tabular} & %
\begin{tabular}{c}
2\tabularnewline
\end{tabular} & %
\begin{tabular}{c}
3.15\tabularnewline
\end{tabular}\tabularnewline
\begin{tabular}{c}
$\mathrm{P3}$\tabularnewline
\end{tabular} & %
\begin{tabular}{c}
$\mathrm{C15}$\tabularnewline
\end{tabular} & %
\begin{tabular}{c}
2\tabularnewline
\end{tabular} & %
\begin{tabular}{c}
3.15\tabularnewline
\end{tabular}\tabularnewline
\begin{tabular}{c}
$\mathrm{P4}$\tabularnewline
\end{tabular} & %
\begin{tabular}{c}
No Convergence\tabularnewline
\end{tabular} & %
\begin{tabular}{c}
2\tabularnewline
\end{tabular} & %
\begin{tabular}{c}
3.55\tabularnewline
\end{tabular}\tabularnewline
\begin{tabular}{c}
$\mathrm{P5}$\tabularnewline
\end{tabular} & %
\begin{tabular}{c}
$\mathrm{N4}$\tabularnewline
\end{tabular} & %
\begin{tabular}{c}
2\tabularnewline
\end{tabular} & %
\begin{tabular}{c}
2.46\tabularnewline
\end{tabular}\tabularnewline
\begin{tabular}{c}
$\mathrm{P6}$\tabularnewline
\end{tabular} & %
\begin{tabular}{c}
$\mathrm{C18}$\tabularnewline
\end{tabular} & %
\begin{tabular}{c}
2\tabularnewline
\end{tabular} & %
\begin{tabular}{c}
3.55\tabularnewline
\end{tabular}\tabularnewline
\bottomrule
\end{tabular}\caption{\label{tab:Vertical_distances-1}The initial and final adsorption
sites of the H atoms along with the distances from the closest atom
of the $3\times3$ penta-CN$_{2}$ QD. Above, a site P$n$ implies
the center of the $n$th pentagon of the QD, while\textquotedblleft unstable\textquotedblright{}
means the presence of imaginary vibrational frequencies.}
\end{table}

\subsection{Formation Energies and the HOMO-LUMO Gaps of Pristine QDs}

Having established the structural stability of the optimized geometries
of penta-CN$_{2}$ QDs using the vibrational frequency analysis, next
we explore their thermodynamic stability by computing their formation
energies. We used Eq. \ref{eq:formation} and the values obtained
in our DFT calculations performed using the B3LYP functional to compute
their formation energies, and the results are presented in Table \ref{tab:all-formation-energies}.
Additionally, in the same table we also present the values of the
HOMO-LUMO gaps ($E_{g}$) obtained in the same calculations performed
on the three structures. We note that formation energies for the $3\times3$,
$3\times4$, and $4\times4$ QDs are -5.529, -5.718, and -5.906 eV,
respectively. The magnitude of formation energies clearly indicate
that the penta-CN$_{2}$ QDs are thermodynamically stable\citep{sarkar2018density}.
Furthermore, increasing magnitudes of the formation energies with
the increasing sizes of the QDs indicate that there should not be
any problems in synthesizing these structures in the laboratory. The
band gaps show oscillatory behavior with respect to the increasing
size, with the gap of the $3\times4$ QD being larger than that of
the smaller ($3\times3$) structure. Nevertheless, for the symmetric
structures we observe the expected trend, i.e., $E_{g}(3\times3)>E_{g}(4\times4)$,
a clear indicator of the finite-size induced quantum confinement.
The band gap of $\mathrm{\mathrm{penta-CN_{2}}}$ nanosheet has been
reported in the literature to be 6.53 eV, computed using the HSE06
functional\citep{zhang2016beyond}. Therefore, the larger gaps of
finite-sized $\mathrm{penta-CN_{2}}$ QDs as compared to the infinite
sheet are also clearly due to quantum confinement.\citep{onyia2018theoretical}
Furthermore, larger H-L gaps of $\mathrm{\mathrm{penta-CN_{2}}}$
QDs as compared to the nanosheet ensure comparatively better kinetic
stability for the predicted QDs\citep{hassanpour2021kinetic}.

\begin{table}[H]
\centering{}%
\begin{tabular}{cccc}
\toprule 
\addlinespace[1mm]
Size & Formula & $E_{f}(eV)$ & $E_{g}(eV)$\tabularnewline\addlinespace[1mm]
\midrule
\addlinespace[1mm]
\addlinespace[1mm]
$3\times3$ & $\mathrm{C_{6}N_{12}H_{14}}$ & -5.529 & \textcolor{black}{7.004}\tabularnewline\addlinespace[1mm]
\addlinespace[1mm]
\addlinespace[1mm]
$3\times4$ & $\mathrm{C_{9}N_{18}H_{18}}$ & -5.718 & 7.164\tabularnewline\addlinespace[1mm]
\addlinespace[1mm]
\addlinespace[1mm]
$4\times4$ & $\mathrm{C_{12}N_{24}H_{20}}$ & -5.906 & 6.919\tabularnewline\addlinespace[1mm]
\bottomrule
\addlinespace[1mm]
\end{tabular}\caption{\label{tab:all-formation-energies}Formation energies ($E_{f}$) and
the HOMO-LUMO gaps ($E_{g}$) of $3\times3$, $3\times4$, $\mbox{ and }4\times4$
$\mathrm{\mathrm{penta-CN_{2}}}$ QDs}
\end{table}

\subsection{Other Electronic Properties}

In order to understand the chemical reactivity and the catalytic properties
of penta-CN$_{2}$ QDs, we discuss their electronic structure next.
The highest occupied molecular orbital (HOMO) and lowest unoccupied
molecular orbital (LUMO) energy values determine a number of other
useful global descriptors such as ionization potential ($IP$), electron
affinity ($EA$), chemical hardness($\mathrm{\eta}$), chemical softness
($S)$, chemical potential ($\mu$), electronegativity ($\chi$),
and electrophilicity ($\omega$)\citep{kumar2011chemical,kumar2013understanding}.
Besides the aforementioned properties, the total density of states
(TDOS), the partial density of states(PDOS), and Mulliken charge analysis
are further investigated to obtain deeper insights into the electronic
structure of $\mathrm{\mathrm{penta-CN_{2}}}$ QDs. Although, for
DFT, the Koopmans' theorem does not hold, nevertheless we have approximated
the $IP$ and $EA$ values of these QDs using that 
\begin{gather}
\begin{array}{cc}
IP & \simeq-E_{HOMO}\\
EA & \simeq-E_{LUMO}
\end{array}\label{eq:HOMO-LUMO}
\end{gather}

The H-L gap $E_{g}=E_{LUMO}-E_{HOMO}$ is a measure of the chemical
hardness of the substance in the sense that its large values imply
reduced chemical activity of the molecule\citep{vijayaraj2009comparison}.
Within DFT, the chemical softness is defined as the inverse of half
of the H-L gap, and therefore is calculated as 
\[
S=\frac{2}{E_{g}}
\]
The smaller values of $S$ imply less polarizability of the molecule.
The chemical potential is defined as
\begin{equation}
\mu=\frac{E_{HOMO}+E_{LUMO}}{2},\label{eq:chemical_potential}
\end{equation}
 whose larger values indicate higher molecular reactivity.

Another important parameter that helps in determining the chemical
reactivity is the global electrophilicity index ($\omega$), which
provides information about energy change when the electron makes a
transition from the HOMO level to the LUMO level
\[
\omega=\frac{\mu^{2}}{E_{g}}
\]
\begin{figure}[H]
\centering{}\includegraphics[scale=0.3]{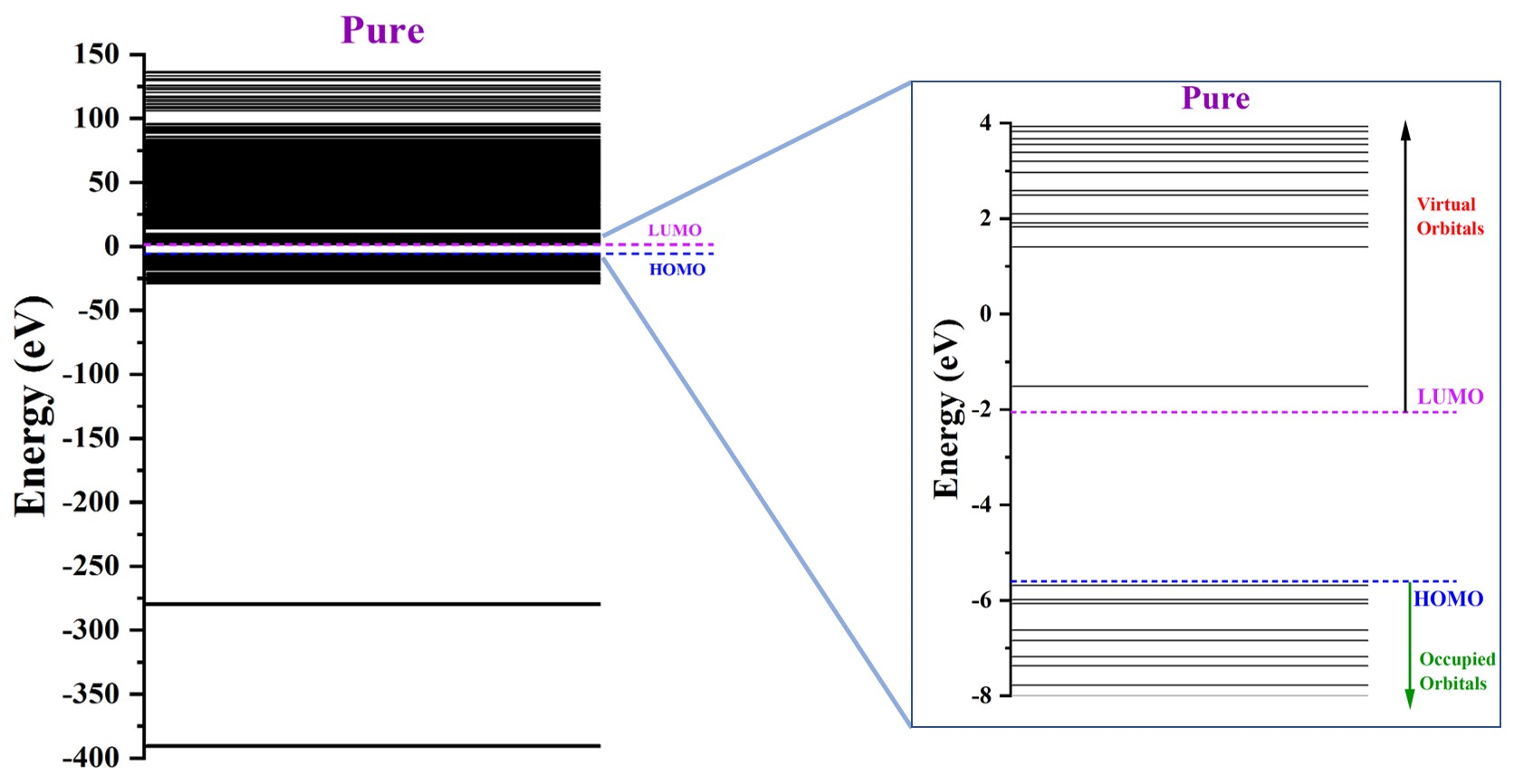}\caption{\label{fig:occupied-virtual-orbitals} The occupied and virtual orbitals
of a $3\times3$ pure $\mathrm{\mathrm{penta-CN_{2}}}$ QD}
\end{figure}

\begin{figure}[H]
\centering{}\includegraphics[scale=0.5]{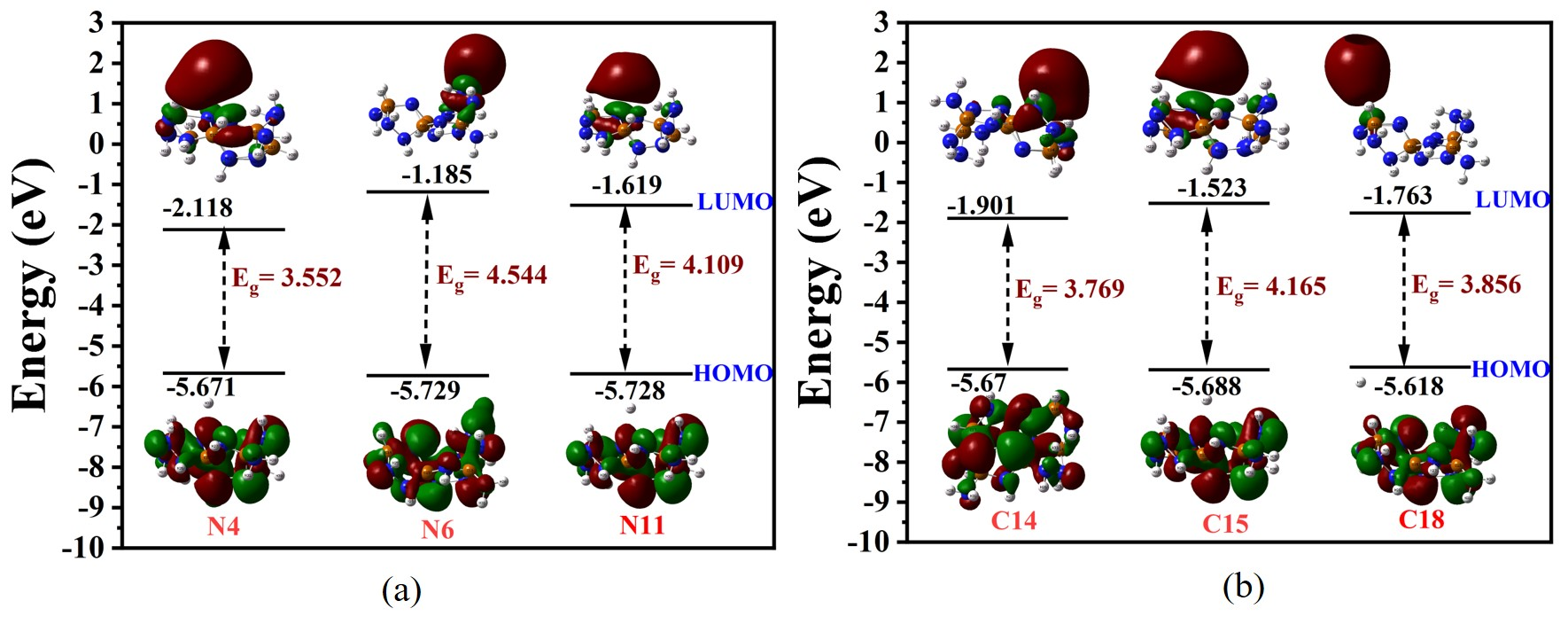}\caption{\label{fig:HOMO-LUMO-of-all}HOMO-LUMO of all configurations of $3\times3$
$\mathrm{penta\:-CN_{2}}$ QD with H adsorbed on the top of N4, N6,
N11, C14, C15 and C18.}
\end{figure}

\textcolor{black}{For the $3\times3$ QD we have plotted the entire
energy-level diagram in Fig. \ref{fig:occupied-virtual-orbitals},
with the levels close to the gap region separately highlighted. The
changes in HOMO values of the H-adsorbed systems are not significant
when compared to that of the pristine penta-CN$_{2}$ QD. The maximum
change after the H adsorption is only 1.8\% when compared to the pristine
penta-CN$_{2}$ QD, which means H adsorption is not affecting the
electron-donor ability of the QDs significantly. However, the LUMO
values reduced immensely after H-adsorption implying that it leads
to significant changes in the electron affinity values as per Eq.
\ref{eq:HOMO-LUMO}. Consequently, enhancing the electron accepting
properties which will be further discussed in subsection }\textbf{\textcolor{black}{3.4.1}}\textcolor{black}{.
Following this, the reduction in the H-L gap ($E_{g}$) of the H-adsorbed
penta-CN$_{2}$ QDs varies in the range 35\%-49\%, as compared to
the pristine QD. The lower values of $E_{g}$, imply higher values
of the electronic conductivity which favors the electrocatalytic behaviour
of the QDs\citep{sharma2022role}. }\\
\textcolor{black}{Table \ref{tab:all_electronic_parameters} illustrates
a substantial enhancement in chemical softness ($S$) for H-adsorbed
systems, signifying increased reactivity and polarizability\citep{vela1990relationship}
compared to the pristine penta-CN$_{2}$ QD (0.286). Notably, N4 (0.563),
C14 (0.531), and C18 (0.519) exhibit comparatively higher $S$ values,
while N6 (0.440), N11 (0.487), and C15 (0.480) show lower $S$ values,
i.e., relatively lower reactivity and polarizability. The chemical
potential values presented in Table \ref{tab:all_electronic_parameters}
demonstrate the influence of the H adsorption on the electropositive
character of penta-CN$_{2}$ QDs\citep{pearson1992chemical}. The
ascending order for the electropositive character can be observed
as N4 < C14 < C18 < N11 < C15 < N6 < pristine. The change in electrophilicity
($\omega$) for the H-adsorbed QDs signifies their tendency for electron
acceptance and engagement in electron transfer processes. The observed
values of 4.269 (N4), 2.630 (N6), 3.284 (N11), 3.802 (C14), 3.121
(C15), and 3.532 (C18) emphasize the enhanced electron-accepting abilities
and reactivity of each H-adsorbed QD, as compared to the pristine
QD which has rather small value of 0.641 eV. $IP$, which is just
the negative of $E_{HOMO}$, refers to the energy required to remove
an electron from the system, and as discussed above, because $E_{HOMO}$
does not change much due to H adsorption, we do not observe any significant
change in the $IP$ values either as provided in SI. From Table \ref{tab:all_electronic_parameters}
it is obvious that $EA$ which is the negative of $E_{LUMO}$, and
represents the electron-accepting abilities of systems, increases
significantly due to H adsorption because of the corresponding changes
in $E_{LUMO}$ discussed above. The plots of the HOMO and LUMO, along
with values of $E_{HOMO}$, $E_{LUMO}$ and $E_{g}$ for the considered
H-adsorbed configurations are presented in Fig. \ref{fig:HOMO-LUMO-of-all}.}

\begin{table}[H]
\centering{}%
\begin{tabular}{cccccccc}
\toprule 
 & $E_{HOMO}$ & $E_{LUMO}$ & $E_{g}$ & $S$ & $\mu$ & $\omega$ & $EA$\tabularnewline
\midrule
\midrule 
pristine & -5.624 & 1.38 & 7.004 & 0.286 & -2.122 & 0.643 & -1.38\tabularnewline
\midrule 
$\mathrm{N4}$ & -5.671 & -2.118 & \textcolor{black}{3.552} & \textcolor{black}{0.563} & -3.895 & 4.269 & 2.118\tabularnewline
\midrule 
$\mathrm{N6}$ & -5.729 & -1.185 & \textcolor{black}{4.544} & \textcolor{black}{0.440} & -3.457 & 2.630 & 1.185\tabularnewline
\midrule 
$\mathrm{N11}$ & -5.728 & -1.619 & 4.109 & 0.487 & -3.674 & 3.284 & 1.619\tabularnewline
\midrule 
$\mathrm{C14}$ & -5.67 & -1.901 & 3.769 & 0.531 & -3.786 & 3.802 & 1.901\tabularnewline
\midrule 
$\mathrm{\mathrm{C15}}$ & -5.688 & -1.523 & 4.165 & 0.480 & -3.606 & 3.121 & 1.523\tabularnewline
\midrule 
$\mathrm{C18}$ & -5.618 & -1.763 & 3.856 & 0.519 & -3.691 & 3.532 & 1.763\tabularnewline
\bottomrule
\end{tabular}\caption{\label{tab:all_electronic_parameters} Global reactivity parameters,
namely, HOMO energy ($E_{HOMO}$), LUMO energy ($E_{LUMO}$), H-L
gap ($E_{g}$), chemical softness ($S$), chemical potential ($\mu$),
electrophilicity index ($\omega$), and electron affinity ($EA$)
for the pristine $3\times3$ penta-CN$_{2}$ QD, and those with H
adsorbed on the sites N4, N6, N11, C14, C15, and C18. Above, $S$
is in the units of eV$^{-1}$, while all other quantities are in eV
units.}
\end{table}

\subsubsection{Density of States \label{subsec:Total-and-Partial}}

\begin{figure}[H]
\centering{}\includegraphics[scale=0.65]{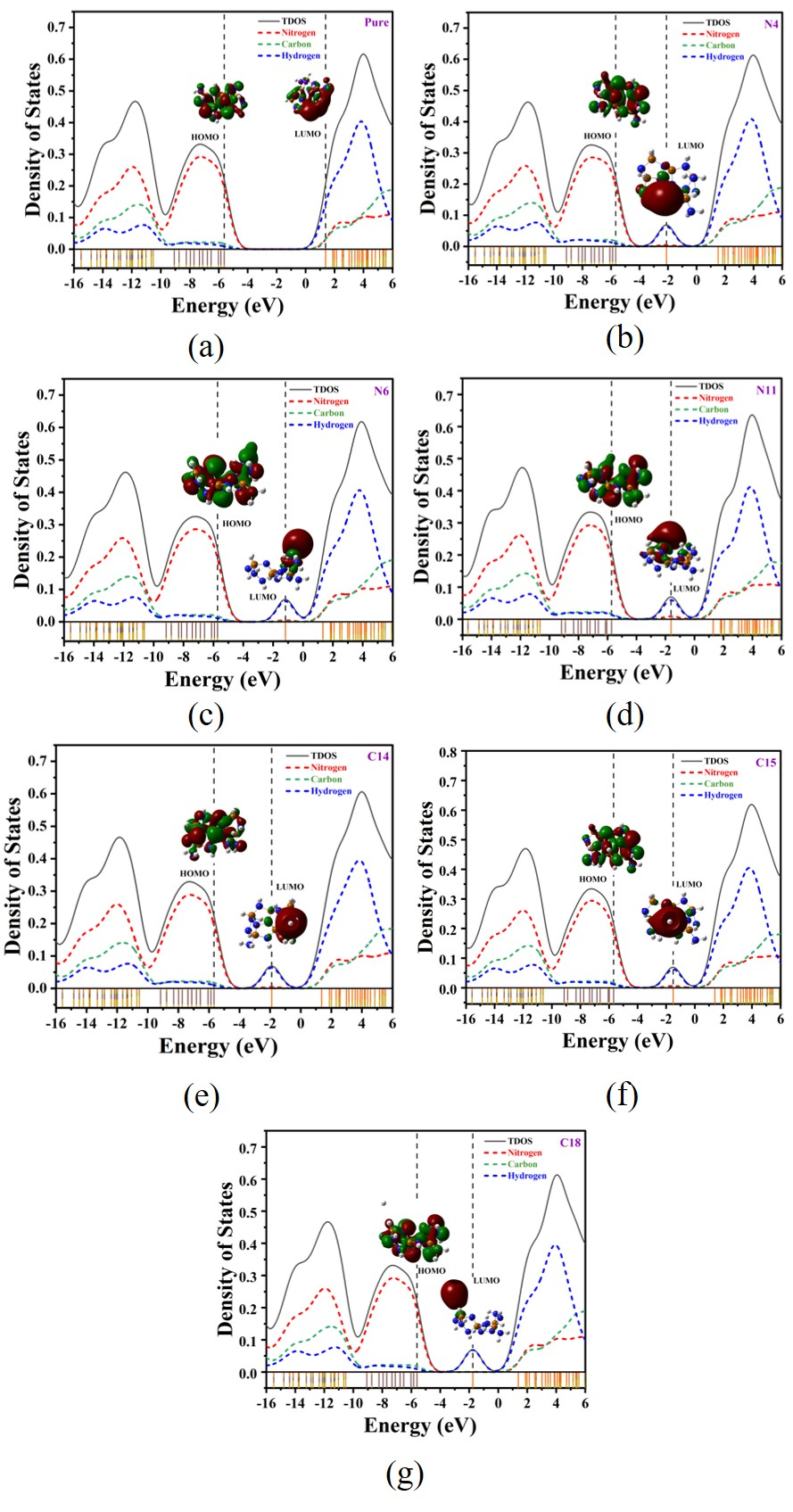}\caption{\label{fig:all_dos} Calculated TDOS and PDOS of $3\times3$ penta-CN$_{2}$
QDs: (a) pristine, and (b)---(g) those with H adsorbed on the sites
N4, N6, N11, C14, C15, and C18, respectively. }
\end{figure}

In order to further understand the changes in the electronic structure
of $3\times3$ penta-CN$_{2}$ QDs due to H adsorption, in this section
we present and discuss the total density of states (TDOS) and the
atom projected density of states (PDOS) of both the pristine as well
as H-adsorbed structures. For all the QDs, DOS was computed using
the multiwfn software\citep{lu2012multiwfn}, and because of the discrete
nature of the energy levels, a Gaussian broadening was employed in
the plots. Nevertheless, in the figures discrete lines indicating
various energy levels are also plotted, with the line height indicating
the level degeneracy. The PDOS helps us understand the contribution
of the individual atoms (C, N, and H), as well as their orbitals (s,
p,...). In Fig. \ref{fig:all_dos}, we represented the TDOS and PDOS
plots for the pristine as well as six unique H-adsorbed configurations
with the adsorption sites N4, N6, N11, C14, C15, and C18.

Although, the DOS plots of the H-adsorbed configurations have features
similar to that of the pristine QD, they do have fingerprints of the
adsorbed H atom. Within the band gap, an extra energy state appears
in $\mathrm{N4}$, $\mathrm{N6}$, $\mathrm{N11}$, $\mathrm{C14}$
$\mathrm{C15}$, and $\mathrm{C18}$ at -2.118 eV, -1.185 eV, -1.619
eV, -1.901 eV, -1.523 eV, and -1.763 eV, respectively. As mentioned
earlier, this indicates an enhancement in the electrocatalytic properties
of the penta-CN$_{2}$ QD. As observed from the orbital diagram of
the LUMO, the antibonding molecular orbital appears due to adsorbed
H labeled H33 in Fig. \ref{fig:all_dos}. Therefore, the adsorbed
H atoms do lead to changes in the H-L gaps, and hence in their electronic
properties.

\subsubsection{Mulliken Charge Analysis \label{subsec:Mulliken-Charge-Analysis}}

\begin{figure}[H]
\centering{}\includegraphics[scale=0.65]{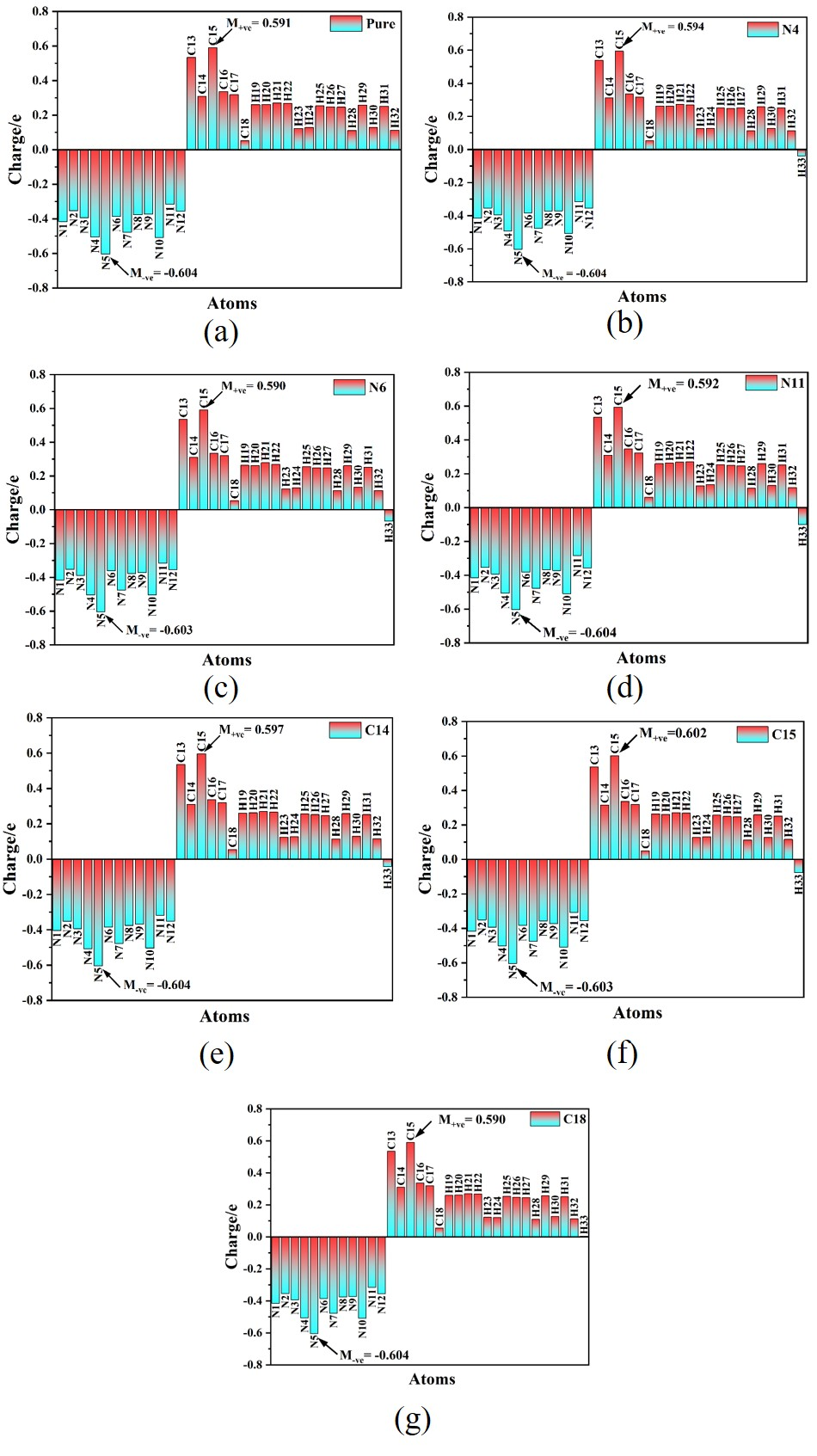}\caption{\label{fig:Mulliken-Charges-all}The Mulliken charges of the $3\times3$
$\mathrm{\mathrm{penta-CN_{2}}}$ (a) pristine QD, (b)---(g) those
with H adsorbed at the sites N4, N6, N11, C15, and C18, respectively.}
\end{figure}

The redistribution of charges occurs after the H adsorption on the
$\mathrm{\mathrm{penta-CN_{2}}}$ QD, and the Mulliken charge analysis
allows us to investigate the charge transfer between the atoms. The
calculated Mulliken charges of the $3\times3$ pristine and H-adsorbed
QDs are presented in Fig. \ref{fig:Mulliken-Charges-all} in which
the first 12 atoms are N, 13--18 are C, 19--32 edge H atoms, while
the 33rd one is the adsorbed H atom. All the edge hydrogens in the
$\mathrm{\mathrm{penta-CN_{2}}}$ QD have positive charges, with those
attached to C (i.e. $\mathrm{H23}$, $\mathrm{H24}$, $\mathrm{H28}$,
$\mathrm{H30}$, $\mathrm{H32}$) have less positive charge in comparison
with the ones attached to N. This clearly is a consequence of the
fact that nitrogen atom is more electronegative as compared to the
carbon atom. The maximum negative Mulliken charge is -0.604$e$ on
N5. All the carbons atoms also have positive charge because of the
charge transfer from them to N atoms to which they are attached. The
value of charge on a C atom depends on the number of $\mathrm{N}$
atoms with which it is bonded. $\mathrm{C13}$ and $\mathrm{C15}$
carbon atoms have the highest positive charges as compared to others
because they are attached to four $\mathrm{N}$ atoms (Zero $\mathrm{H}$),
followed by $\mathrm{C14}$ , $\mathrm{C16}$, and $\mathrm{C17}$
which attached to three $\mathrm{N}$ atoms each, while $\mathrm{C18}$
has the least positive charge because it is attached to just two $\mathrm{N}$
atoms. From Fig.\ref{fig:Mulliken-Charges-all} it is indicated that
the maximum positive charge $\mathrm{M_{+ve}}$ observed on $\mathrm{C15}$
varies depending on the position of the adsorbed $\mathrm{H}$. The
increase in the charge on $\mathrm{C15}$ when H is adsorbed confirms
that the source of this additional charge is the adsorbed H atom ($\mathrm{H33}$).
Therefore, we can conclude that the bonds in penta-CN$_{2}$ QDs are
not purely covalent because of the presence of three types of atoms
with varied electronegativities.

\subsection{HER Activity And \textcolor{black}{Stability}}

In this section, we analyze the HER catalytic activity for the $\mathrm{\mathrm{penta-CN_{2}}}$
QD using the well-known descriptors: (a) adsorption energy, (b) Gibbs
free energy, (c) over potential, and (d) exchange current density,
presented in Table \ref{tab:HER-descriptors-1}. The negative adsorption
energies for the H atom on all the six sites confirm that they are
catalytically active.

\begin{table}[H]
\centering{}%
\begin{tabular}{ccccccc}
\toprule 
H-adsorbed  & \multirow{2}{*}{Functional} & \multirow{2}{*}{Basis-Set} & \multirow{2}{*}{$\Delta E_{ads}$(eV)} & \multirow{2}{*}{$\Delta G$(eV)} & \multirow{2}{*}{$\varphi$(mV)} & \multirow{2}{*}{$i_{0}(\mathrm{Acm^{-2}}$)}\tabularnewline
configurations &  &  &  &  &  & \tabularnewline
\midrule
\midrule 
N4 & \multirow{6}{*}{B3LYP} & \multirow{6}{*}{6-31g(d,p)} & -0.044 & 0.196 & 196 & $5.17\times10^{-4}$\tabularnewline
\cmidrule{1-1} \cmidrule{4-7} \cmidrule{5-7} \cmidrule{6-7} \cmidrule{7-7} 
N6 &  &  & -0.049 & 0.191 & 191 & $6.27\times10^{-4}$\tabularnewline
\cmidrule{1-1} \cmidrule{4-7} \cmidrule{5-7} \cmidrule{6-7} \cmidrule{7-7} 
N11 &  &  & -0.082 & 0.158 & 158 & $2.24\times10^{-3}$\tabularnewline
\cmidrule{1-1} \cmidrule{4-7} \cmidrule{5-7} \cmidrule{6-7} \cmidrule{7-7} 
C14 &  &  & -0.038 & 0.202 & 202 & $4.10\times10^{-4}$\tabularnewline
\cmidrule{1-1} \cmidrule{4-7} \cmidrule{5-7} \cmidrule{6-7} \cmidrule{7-7} 
C15 &  &  & -0.057 & 0.183 & 183 & $8.53\times10^{-4}$\tabularnewline
\cmidrule{1-1} \cmidrule{4-7} \cmidrule{5-7} \cmidrule{6-7} \cmidrule{7-7} 
C18 &  &  & -0.009 & 0.231 & 231 & $1.34\times10^{-4}$\tabularnewline
\bottomrule
\end{tabular}\caption{\label{tab:HER-descriptors-1}Values of adsorbtion energy($\Delta E_{ads}$),
Gibbs free energy($\Delta G$), over potential($\varphi$), and the
exchange current($i_{0}$) density for all the adsorption sites for
the $3\times3$ penta-CN$_{2}$ QD.}
\end{table}

\begin{figure}[H]
\centering{}\includegraphics[scale=0.5]{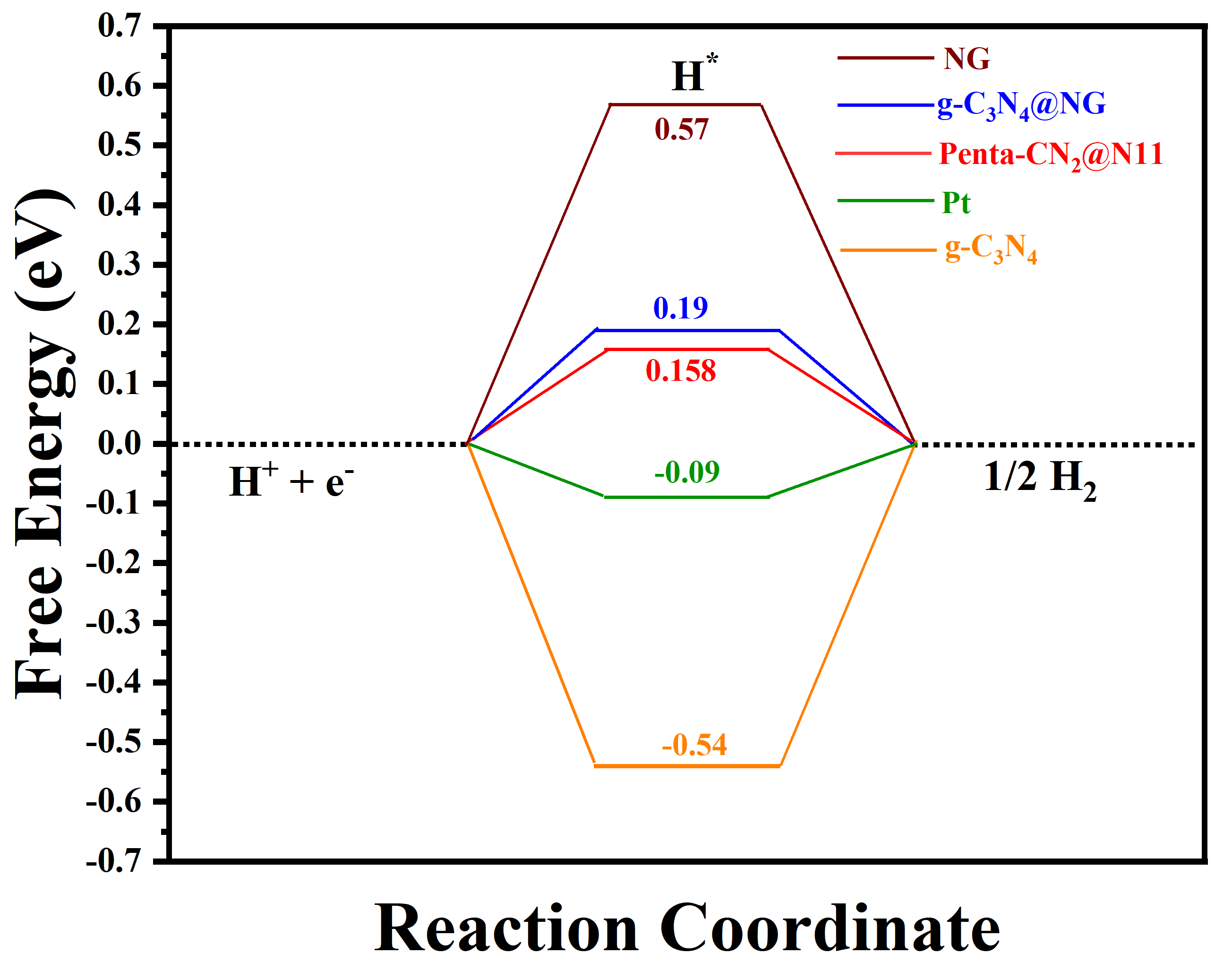}\caption{\label{fig:Free-energy-diagram-of}Free-energy diagram of $3\times3$
$\mathrm{\mathrm{penta-CN_{2}}}$ QD}
\end{figure}

As per the Sabatier principle, the optimal adsorption energy for the
H atom on penta-CN$_{2}$ QD ensures smooth adsorption and desorption
of the H during the HER \citep{sabatier1911hydrogenations}, \citep{greeley2006computational}.
It is well known that HER is a two electron transfer process that
occurs in two main steps \citep{bockris1952mechanism}, the first
step involving the discharging of protons is quite fast and is called
the Volmer reaction $\mathrm{H^{+}+e^{-}+*\rightarrow H_{ad}^{*}}$.
The second step has two possibilities either through Tafel reaction
$\mathrm{2H_{ad}^{*}\rightarrow H_{2}+*}$\citep{tafel1905polarisation}or
via Heyrovsky reaction $\mathrm{H_{ad}^{*}+H^{+}+e^{-}\rightarrow H_{2}+*}$\citep{lasia2019mechanism}.
The H adsorption ability of $\mathrm{\mathrm{penta-CN_{2}}}$ QDs
in the Volmer reaction, and desorption ability in the Tafel or Heyrovky
reactions determine their HER catalytic activity. The overall reaction
rate is heavily influenced by the Gibbs free energy, $\Delta G$\citep{parsons1958rate},
\citep{norskov2005trends} whose ideal value is zero for the perfect
catalytic activity. However, under realistic conditions it is impossible
to achieve $\Delta G=0$, therefore, one looks for as small values
as possible. In that sense site N11 emerges as the most promising
for achieving high performance in HER, while C18 being the least preferable
for H-adsorption, is expected to yield the least favorable performance
among the six sites. The comparison of $\Delta G=0$ values for site
N11 with platinum and different metal-free catalysts (NG, $\mathrm{g}-\mathrm{C}_{3}\mathrm{N}_{4}$@NG,
and $\mathrm{g}-\mathrm{C}_{3}\mathrm{N}_{4}$) is presented in Fig.
\ref{fig:Free-energy-diagram-of}. As far as remaining 4 sites are
concerned, C15, N6, N4, and C14 will offer HER performances in the
descending order.\textcolor{blue}{{} }\textcolor{black}{Gibbs free energy
calculations were also carried out for the $3\times4$ and $4\times4$
penta-CN$_{2}$ QDs, and the results can be found in Tables S7 and
S8, respectively, of the SI. In case of the $3\times4$ QD, hydrogen
adsorption for four distinct sites exhibited favorable values of $\Delta G$:
N3 (0.211 eV), N15 (0.208 eV), N21 (0.212 eV), and N22 (0.196 eV).
These findings suggest that hydrogen adsorption on the top of the
N atom is preferred more, with N22 demonstrating the highest catalytic
activity. However, the $4\times4$ penta--CN$_{2}$ QD showed low
values of $\Delta G$ for three preferred hydrogen adsorption sites:
N4 (0.205 eV), N27 (0.210 eV), and N28 (0.191 eV). Among these sites,
N28 displayed the best catalytic activity. Thus, penta--CN$_{2}$
QDs demonstrate limited variation in $\Delta G$ values as a function
of size, indicating their strong potential as electrocatalysts.}

\textcolor{black}{For each penta--CN$_{2}$ QD, we also calculated
the average of the $\Delta G$ values of all its active sites, which
indicates the net HER activity of the QD. The average $\Delta G$
values for three different sizes of penta--CN$_{2}$ QDs, denoted
as $\Delta G^{(av)}$ are as follows: 0.194 eV for the $3\times3$
penta CN$_{2}$ QD, 0.207 eV for the $3\times4$ penta CN$_{2}$ QD,
and 0.202 eV for the $4\times4$ penta--CN$_{2}$ QD. These results
indicate that the $3\times3$ QD exhibits the most preferable HER
activity, with the performance of the bigger QDs slightly less favorable.}

\textcolor{black}{With reference to the data listed in the table \ref{tab:Reported-experimental-data},
all the three penta--CN$_{2}$ QDs considered in this work have lower
$\Delta G^{(av)}$ (or equivalently overpotential) values except for
the two materials, namely, $\mathrm{CNQDs@G}$ reported by Zhong et
al.\citep{zhong2018engineering}, and $\mathrm{CNF-G}$ reported by
Meng et al. \citep{meng2017carbon}. This strongly suggests that penta
CN$_{2}$ QDs are a highly competitive catalysts for HER, when it
comes to the state of the art in this field. Because the overpotential
is just a multiple of the Gibbs free energy, therefore, our calculations
imply that penta--CN$_{2}$ QDs are highly favorable on that count
as well, when it comes to the comparison with other metal-free catalysts
(see Table \ref{tab:Reported-experimental-data})}

\begin{table}[H]
\centering{}%
\begin{tabular}{ccccc}
\toprule 
\begin{tabular}{c}
Metal free\tabularnewline
\end{tabular} & Experimental &  &  & References\tabularnewline
\begin{tabular}{c}
Catalysts\tabularnewline
\end{tabular} & Overpotential (mV) &  &  & \tabularnewline
\midrule
\midrule 
$\mathrm{g-C_{3}N_{4}@NG}$ & 240 &  &  & Zheng et al. \citep{zheng2014hydrogen}\tabularnewline
\midrule 
$\mathrm{g-C_{3}N_{4}}$+ $\mathrm{graphene}$ & 207\LyXFourPerEmSpace{} &  &  & Zhao et al. \citep{zhao2014graphitic}\tabularnewline
\midrule 
$\mathrm{g-C_{3}N_{4}@S\text{–}Se-pGr}$ & 300 &  &  & Shinde et al.\citep{shinde2015electrocatalytic}\tabularnewline
\midrule 
$\mathrm{CNF-G}$ & 149 &  &  & Meng et al.\citep{meng2017carbon}\tabularnewline
\midrule 
$\mathrm{CNQDs@G}$ & 110 &  &  & Zhong et al. \citep{zhong2018engineering}\tabularnewline
\midrule 
$\mathrm{\mathrm{g-C3N4@P-pGr}}$ & 340 &  &  & Shinde et al. \citep{shinde2015nitrogen}\tabularnewline
\midrule 
$\mathrm{MGCN-0.3}$ & 272 &  &  & Idris and Devaraj\citep{idris2019mesoporous}\tabularnewline
\bottomrule
\end{tabular}\caption{\label{tab:Reported-experimental-data}Reported experimental data
for metal free carbon nitride hybrid structures}
\end{table}

\begin{figure}[H]
\centering{}\includegraphics[scale=0.5]{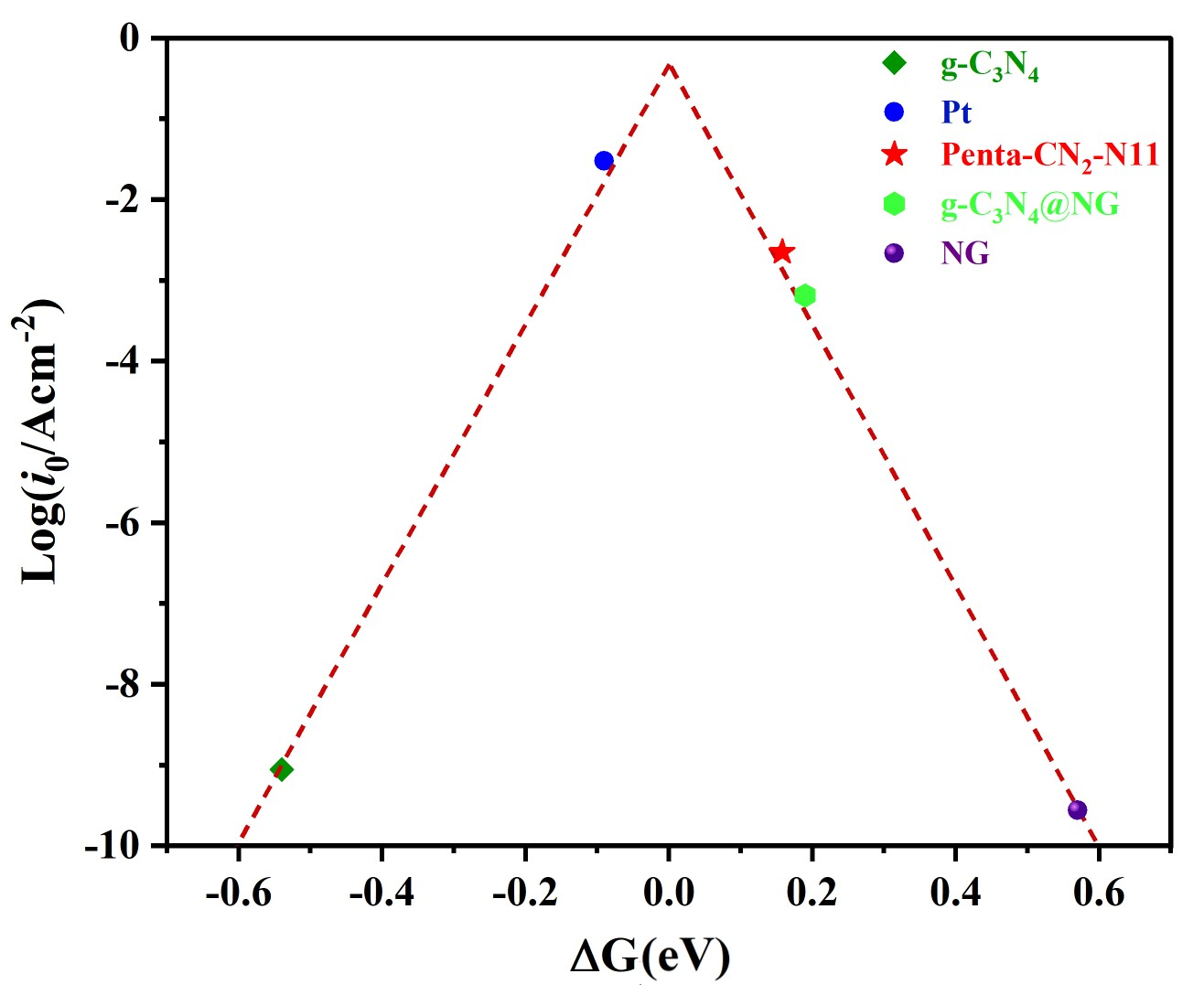}\caption{\label{fig:Volcano-curve}Volcano curve of exchange current density
($\mathrm{log(}i_{0})$) vs the Gibbs free energy values of indicated
systems.}
\end{figure}

The exchange current density measures how quickly charges are transferred
in either direction, when the cathodic and anodic half-cell reactions
are in equilibrium, and is calculated using Eq. \ref{eq:exchange current}.
The activation energy required for the charge transfer between the
electrolyte and electrode surface will be less in case of high exchange
current density. The logarithmic values of the exchange current densities
are plotted as a function of Gibbs free energies in Fig. \ref{fig:Volcano-curve}.
The better catalyst remains on the top of the volcano plot, e.g.,
Pt (blue circled) has Gibbs free energy closest to zero(-0.09 eV)
providing the best catalytic results. The value of exchange current
in $\mathrm{\mathrm{penta-CN_{2}}}$ QDs is also lies close to the
top of the volcano plot (indicated by red triangle) indicating better
catalytic activity as compared to the previously reported values of
hybrid $\mathrm{g-C_{3}N_{4}@NG}$ and single $\mathrm{NG}$, $\mathrm{\mathrm{g-C_{3}N_{4}}}$
metal-free catalysts. The highest value of exchange current density
of $\mathrm{2.24\times10^{-3}}\:\mathrm{A-cm^{-2}}$ for the $\mathrm{\mathrm{3\times3\;penta-CN_{2}}}$
QD at N11 positions is expected to generate excellent results towards
$\mathrm{H_{2}}$ production.

\section{Conclusion}

\textcolor{black}{By means of first-principles DFT calculations we
have investigated the potential of metal-free penta--CN$_{2}$ QDs
as catalysts for HER. First, by performing geometry optimization and
subsequent vibrational analysis, we confirmed the stability of penta--CN$_{2}$
QDs. To confirm the potential of considered QDs for HER, we carefully
determined the atomic sites most suitable for H adsorption (N11 for
$3\times3$ QD). A detailed investigation of the electronic properties
of both pristine and H-adsorbed penta--CN$_{2}$ QDs are performed
by examining their TDOS, PDOS, and other global descriptors. The catalytic
potential for HER of these QDs was further supported by net negative
adsorption energies and the lowest overpotential values, specifically
158 mV for $3\times3$, 196 mV for $3\times4$, and 191 mV for $4\times4$
QD sizes. As far as size dependence of the catalytic performance is
concerned, the value of $\Delta G^{(av)}$ is minimum for $3\times3$
penta--CN$_{2}$ QD, with those of $3\times4$ and $4\times4$ QDs
being marginally larger. Even though the smallest penta-CN$_{2}$
QD demonstrates the best catalytic activity, it is worth noting that
larger QDs provide a greater number of catalytic sites in comparison
to the smaller ones. We also compared our calculated values of Gibbs
energy changes and the overpotential values with those of several
other metal free C-N based catalysts, and found that penta--CN$_{2}$
QDs outperform a majority of those. In conclusion, our study emphasizes
that $\mathrm{\mathrm{penta-CN_{2}}}$ QDs can prove to be immensely
advantageous in increasing the slow kinetics of the HER, and should
be considered for synthesis in the field of catalysis.}

\bibliographystyle{achemso}
\bibliography{penta-CN2}

\providecommand{\latin}[1]{#1}
\makeatletter
\providecommand{\doi}
  {\begingroup\let\do\@makeother\dospecials
  \catcode`\{=1 \catcode`\}=2 \doi@aux}
\providecommand{\doi@aux}[1]{\endgroup\texttt{#1}}
\makeatother
\providecommand*\mcitethebibliography{\thebibliography}
\csname @ifundefined\endcsname{endmcitethebibliography}
  {\let\endmcitethebibliography\endthebibliography}{}
\begin{mcitethebibliography}{90}
\providecommand*\natexlab[1]{#1}
\providecommand*\mciteSetBstSublistMode[1]{}
\providecommand*\mciteSetBstMaxWidthForm[2]{}
\providecommand*\mciteBstWouldAddEndPuncttrue
  {\def\EndOfBibitem{\unskip.}}
\providecommand*\mciteBstWouldAddEndPunctfalse
  {\let\EndOfBibitem\relax}
\providecommand*\mciteSetBstMidEndSepPunct[3]{}
\providecommand*\mciteSetBstSublistLabelBeginEnd[3]{}
\providecommand*\EndOfBibitem{}
\mciteSetBstSublistMode{f}
\mciteSetBstMaxWidthForm{subitem}{(\alph{mcitesubitemcount})}
\mciteSetBstSublistLabelBeginEnd
  {\mcitemaxwidthsubitemform\space}
  {\relax}
  {\relax}

\bibitem[Lewis and Nocera(2006)Lewis, and Nocera]{lewis2006powering}
Lewis,~N.~S.; Nocera,~D.~G. Powering the planet: Chemical challenges in solar
  energy utilization. \emph{Proceedings of the National Academy of Sciences}
  \textbf{2006}, \emph{103}, 15729--15735\relax
\mciteBstWouldAddEndPuncttrue
\mciteSetBstMidEndSepPunct{\mcitedefaultmidpunct}
{\mcitedefaultendpunct}{\mcitedefaultseppunct}\relax
\EndOfBibitem
\bibitem[Wang \latin{et~al.}(2019)Wang, Kleme{\v{s}}, Dong, Fan, Xu, Wang, and
  Varbanov]{wang2019air}
Wang,~X.-C.; Kleme{\v{s}},~J.~J.; Dong,~X.; Fan,~W.; Xu,~Z.; Wang,~Y.;
  Varbanov,~P.~S. Air pollution terrain nexus: A review considering energy
  generation and consumption. \emph{Renewable and Sustainable Energy Reviews}
  \textbf{2019}, \emph{105}, 71--85\relax
\mciteBstWouldAddEndPuncttrue
\mciteSetBstMidEndSepPunct{\mcitedefaultmidpunct}
{\mcitedefaultendpunct}{\mcitedefaultseppunct}\relax
\EndOfBibitem
\bibitem[Levin and Chahine(2010)Levin, and Chahine]{levin2010challenges}
Levin,~D.~B.; Chahine,~R. Challenges for renewable hydrogen production from
  biomass. \emph{International Journal of Hydrogen Energy} \textbf{2010},
  \emph{35}, 4962--4969\relax
\mciteBstWouldAddEndPuncttrue
\mciteSetBstMidEndSepPunct{\mcitedefaultmidpunct}
{\mcitedefaultendpunct}{\mcitedefaultseppunct}\relax
\EndOfBibitem
\bibitem[Hasani \latin{et~al.}(2019)Hasani, Tekalgne, Van~Le, Jang, and
  Kim]{hasani2019two}
Hasani,~A.; Tekalgne,~M.; Van~Le,~Q.; Jang,~H.~W.; Kim,~S.~Y. Two-dimensional
  materials as catalysts for solar fuels: hydrogen evolution reaction and CO 2
  reduction. \emph{Journal of Materials Chemistry A} \textbf{2019}, \emph{7},
  430--454\relax
\mciteBstWouldAddEndPuncttrue
\mciteSetBstMidEndSepPunct{\mcitedefaultmidpunct}
{\mcitedefaultendpunct}{\mcitedefaultseppunct}\relax
\EndOfBibitem
\bibitem[Dincer(2012)]{dincer2012green}
Dincer,~I. Green methods for hydrogen production. \emph{International journal
  of hydrogen energy} \textbf{2012}, \emph{37}, 1954--1971\relax
\mciteBstWouldAddEndPuncttrue
\mciteSetBstMidEndSepPunct{\mcitedefaultmidpunct}
{\mcitedefaultendpunct}{\mcitedefaultseppunct}\relax
\EndOfBibitem
\bibitem[Zhu \latin{et~al.}(2020)Zhu, Qu, Liu, Ng, and Pan]{zhu2020two}
Zhu,~Q.; Qu,~Y.; Liu,~D.; Ng,~K.~W.; Pan,~H. Two-dimensional layered materials:
  high-efficient electrocatalysts for hydrogen evolution reaction. \emph{ACS
  Applied Nano Materials} \textbf{2020}, \emph{3}, 6270--6296\relax
\mciteBstWouldAddEndPuncttrue
\mciteSetBstMidEndSepPunct{\mcitedefaultmidpunct}
{\mcitedefaultendpunct}{\mcitedefaultseppunct}\relax
\EndOfBibitem
\bibitem[Olabi \latin{et~al.}(2021)Olabi, Abdelghafar, Baroutaji, Sayed, Alami,
  Rezk, Abdelkareem, \latin{et~al.} others]{olabi2021large}
Olabi,~A.; Abdelghafar,~A.~A.; Baroutaji,~A.; Sayed,~E.~T.; Alami,~A.~H.;
  Rezk,~H.; Abdelkareem,~M.~A.; others Large-vscale hydrogen production and
  storage technologies: Current status and future directions.
  \emph{International Journal of Hydrogen Energy} \textbf{2021}, \emph{46},
  23498--23528\relax
\mciteBstWouldAddEndPuncttrue
\mciteSetBstMidEndSepPunct{\mcitedefaultmidpunct}
{\mcitedefaultendpunct}{\mcitedefaultseppunct}\relax
\EndOfBibitem
\bibitem[Turn \latin{et~al.}(1998)Turn, Kinoshita, Zhang, Ishimura, and
  Zhou]{turn1998experimental}
Turn,~S.; Kinoshita,~C.; Zhang,~Z.; Ishimura,~D.; Zhou,~J. An experimental
  investigation of hydrogen production from biomass gasification.
  \emph{International journal of hydrogen energy} \textbf{1998}, \emph{23},
  641--648\relax
\mciteBstWouldAddEndPuncttrue
\mciteSetBstMidEndSepPunct{\mcitedefaultmidpunct}
{\mcitedefaultendpunct}{\mcitedefaultseppunct}\relax
\EndOfBibitem
\bibitem[Cilogullar{\i} \latin{et~al.}(2017)Cilogullar{\i}, Erden, Karakilcik,
  and Dincer]{cilogullari2017investigation}
Cilogullar{\i},~M.; Erden,~M.; Karakilcik,~M.; Dincer,~I. Investigation of
  hydrogen production performance of a Photovoltaic and Thermal System.
  \emph{International journal of hydrogen energy} \textbf{2017}, \emph{42},
  2547--2552\relax
\mciteBstWouldAddEndPuncttrue
\mciteSetBstMidEndSepPunct{\mcitedefaultmidpunct}
{\mcitedefaultendpunct}{\mcitedefaultseppunct}\relax
\EndOfBibitem
\bibitem[Hughes \latin{et~al.}(2021)Hughes, Clipsham, Chavushoglu,
  Rowley-Neale, and Banks]{hughes2021polymer}
Hughes,~J.; Clipsham,~J.; Chavushoglu,~H.; Rowley-Neale,~S.; Banks,~C. Polymer
  electrolyte electrolysis: A review of the activity and stability of
  non-precious metal hydrogen evolution reaction and oxygen evolution reaction
  catalysts. \emph{Renewable and Sustainable Energy Reviews} \textbf{2021},
  \emph{139}, 110709\relax
\mciteBstWouldAddEndPuncttrue
\mciteSetBstMidEndSepPunct{\mcitedefaultmidpunct}
{\mcitedefaultendpunct}{\mcitedefaultseppunct}\relax
\EndOfBibitem
\bibitem[Mahmood \latin{et~al.}(2017)Mahmood, Li, Jung, Okyay, Ahmad, Kim,
  Park, Jeong, and Baek]{mahmood2017efficient}
Mahmood,~J.; Li,~F.; Jung,~S.-M.; Okyay,~M.~S.; Ahmad,~I.; Kim,~S.-J.;
  Park,~N.; Jeong,~H.~Y.; Baek,~J.-B. An efficient and pH-universal
  ruthenium-based catalyst for the hydrogen evolution reaction. \emph{Nature
  nanotechnology} \textbf{2017}, \emph{12}, 441--446\relax
\mciteBstWouldAddEndPuncttrue
\mciteSetBstMidEndSepPunct{\mcitedefaultmidpunct}
{\mcitedefaultendpunct}{\mcitedefaultseppunct}\relax
\EndOfBibitem
\bibitem[Fujishima and Honda(1972)Fujishima, and
  Honda]{fujishima1972electrochemical}
Fujishima,~A.; Honda,~K. Electrochemical photolysis of water at a semiconductor
  electrode. \emph{nature} \textbf{1972}, \emph{238}, 37--38\relax
\mciteBstWouldAddEndPuncttrue
\mciteSetBstMidEndSepPunct{\mcitedefaultmidpunct}
{\mcitedefaultendpunct}{\mcitedefaultseppunct}\relax
\EndOfBibitem
\bibitem[Kalamaras and Efstathiou(2013)Kalamaras, and
  Efstathiou]{kalamaras2013hydrogen}
Kalamaras,~C.~M.; Efstathiou,~A.~M. Hydrogen production technologies: current
  state and future developments. Conference papers in science. 2013\relax
\mciteBstWouldAddEndPuncttrue
\mciteSetBstMidEndSepPunct{\mcitedefaultmidpunct}
{\mcitedefaultendpunct}{\mcitedefaultseppunct}\relax
\EndOfBibitem
\bibitem[Turner \latin{et~al.}(2008)Turner, Sverdrup, Mann, Maness, Kroposki,
  Ghirardi, Evans, and Blake]{turner2008renewable}
Turner,~J.; Sverdrup,~G.; Mann,~M.~K.; Maness,~P.-C.; Kroposki,~B.;
  Ghirardi,~M.; Evans,~R.~J.; Blake,~D. Renewable hydrogen production.
  \emph{International journal of energy research} \textbf{2008}, \emph{32},
  379--407\relax
\mciteBstWouldAddEndPuncttrue
\mciteSetBstMidEndSepPunct{\mcitedefaultmidpunct}
{\mcitedefaultendpunct}{\mcitedefaultseppunct}\relax
\EndOfBibitem
\bibitem[Dubouis and Grimaud(2019)Dubouis, and Grimaud]{dubouis2019hydrogen}
Dubouis,~N.; Grimaud,~A. The hydrogen evolution reaction: from material to
  interfacial descriptors. \emph{Chemical Science} \textbf{2019}, \emph{10},
  9165--9181\relax
\mciteBstWouldAddEndPuncttrue
\mciteSetBstMidEndSepPunct{\mcitedefaultmidpunct}
{\mcitedefaultendpunct}{\mcitedefaultseppunct}\relax
\EndOfBibitem
\bibitem[Yu \latin{et~al.}(2020)Yu, Lang, Yin, Feng, Xia, Tan, Zhu, Zhong,
  Kang, and Li]{yu2020pt}
Yu,~F.-Y.; Lang,~Z.-L.; Yin,~L.-Y.; Feng,~K.; Xia,~Y.-J.; Tan,~H.-Q.;
  Zhu,~H.-T.; Zhong,~J.; Kang,~Z.-H.; Li,~Y.-G. Pt-O bond as an active site
  superior to Pt0 in hydrogen evolution reaction. \emph{Nature communications}
  \textbf{2020}, \emph{11}, 1--7\relax
\mciteBstWouldAddEndPuncttrue
\mciteSetBstMidEndSepPunct{\mcitedefaultmidpunct}
{\mcitedefaultendpunct}{\mcitedefaultseppunct}\relax
\EndOfBibitem
\bibitem[Shao \latin{et~al.}(2018)Shao, Li, Chen, and Huang]{shao2018advanced}
Shao,~Q.; Li,~F.; Chen,~Y.; Huang,~X. The Advanced Designs of High-Performance
  Platinum-Based Electrocatalysts: Recent Progresses and Challenges.
  \emph{Advanced Materials Interfaces} \textbf{2018}, \emph{5}, 1800486\relax
\mciteBstWouldAddEndPuncttrue
\mciteSetBstMidEndSepPunct{\mcitedefaultmidpunct}
{\mcitedefaultendpunct}{\mcitedefaultseppunct}\relax
\EndOfBibitem
\bibitem[Tsai \latin{et~al.}(2014)Tsai, Abild-Pedersen, and
  N{\o}rskov]{tsai2014tuning}
Tsai,~C.; Abild-Pedersen,~F.; N{\o}rskov,~J.~K. Tuning the MoS2 edge-site
  activity for hydrogen evolution via support interactions. \emph{Nano letters}
  \textbf{2014}, \emph{14}, 1381--1387\relax
\mciteBstWouldAddEndPuncttrue
\mciteSetBstMidEndSepPunct{\mcitedefaultmidpunct}
{\mcitedefaultendpunct}{\mcitedefaultseppunct}\relax
\EndOfBibitem
\bibitem[Er \latin{et~al.}(2018)Er, Ye, Frey, Kumar, Lou, and
  Shenoy]{er2018prediction}
Er,~D.; Ye,~H.; Frey,~N.~C.; Kumar,~H.; Lou,~J.; Shenoy,~V.~B. Prediction of
  enhanced catalytic activity for hydrogen evolution reaction in Janus
  transition metal dichalcogenides. \emph{Nano letters} \textbf{2018},
  \emph{18}, 3943--3949\relax
\mciteBstWouldAddEndPuncttrue
\mciteSetBstMidEndSepPunct{\mcitedefaultmidpunct}
{\mcitedefaultendpunct}{\mcitedefaultseppunct}\relax
\EndOfBibitem
\bibitem[Govindaraju \latin{et~al.}(2021)Govindaraju, Sureshkumar,
  Ramakrishnappa, Muralikrishna, Samrat, Pai, Kumar, Vikrant, and
  Kim]{govindaraju2021graphitic}
Govindaraju,~V.~R.; Sureshkumar,~K.; Ramakrishnappa,~T.; Muralikrishna,~S.;
  Samrat,~D.; Pai,~R.~K.; Kumar,~V.; Vikrant,~K.; Kim,~K.-H. Graphitic carbon
  nitride composites as electro catalysts: Applications in energy
  conversion/storage and sensing system. \emph{Journal of Cleaner Production}
  \textbf{2021}, \emph{320}, 128693\relax
\mciteBstWouldAddEndPuncttrue
\mciteSetBstMidEndSepPunct{\mcitedefaultmidpunct}
{\mcitedefaultendpunct}{\mcitedefaultseppunct}\relax
\EndOfBibitem
\bibitem[Liu and Dai(2016)Liu, and Dai]{liu2016carbon}
Liu,~X.; Dai,~L. Carbon-based metal-free catalysts. \emph{Nature Reviews
  Materials} \textbf{2016}, \emph{1}, 1--12\relax
\mciteBstWouldAddEndPuncttrue
\mciteSetBstMidEndSepPunct{\mcitedefaultmidpunct}
{\mcitedefaultendpunct}{\mcitedefaultseppunct}\relax
\EndOfBibitem
\bibitem[Conway and Tilak(2002)Conway, and Tilak]{conway2002interfacial}
Conway,~B.; Tilak,~B. Interfacial processes involving electrocatalytic
  evolution and oxidation of H2, and the role of chemisorbed H.
  \emph{Electrochimica acta} \textbf{2002}, \emph{47}, 3571--3594\relax
\mciteBstWouldAddEndPuncttrue
\mciteSetBstMidEndSepPunct{\mcitedefaultmidpunct}
{\mcitedefaultendpunct}{\mcitedefaultseppunct}\relax
\EndOfBibitem
\bibitem[Frankel \latin{et~al.}(2013)Frankel, Samaniego, and
  Birbilis]{frankel2013evolution}
Frankel,~G.~S.; Samaniego,~A.; Birbilis,~N. Evolution of hydrogen at dissolving
  magnesium surfaces. \emph{Corrosion Science} \textbf{2013}, \emph{70},
  104--111\relax
\mciteBstWouldAddEndPuncttrue
\mciteSetBstMidEndSepPunct{\mcitedefaultmidpunct}
{\mcitedefaultendpunct}{\mcitedefaultseppunct}\relax
\EndOfBibitem
\bibitem[Frankel \latin{et~al.}(2015)Frankel, Fajardo, and
  Lynch]{frankel2015introductory}
Frankel,~G.; Fajardo,~S.; Lynch,~B. Introductory lecture on corrosion
  chemistry: a focus on anodic hydrogen evolution on Al and Mg. \emph{faraday
  discussions} \textbf{2015}, \emph{180}, 11--33\relax
\mciteBstWouldAddEndPuncttrue
\mciteSetBstMidEndSepPunct{\mcitedefaultmidpunct}
{\mcitedefaultendpunct}{\mcitedefaultseppunct}\relax
\EndOfBibitem
\bibitem[Jun \latin{et~al.}(2013)Jun, Park, Lee, Thomas, Hong, and
  Stucky]{jun2013three}
Jun,~Y.-S.; Park,~J.; Lee,~S.~U.; Thomas,~A.; Hong,~W.~H.; Stucky,~G.~D.
  Three-dimensional macroscopic assemblies of low-dimensional carbon nitrides
  for enhanced hydrogen evolution. \emph{Angewandte Chemie International
  Edition} \textbf{2013}, \emph{52}, 11083--11087\relax
\mciteBstWouldAddEndPuncttrue
\mciteSetBstMidEndSepPunct{\mcitedefaultmidpunct}
{\mcitedefaultendpunct}{\mcitedefaultseppunct}\relax
\EndOfBibitem
\bibitem[Voiry \latin{et~al.}(2016)Voiry, Yang, and Chhowalla]{voiry2016recent}
Voiry,~D.; Yang,~J.; Chhowalla,~M. Recent strategies for improving the
  catalytic activity of 2D TMD nanosheets toward the hydrogen evolution
  reaction. \emph{Advanced materials} \textbf{2016}, \emph{28},
  6197--6206\relax
\mciteBstWouldAddEndPuncttrue
\mciteSetBstMidEndSepPunct{\mcitedefaultmidpunct}
{\mcitedefaultendpunct}{\mcitedefaultseppunct}\relax
\EndOfBibitem
\bibitem[Zheng \latin{et~al.}(2023)Zheng, Nan, Lu, Wang, and
  Wang]{zheng2023core}
Zheng,~H.; Nan,~K.; Lu,~Z.; Wang,~N.; Wang,~Y. Core-shell FeCo@ carbon
  nanocages encapsulated in biomass-derived carbon aerogel: Architecture design
  and interface engineering of lightweight, anti-corrosion and superior
  microwave absorption. \emph{Journal of Colloid and Interface Science}
  \textbf{2023}, \emph{646}, 555--566\relax
\mciteBstWouldAddEndPuncttrue
\mciteSetBstMidEndSepPunct{\mcitedefaultmidpunct}
{\mcitedefaultendpunct}{\mcitedefaultseppunct}\relax
\EndOfBibitem
\bibitem[Wang \latin{et~al.}(2023)Wang, Nan, Zheng, Li, and Wang]{wang2023ion}
Wang,~W.; Nan,~K.; Zheng,~H.; Li,~Q.; Wang,~Y. Ion-exchange reaction
  construction of carbon nanotube-modified CoNi@ MoO2/C composite for
  ultra-intense and broad electromagnetic wave absorption. \emph{Carbon}
  \textbf{2023}, \emph{210}, 118074\relax
\mciteBstWouldAddEndPuncttrue
\mciteSetBstMidEndSepPunct{\mcitedefaultmidpunct}
{\mcitedefaultendpunct}{\mcitedefaultseppunct}\relax
\EndOfBibitem
\bibitem[Wang \latin{et~al.}(2023)Wang, Cheng, Cui, Lu, Yang, Pan, and
  Che]{wang2023heterostructure}
Wang,~Y.; Cheng,~R.; Cui,~W.-G.; Lu,~Z.; Yang,~Y.; Pan,~H.; Che,~R.
  Heterostructure design of 3D hydrangea-like Fe3O4/Fe7S8@ C core-shell
  composite as a high-efficiency microwave absorber. \emph{Carbon}
  \textbf{2023}, \emph{210}, 118043\relax
\mciteBstWouldAddEndPuncttrue
\mciteSetBstMidEndSepPunct{\mcitedefaultmidpunct}
{\mcitedefaultendpunct}{\mcitedefaultseppunct}\relax
\EndOfBibitem
\bibitem[Lim \latin{et~al.}(2015)Lim, Shen, and Gao]{lim2015carbon}
Lim,~S.~Y.; Shen,~W.; Gao,~Z. Carbon quantum dots and their applications.
  \emph{Chemical Society Reviews} \textbf{2015}, \emph{44}, 362--381\relax
\mciteBstWouldAddEndPuncttrue
\mciteSetBstMidEndSepPunct{\mcitedefaultmidpunct}
{\mcitedefaultendpunct}{\mcitedefaultseppunct}\relax
\EndOfBibitem
\bibitem[Paunovi{\'c} \latin{et~al.}(2007)Paunovi{\'c}, Dimitrov, Popovski,
  Slavkov, and Jordanov]{paunovic2007effect}
Paunovi{\'c},~P.; Dimitrov,~A.~T.; Popovski,~O.; Slavkov,~D.; Jordanov,~S.~H.
  Effect of carbon nanotubes support in improving the performance of mixed
  electrocatalysts for hydrogen evolution. \emph{Macedonian Journal of
  Chemistry and Chemical Engineering} \textbf{2007}, \emph{26}, 87--93\relax
\mciteBstWouldAddEndPuncttrue
\mciteSetBstMidEndSepPunct{\mcitedefaultmidpunct}
{\mcitedefaultendpunct}{\mcitedefaultseppunct}\relax
\EndOfBibitem
\bibitem[Pachaiappan \latin{et~al.}(2021)Pachaiappan, Rajendran, Kumar, Vo,
  Hoang, and Cornejo-Ponce]{pachaiappan2021recent}
Pachaiappan,~R.; Rajendran,~S.; Kumar,~P.~S.; Vo,~D.-V.~N.; Hoang,~T.~K.;
  Cornejo-Ponce,~L. Recent advances in carbon nitride-based nanomaterials for
  hydrogen production and storage. \emph{International Journal of Hydrogen
  Energy} \textbf{2021}, \relax
\mciteBstWouldAddEndPunctfalse
\mciteSetBstMidEndSepPunct{\mcitedefaultmidpunct}
{}{\mcitedefaultseppunct}\relax
\EndOfBibitem
\bibitem[Zheng \latin{et~al.}(2014)Zheng, Jiao, Zhu, Li, Han, Chen, Du,
  Jaroniec, and Qiao]{zheng2014hydrogen}
Zheng,~Y.; Jiao,~Y.; Zhu,~Y.; Li,~L.~H.; Han,~Y.; Chen,~Y.; Du,~A.;
  Jaroniec,~M.; Qiao,~S.~Z. Hydrogen evolution by a metal-free electrocatalyst.
  \emph{Nature communications} \textbf{2014}, \emph{5}, 1--8\relax
\mciteBstWouldAddEndPuncttrue
\mciteSetBstMidEndSepPunct{\mcitedefaultmidpunct}
{\mcitedefaultendpunct}{\mcitedefaultseppunct}\relax
\EndOfBibitem
\bibitem[Chen and Bai(2020)Chen, and Bai]{chen2020review}
Chen,~Y.; Bai,~X. A review on quantum dots modified g-C3N4-based photocatalysts
  with improved photocatalytic activity. \emph{Catalysts} \textbf{2020},
  \emph{10}, 142\relax
\mciteBstWouldAddEndPuncttrue
\mciteSetBstMidEndSepPunct{\mcitedefaultmidpunct}
{\mcitedefaultendpunct}{\mcitedefaultseppunct}\relax
\EndOfBibitem
\bibitem[Hong \latin{et~al.}(2017)Hong, Li, Fang, Luo, and
  Shi]{hong2017rational}
Hong,~Y.; Li,~C.; Fang,~Z.; Luo,~B.; Shi,~W. Rational synthesis of ultrathin
  graphitic carbon nitride nanosheets for efficient photocatalytic hydrogen
  evolution. \emph{Carbon} \textbf{2017}, \emph{121}, 463--471\relax
\mciteBstWouldAddEndPuncttrue
\mciteSetBstMidEndSepPunct{\mcitedefaultmidpunct}
{\mcitedefaultendpunct}{\mcitedefaultseppunct}\relax
\EndOfBibitem
\bibitem[Han \latin{et~al.}(2017)Han, Chen, Zhang, and Qu]{han2017graphene}
Han,~Q.; Chen,~N.; Zhang,~J.; Qu,~L. Graphene/graphitic carbon nitride hybrids
  for catalysis. \emph{Materials Horizons} \textbf{2017}, \emph{4},
  832--850\relax
\mciteBstWouldAddEndPuncttrue
\mciteSetBstMidEndSepPunct{\mcitedefaultmidpunct}
{\mcitedefaultendpunct}{\mcitedefaultseppunct}\relax
\EndOfBibitem
\bibitem[Brus(1984)]{brus1984electron}
Brus,~L.~E. Electron--electron and electron-hole interactions in small
  semiconductor crystallites: The size dependence of the lowest excited
  electronic state. \emph{The Journal of chemical physics} \textbf{1984},
  \emph{80}, 4403--4409\relax
\mciteBstWouldAddEndPuncttrue
\mciteSetBstMidEndSepPunct{\mcitedefaultmidpunct}
{\mcitedefaultendpunct}{\mcitedefaultseppunct}\relax
\EndOfBibitem
\bibitem[Jindal \latin{et~al.}(2023)Jindal, Roondhe, and
  Shukla]{jindal2023first}
Jindal,~R.; Roondhe,~V.; Shukla,~A. A first-principles study of the electronic,
  vibrational, and optical properties of planar SiC quantum dots. \emph{Journal
  of Physics D: Applied Physics} \textbf{2023}, \emph{57}, 065103\relax
\mciteBstWouldAddEndPuncttrue
\mciteSetBstMidEndSepPunct{\mcitedefaultmidpunct}
{\mcitedefaultendpunct}{\mcitedefaultseppunct}\relax
\EndOfBibitem
\bibitem[Wu and Lian(2016)Wu, and Lian]{wu2016quantum}
Wu,~K.; Lian,~T. Quantum confined colloidal nanorod heterostructures for
  solar-to-fuel conversion. \emph{Chemical Society Reviews} \textbf{2016},
  \emph{45}, 3781--3810\relax
\mciteBstWouldAddEndPuncttrue
\mciteSetBstMidEndSepPunct{\mcitedefaultmidpunct}
{\mcitedefaultendpunct}{\mcitedefaultseppunct}\relax
\EndOfBibitem
\bibitem[Fan \latin{et~al.}(2019)Fan, Yu, Hou, and Kim]{fan2019quantum}
Fan,~X.-B.; Yu,~S.; Hou,~B.; Kim,~J.~M. Quantum dots based photocatalytic
  hydrogen evolution. \emph{Israel Journal of Chemistry} \textbf{2019},
  \emph{59}, 762--773\relax
\mciteBstWouldAddEndPuncttrue
\mciteSetBstMidEndSepPunct{\mcitedefaultmidpunct}
{\mcitedefaultendpunct}{\mcitedefaultseppunct}\relax
\EndOfBibitem
\bibitem[Liu \latin{et~al.}(2021)Liu, Ali, Ma, Jiao, Yin, Mu, and
  Jian]{liu2021graphene}
Liu,~Y.; Ali,~R.; Ma,~J.; Jiao,~W.; Yin,~L.; Mu,~C.; Jian,~X.
  Graphene-Decorated Boron--Carbon--Nitride-Based Metal-Free Catalysts for an
  Enhanced Hydrogen Evolution Reaction. \emph{ACS Applied Energy Materials}
  \textbf{2021}, \emph{4}, 3861--3868\relax
\mciteBstWouldAddEndPuncttrue
\mciteSetBstMidEndSepPunct{\mcitedefaultmidpunct}
{\mcitedefaultendpunct}{\mcitedefaultseppunct}\relax
\EndOfBibitem
\bibitem[Mohanty \latin{et~al.}(2021)Mohanty, Giri, and Jena]{mohanty2021mxene}
Mohanty,~B.; Giri,~L.; Jena,~B.~K. MXene-derived quantum dots for energy
  conversion and storage applications. \emph{Energy \& Fuels} \textbf{2021},
  \emph{35}, 14304--14324\relax
\mciteBstWouldAddEndPuncttrue
\mciteSetBstMidEndSepPunct{\mcitedefaultmidpunct}
{\mcitedefaultendpunct}{\mcitedefaultseppunct}\relax
\EndOfBibitem
\bibitem[Molaei(2020)]{molaei2020optical}
Molaei,~M.~J. The optical properties and solar energy conversion applications
  of carbon quantum dots: A review. \emph{Solar Energy} \textbf{2020},
  \emph{196}, 549--566\relax
\mciteBstWouldAddEndPuncttrue
\mciteSetBstMidEndSepPunct{\mcitedefaultmidpunct}
{\mcitedefaultendpunct}{\mcitedefaultseppunct}\relax
\EndOfBibitem
\bibitem[Arakawa(2002)]{arakawa2002progress}
Arakawa,~Y. Progress in GaN-based quantum dots for optoelectronics
  applications. \emph{IEEE journal of selected topics in quantum electronics}
  \textbf{2002}, \emph{8}, 823--832\relax
\mciteBstWouldAddEndPuncttrue
\mciteSetBstMidEndSepPunct{\mcitedefaultmidpunct}
{\mcitedefaultendpunct}{\mcitedefaultseppunct}\relax
\EndOfBibitem
\bibitem[Litvin \latin{et~al.}(2017)Litvin, Martynenko, Purcell-Milton,
  Baranov, Fedorov, and Gun'Ko]{litvin2017colloidal}
Litvin,~A.; Martynenko,~I.; Purcell-Milton,~F.; Baranov,~A.; Fedorov,~A.;
  Gun'Ko,~Y. Colloidal quantum dots for optoelectronics. \emph{Journal of
  Materials Chemistry A} \textbf{2017}, \emph{5}, 13252--13275\relax
\mciteBstWouldAddEndPuncttrue
\mciteSetBstMidEndSepPunct{\mcitedefaultmidpunct}
{\mcitedefaultendpunct}{\mcitedefaultseppunct}\relax
\EndOfBibitem
\bibitem[Frasco and Chaniotakis(2009)Frasco, and
  Chaniotakis]{frasco2009semiconductor}
Frasco,~M.~F.; Chaniotakis,~N. Semiconductor quantum dots in chemical sensors
  and biosensors. \emph{Sensors} \textbf{2009}, \emph{9}, 7266--7286\relax
\mciteBstWouldAddEndPuncttrue
\mciteSetBstMidEndSepPunct{\mcitedefaultmidpunct}
{\mcitedefaultendpunct}{\mcitedefaultseppunct}\relax
\EndOfBibitem
\bibitem[Raeyani \latin{et~al.}(2020)Raeyani, Shojaei, and
  Ahmadi-Kandjani]{raeyani2020optical}
Raeyani,~D.; Shojaei,~S.; Ahmadi-Kandjani,~S. Optical graphene quantum dots gas
  sensors: Experimental study. \emph{Materials Research Express} \textbf{2020},
  \emph{7}, 015608\relax
\mciteBstWouldAddEndPuncttrue
\mciteSetBstMidEndSepPunct{\mcitedefaultmidpunct}
{\mcitedefaultendpunct}{\mcitedefaultseppunct}\relax
\EndOfBibitem
\bibitem[Lesiak \latin{et~al.}(2019)Lesiak, Drzozga, Cabaj, Ba{\'n}ski,
  Malecha, and Podhorodecki]{lesiak2019optical}
Lesiak,~A.; Drzozga,~K.; Cabaj,~J.; Ba{\'n}ski,~M.; Malecha,~K.;
  Podhorodecki,~A. Optical sensors based on II-VI quantum dots.
  \emph{Nanomaterials} \textbf{2019}, \emph{9}, 192\relax
\mciteBstWouldAddEndPuncttrue
\mciteSetBstMidEndSepPunct{\mcitedefaultmidpunct}
{\mcitedefaultendpunct}{\mcitedefaultseppunct}\relax
\EndOfBibitem
\bibitem[Li \latin{et~al.}(2012)Li, Kang, Liu, and Lee]{li2012carbon}
Li,~H.; Kang,~Z.; Liu,~Y.; Lee,~S.-T. Carbon nanodots: synthesis, properties
  and applications. \emph{Journal of materials chemistry} \textbf{2012},
  \emph{22}, 24230--24253\relax
\mciteBstWouldAddEndPuncttrue
\mciteSetBstMidEndSepPunct{\mcitedefaultmidpunct}
{\mcitedefaultendpunct}{\mcitedefaultseppunct}\relax
\EndOfBibitem
\bibitem[Jindal \latin{et~al.}(2022)Jindal, Sharma, and
  Shukla]{jindal2022density}
Jindal,~R.; Sharma,~V.; Shukla,~A. Density functional theory study of the
  hydrogen evolution reaction in haeckelite boron nitride quantum dots.
  \emph{International Journal of Hydrogen Energy} \textbf{2022}, \relax
\mciteBstWouldAddEndPunctfalse
\mciteSetBstMidEndSepPunct{\mcitedefaultmidpunct}
{}{\mcitedefaultseppunct}\relax
\EndOfBibitem
\bibitem[Mohanty \latin{et~al.}(2020)Mohanty, Mitra, Jena, and
  Jena]{mohanty2020mos2}
Mohanty,~B.; Mitra,~A.; Jena,~B.; Jena,~B.~K. MoS2 quantum dots as efficient
  electrocatalyst for hydrogen evolution reaction over a wide pH range.
  \emph{Energy \& Fuels} \textbf{2020}, \emph{34}, 10268--10275\relax
\mciteBstWouldAddEndPuncttrue
\mciteSetBstMidEndSepPunct{\mcitedefaultmidpunct}
{\mcitedefaultendpunct}{\mcitedefaultseppunct}\relax
\EndOfBibitem
\bibitem[Sargin \latin{et~al.}(2019)Sargin, Yanalak, Arslan, and
  Patir]{sargin2019green}
Sargin,~I.; Yanalak,~G.; Arslan,~G.; Patir,~I.~H. Green synthesized carbon
  quantum dots as TiO2 sensitizers for photocatalytic hydrogen evolution.
  \emph{International Journal of Hydrogen Energy} \textbf{2019}, \emph{44},
  21781--21789\relax
\mciteBstWouldAddEndPuncttrue
\mciteSetBstMidEndSepPunct{\mcitedefaultmidpunct}
{\mcitedefaultendpunct}{\mcitedefaultseppunct}\relax
\EndOfBibitem
\bibitem[Gogoi \latin{et~al.}(2020)Gogoi, Koyani, Golder, and
  Peela]{gogoi2020enhanced}
Gogoi,~D.; Koyani,~R.; Golder,~A.~K.; Peela,~N.~R. Enhanced photocatalytic
  hydrogen evolution using green carbon quantum dots modified 1-D CdS nanowires
  under visible light irradiation. \emph{Solar Energy} \textbf{2020},
  \emph{208}, 966--977\relax
\mciteBstWouldAddEndPuncttrue
\mciteSetBstMidEndSepPunct{\mcitedefaultmidpunct}
{\mcitedefaultendpunct}{\mcitedefaultseppunct}\relax
\EndOfBibitem
\bibitem[Wang \latin{et~al.}(2019)Wang, Chen, Liu, Xi, Li, Geng, Jiang, and
  Zhao]{wang2019novel}
Wang,~Y.; Chen,~J.; Liu,~L.; Xi,~X.; Li,~Y.; Geng,~Z.; Jiang,~G.; Zhao,~Z.
  Novel metal doped carbon quantum dots/CdS composites for efficient
  photocatalytic hydrogen evolution. \emph{Nanoscale} \textbf{2019}, \emph{11},
  1618--1625\relax
\mciteBstWouldAddEndPuncttrue
\mciteSetBstMidEndSepPunct{\mcitedefaultmidpunct}
{\mcitedefaultendpunct}{\mcitedefaultseppunct}\relax
\EndOfBibitem
\bibitem[Meng \latin{et~al.}(2020)Meng, Zhang, Dong, Sun, Ji, and
  Ding]{meng2020carbon}
Meng,~X.; Zhang,~C.; Dong,~C.; Sun,~W.; Ji,~D.; Ding,~Y. Carbon quantum dots
  assisted strategy to synthesize Co@ NC for boosting photocatalytic hydrogen
  evolution performance of CdS. \emph{Chemical Engineering Journal}
  \textbf{2020}, \emph{389}, 124432\relax
\mciteBstWouldAddEndPuncttrue
\mciteSetBstMidEndSepPunct{\mcitedefaultmidpunct}
{\mcitedefaultendpunct}{\mcitedefaultseppunct}\relax
\EndOfBibitem
\bibitem[Wang \latin{et~al.}(2021)Wang, Wang, Liu, Zhang, Han, Vomiero, and
  Zhao]{wang2021colloidal}
Wang,~X.; Wang,~M.; Liu,~G.; Zhang,~Y.; Han,~G.; Vomiero,~A.; Zhao,~H.
  Colloidal carbon quantum dots as light absorber for efficient and stable
  ecofriendly photoelectrochemical hydrogen generation. \emph{Nano Energy}
  \textbf{2021}, \emph{86}, 106122\relax
\mciteBstWouldAddEndPuncttrue
\mciteSetBstMidEndSepPunct{\mcitedefaultmidpunct}
{\mcitedefaultendpunct}{\mcitedefaultseppunct}\relax
\EndOfBibitem
\bibitem[Wang \latin{et~al.}(2023)Wang, Han, Li, Li, Ren, Wang, Han, Yu, Zhang,
  and Zhao]{wang2023doped}
Wang,~X.; Han,~Y.; Li,~W.; Li,~J.; Ren,~S.; Wang,~M.; Han,~G.; Yu,~J.;
  Zhang,~Y.; Zhao,~H. Doped Carbon Dots Enable Highly Efficient Multiple-Color
  Room Temperature Phosphorescence. \emph{Advanced Optical Materials}
  \textbf{2023}, 2301962\relax
\mciteBstWouldAddEndPuncttrue
\mciteSetBstMidEndSepPunct{\mcitedefaultmidpunct}
{\mcitedefaultendpunct}{\mcitedefaultseppunct}\relax
\EndOfBibitem
\bibitem[Meng \latin{et~al.}(2023)Meng, Song, Jing, and Zhao]{meng2023self}
Meng,~X.; Song,~Y.; Jing,~Q.; Zhao,~H. Self-precipitation of highly purified
  red emitting carbon dots as red phosphors. \emph{The Journal of Physical
  Chemistry Letters} \textbf{2023}, \emph{14}, 9176--9182\relax
\mciteBstWouldAddEndPuncttrue
\mciteSetBstMidEndSepPunct{\mcitedefaultmidpunct}
{\mcitedefaultendpunct}{\mcitedefaultseppunct}\relax
\EndOfBibitem
\bibitem[Zhang \latin{et~al.}(2016)Zhang, Zhou, Wang, and
  Jena]{zhang2016beyond}
Zhang,~S.; Zhou,~J.; Wang,~Q.; Jena,~P. Beyond graphitic carbon nitride:
  nitrogen-rich penta-CN2 sheet. \emph{The Journal of Physical Chemistry C}
  \textbf{2016}, \emph{120}, 3993--3998\relax
\mciteBstWouldAddEndPuncttrue
\mciteSetBstMidEndSepPunct{\mcitedefaultmidpunct}
{\mcitedefaultendpunct}{\mcitedefaultseppunct}\relax
\EndOfBibitem
\bibitem[Cheng and He(2018)Cheng, and He]{cheng2018electronic}
Cheng,~F.; He,~B. Electronic properties of pentagonal CN2 nanoribbon under
  external electric field and tensile strain. \emph{Physica E: Low-dimensional
  Systems and Nanostructures} \textbf{2018}, \emph{104}, 6--10\relax
\mciteBstWouldAddEndPuncttrue
\mciteSetBstMidEndSepPunct{\mcitedefaultmidpunct}
{\mcitedefaultendpunct}{\mcitedefaultseppunct}\relax
\EndOfBibitem
\bibitem[Smith and Nie(2010)Smith, and Nie]{smith2010semiconductor}
Smith,~A.~M.; Nie,~S. Semiconductor nanocrystals: structure, properties, and
  band gap engineering. \emph{Accounts of chemical research} \textbf{2010},
  \emph{43}, 190--200\relax
\mciteBstWouldAddEndPuncttrue
\mciteSetBstMidEndSepPunct{\mcitedefaultmidpunct}
{\mcitedefaultendpunct}{\mcitedefaultseppunct}\relax
\EndOfBibitem
\bibitem[Zhao \latin{et~al.}(2014)Zhao, Zhao, Wang, Xu, Zhang, Shi, and
  Qu]{zhao2014graphitic}
Zhao,~Y.; Zhao,~F.; Wang,~X.; Xu,~C.; Zhang,~Z.; Shi,~G.; Qu,~L. Graphitic
  carbon nitride nanoribbons: graphene-assisted formation and synergic function
  for highly efficient hydrogen evolution. \emph{Angewandte Chemie
  International Edition} \textbf{2014}, \emph{53}, 13934--13939\relax
\mciteBstWouldAddEndPuncttrue
\mciteSetBstMidEndSepPunct{\mcitedefaultmidpunct}
{\mcitedefaultendpunct}{\mcitedefaultseppunct}\relax
\EndOfBibitem
\bibitem[Shabnam \latin{et~al.}(2020)Shabnam, Faisal, Martucci, Gomes,
  \latin{et~al.} others]{shabnam2020doping}
Shabnam,~L.; Faisal,~S.~N.; Martucci,~A.; Gomes,~V.~G.; others Doping reduced
  graphene oxide and graphitic carbon nitride hybrid for dual functionality:
  High performance supercapacitance and hydrogen evolution reaction.
  \emph{Journal of Electroanalytical Chemistry} \textbf{2020}, \emph{856},
  113503\relax
\mciteBstWouldAddEndPuncttrue
\mciteSetBstMidEndSepPunct{\mcitedefaultmidpunct}
{\mcitedefaultendpunct}{\mcitedefaultseppunct}\relax
\EndOfBibitem
\bibitem[Idris and Devaraj(2019)Idris, and Devaraj]{idris2019mesoporous}
Idris,~M.~B.; Devaraj,~S. Mesoporous graphitic carbon nitride synthesized using
  biotemplate as a high-performance electrode material for supercapacitor and
  electrocatalyst for hydrogen evolution reaction in acidic medium.
  \emph{Journal of Energy Storage} \textbf{2019}, \emph{26}, 101032\relax
\mciteBstWouldAddEndPuncttrue
\mciteSetBstMidEndSepPunct{\mcitedefaultmidpunct}
{\mcitedefaultendpunct}{\mcitedefaultseppunct}\relax
\EndOfBibitem
\bibitem[Frisch \latin{et~al.}(2016)Frisch, Trucks, Schlegel, Scuseria, Robb,
  Cheeseman, Scalmani, Barone, Petersson, Nakatsuji, \latin{et~al.}
  others]{frisch2016gaussian}
Frisch,~M.~e.; Trucks,~G.; Schlegel,~H.; Scuseria,~G.; Robb,~M.; Cheeseman,~J.;
  Scalmani,~G.; Barone,~V.; Petersson,~G.; Nakatsuji,~H.; others Gaussian 16.
  2016\relax
\mciteBstWouldAddEndPuncttrue
\mciteSetBstMidEndSepPunct{\mcitedefaultmidpunct}
{\mcitedefaultendpunct}{\mcitedefaultseppunct}\relax
\EndOfBibitem
\bibitem[Becke(1988)]{becke1988density}
Becke,~A.~D. Density-functional exchange-energy approximation with correct
  asymptotic behavior. \emph{Physical review A} \textbf{1988}, \emph{38},
  3098\relax
\mciteBstWouldAddEndPuncttrue
\mciteSetBstMidEndSepPunct{\mcitedefaultmidpunct}
{\mcitedefaultendpunct}{\mcitedefaultseppunct}\relax
\EndOfBibitem
\bibitem[Lee \latin{et~al.}(1988)Lee, Yang, and Parr]{lee1988development}
Lee,~C.; Yang,~W.; Parr,~R.~G. Development of the Colle-Salvetti
  correlation-energy formula into a functional of the electron density.
  \emph{Physical review B} \textbf{1988}, \emph{37}, 785\relax
\mciteBstWouldAddEndPuncttrue
\mciteSetBstMidEndSepPunct{\mcitedefaultmidpunct}
{\mcitedefaultendpunct}{\mcitedefaultseppunct}\relax
\EndOfBibitem
\bibitem[Dennington \latin{et~al.}(2016)Dennington, Keith, Millam,
  \latin{et~al.} others]{dennington2016gaussview}
Dennington,~R.; Keith,~T.; Millam,~J.; others GaussView, version 6. 2016\relax
\mciteBstWouldAddEndPuncttrue
\mciteSetBstMidEndSepPunct{\mcitedefaultmidpunct}
{\mcitedefaultendpunct}{\mcitedefaultseppunct}\relax
\EndOfBibitem
\bibitem[Lu and Chen(2012)Lu, and Chen]{lu2012multiwfn}
Lu,~T.; Chen,~F. Multiwfn: a multifunctional wavefunction analyzer.
  \emph{Journal of computational chemistry} \textbf{2012}, \emph{33},
  580--592\relax
\mciteBstWouldAddEndPuncttrue
\mciteSetBstMidEndSepPunct{\mcitedefaultmidpunct}
{\mcitedefaultendpunct}{\mcitedefaultseppunct}\relax
\EndOfBibitem
\bibitem[N{\o}rskov \latin{et~al.}(2005)N{\o}rskov, Bligaard, Logadottir,
  Kitchin, Chen, Pandelov, and Stimming]{norskov2005trends}
N{\o}rskov,~J.~K.; Bligaard,~T.; Logadottir,~A.; Kitchin,~J.; Chen,~J.~G.;
  Pandelov,~S.; Stimming,~U. Trends in the exchange current for hydrogen
  evolution. \emph{Journal of The Electrochemical Society} \textbf{2005},
  \emph{152}, J23\relax
\mciteBstWouldAddEndPuncttrue
\mciteSetBstMidEndSepPunct{\mcitedefaultmidpunct}
{\mcitedefaultendpunct}{\mcitedefaultseppunct}\relax
\EndOfBibitem
\bibitem[Sabatier(1911)]{sabatier1911hydrogenations}
Sabatier,~P. Hydrog{\'e}nations et d{\'e}shydrog{\'e}nations par catalyse.
  \emph{Berichte der deutschen chemischen Gesellschaft} \textbf{1911},
  \emph{44}, 1984--2001\relax
\mciteBstWouldAddEndPuncttrue
\mciteSetBstMidEndSepPunct{\mcitedefaultmidpunct}
{\mcitedefaultendpunct}{\mcitedefaultseppunct}\relax
\EndOfBibitem
\bibitem[Sarkar and Kundu(2018)Sarkar, and Kundu]{sarkar2018density}
Sarkar,~R.; Kundu,~T. Density functional theory studies on PVDF/ionic liquid
  composite systems. \emph{Journal of Chemical Sciences} \textbf{2018},
  \emph{130}, 1--18\relax
\mciteBstWouldAddEndPuncttrue
\mciteSetBstMidEndSepPunct{\mcitedefaultmidpunct}
{\mcitedefaultendpunct}{\mcitedefaultseppunct}\relax
\EndOfBibitem
\bibitem[Onyia \latin{et~al.}(2018)Onyia, Ikeri, and
  Nwobodo]{onyia2018theoretical}
Onyia,~A.; Ikeri,~H.; Nwobodo,~A. Theoretical study of the quantum confinement
  effects on quantum dots using particle in a box model. \emph{Journal of
  Ovonic Research} \textbf{2018}, \emph{14}, 49--54\relax
\mciteBstWouldAddEndPuncttrue
\mciteSetBstMidEndSepPunct{\mcitedefaultmidpunct}
{\mcitedefaultendpunct}{\mcitedefaultseppunct}\relax
\EndOfBibitem
\bibitem[Hassanpour \latin{et~al.}(2021)Hassanpour, Youseftabar-Miri, Nezhad,
  Ahmadi, and Ebrahimiasl]{hassanpour2021kinetic}
Hassanpour,~A.; Youseftabar-Miri,~L.; Nezhad,~P. D.~K.; Ahmadi,~S.;
  Ebrahimiasl,~S. Kinetic stability and NBO analysis of the C20-nAln nanocages
  (n= 1--5) using DFT investigation. \emph{Journal of Molecular Structure}
  \textbf{2021}, \emph{1233}, 130079\relax
\mciteBstWouldAddEndPuncttrue
\mciteSetBstMidEndSepPunct{\mcitedefaultmidpunct}
{\mcitedefaultendpunct}{\mcitedefaultseppunct}\relax
\EndOfBibitem
\bibitem[Kumar \latin{et~al.}(2011)Kumar, Jain, Kishor, and
  Ramaniah]{kumar2011chemical}
Kumar,~V.; Jain,~G.; Kishor,~S.; Ramaniah,~L.~M. Chemical reactivity analysis
  of some alkylating drug molecules--A density functional theory approach.
  \emph{Computational and Theoretical Chemistry} \textbf{2011}, \emph{968},
  18--25\relax
\mciteBstWouldAddEndPuncttrue
\mciteSetBstMidEndSepPunct{\mcitedefaultmidpunct}
{\mcitedefaultendpunct}{\mcitedefaultseppunct}\relax
\EndOfBibitem
\bibitem[Kumar \latin{et~al.}(2013)Kumar, Kishor, and
  Ramaniah]{kumar2013understanding}
Kumar,~V.; Kishor,~S.; Ramaniah,~L.~M. Understanding the antioxidant behavior
  of some vitamin molecules: a first-principles density functional approach.
  \emph{Journal of molecular modeling} \textbf{2013}, \emph{19},
  3175--3186\relax
\mciteBstWouldAddEndPuncttrue
\mciteSetBstMidEndSepPunct{\mcitedefaultmidpunct}
{\mcitedefaultendpunct}{\mcitedefaultseppunct}\relax
\EndOfBibitem
\bibitem[Vijayaraj \latin{et~al.}(2009)Vijayaraj, Subramanian, and
  Chattaraj]{vijayaraj2009comparison}
Vijayaraj,~R.; Subramanian,~V.; Chattaraj,~P. Comparison of global reactivity
  descriptors calculated using various density functionals: a QSAR perspective.
  \emph{Journal of chemical theory and computation} \textbf{2009}, \emph{5},
  2744--2753\relax
\mciteBstWouldAddEndPuncttrue
\mciteSetBstMidEndSepPunct{\mcitedefaultmidpunct}
{\mcitedefaultendpunct}{\mcitedefaultseppunct}\relax
\EndOfBibitem
\bibitem[Sharma \latin{et~al.}(2022)Sharma, Roondhe, Saxena, and
  Shukla]{sharma2022role}
Sharma,~V.; Roondhe,~B.; Saxena,~S.; Shukla,~A. Role of functionalized graphene
  quantum dots in hydrogen evolution reaction: A density functional theory
  study. \emph{International Journal of Hydrogen Energy} \textbf{2022},
  \emph{47}, 41748--41758\relax
\mciteBstWouldAddEndPuncttrue
\mciteSetBstMidEndSepPunct{\mcitedefaultmidpunct}
{\mcitedefaultendpunct}{\mcitedefaultseppunct}\relax
\EndOfBibitem
\bibitem[Vela and Gazquez(1990)Vela, and Gazquez]{vela1990relationship}
Vela,~A.; Gazquez,~J.~L. A relationship between the static dipole
  polarizability, the global softness, and the fukui function. \emph{Journal of
  the American Chemical Society} \textbf{1990}, \emph{112}, 1490--1492\relax
\mciteBstWouldAddEndPuncttrue
\mciteSetBstMidEndSepPunct{\mcitedefaultmidpunct}
{\mcitedefaultendpunct}{\mcitedefaultseppunct}\relax
\EndOfBibitem
\bibitem[Pearson(1992)]{pearson1992chemical}
Pearson,~R.~G. Chemical hardness and the electronic chemical potential.
  \emph{Inorganica chimica acta} \textbf{1992}, \emph{198}, 781--786\relax
\mciteBstWouldAddEndPuncttrue
\mciteSetBstMidEndSepPunct{\mcitedefaultmidpunct}
{\mcitedefaultendpunct}{\mcitedefaultseppunct}\relax
\EndOfBibitem
\bibitem[Greeley \latin{et~al.}(2006)Greeley, Jaramillo, Bonde, Chorkendorff,
  and N{\o}rskov]{greeley2006computational}
Greeley,~J.; Jaramillo,~T.~F.; Bonde,~J.; Chorkendorff,~I.; N{\o}rskov,~J.~K.
  Computational high-throughput screening of electrocatalytic materials for
  hydrogen evolution. \emph{Nature materials} \textbf{2006}, \emph{5},
  909--913\relax
\mciteBstWouldAddEndPuncttrue
\mciteSetBstMidEndSepPunct{\mcitedefaultmidpunct}
{\mcitedefaultendpunct}{\mcitedefaultseppunct}\relax
\EndOfBibitem
\bibitem[Bockris and Potter(1952)Bockris, and Potter]{bockris1952mechanism}
Bockris,~J.; Potter,~E. The mechanism of the cathodic hydrogen evolution
  reaction. \emph{Journal of The Electrochemical Society} \textbf{1952},
  \emph{99}, 169\relax
\mciteBstWouldAddEndPuncttrue
\mciteSetBstMidEndSepPunct{\mcitedefaultmidpunct}
{\mcitedefaultendpunct}{\mcitedefaultseppunct}\relax
\EndOfBibitem
\bibitem[Tafel(1905)]{tafel1905polarisation}
Tafel,~J. {\"U}ber die Polarisation bei kathodischer Wasserstoffentwicklung.
  \emph{Zeitschrift f{\"u}r physikalische Chemie} \textbf{1905}, \emph{50},
  641--712\relax
\mciteBstWouldAddEndPuncttrue
\mciteSetBstMidEndSepPunct{\mcitedefaultmidpunct}
{\mcitedefaultendpunct}{\mcitedefaultseppunct}\relax
\EndOfBibitem
\bibitem[Lasia(2019)]{lasia2019mechanism}
Lasia,~A. Mechanism and kinetics of the hydrogen evolution reaction.
  \emph{International journal of hydrogen energy} \textbf{2019}, \emph{44},
  19484--19518\relax
\mciteBstWouldAddEndPuncttrue
\mciteSetBstMidEndSepPunct{\mcitedefaultmidpunct}
{\mcitedefaultendpunct}{\mcitedefaultseppunct}\relax
\EndOfBibitem
\bibitem[Parsons(1958)]{parsons1958rate}
Parsons,~R. The rate of electrolytic hydrogen evolution and the heat of
  adsorption of hydrogen. \emph{Transactions of the Faraday Society}
  \textbf{1958}, \emph{54}, 1053--1063\relax
\mciteBstWouldAddEndPuncttrue
\mciteSetBstMidEndSepPunct{\mcitedefaultmidpunct}
{\mcitedefaultendpunct}{\mcitedefaultseppunct}\relax
\EndOfBibitem
\bibitem[Zhong \latin{et~al.}(2018)Zhong, Zhang, Wang, Zhang, Wei, Wu, Li,
  Meng, Bao, and Yan]{zhong2018engineering}
Zhong,~H.-x.; Zhang,~Q.; Wang,~J.; Zhang,~X.-b.; Wei,~X.-l.; Wu,~Z.-j.; Li,~K.;
  Meng,~F.-l.; Bao,~D.; Yan,~J.-m. Engineering ultrathin C3N4 quantum dots on
  graphene as a metal-free water reduction electrocatalyst. \emph{Acs
  Catalysis} \textbf{2018}, \emph{8}, 3965--3970\relax
\mciteBstWouldAddEndPuncttrue
\mciteSetBstMidEndSepPunct{\mcitedefaultmidpunct}
{\mcitedefaultendpunct}{\mcitedefaultseppunct}\relax
\EndOfBibitem
\bibitem[Meng \latin{et~al.}(2017)Meng, Li, Li, and Yang]{meng2017carbon}
Meng,~S.; Li,~B.; Li,~S.; Yang,~S. Carbon nitride frameworks padded with
  graphene as efficient metal-free catalyst for HER in acidic and alkali
  electrolytes. \emph{Materials Research Express} \textbf{2017}, \emph{4},
  055602\relax
\mciteBstWouldAddEndPuncttrue
\mciteSetBstMidEndSepPunct{\mcitedefaultmidpunct}
{\mcitedefaultendpunct}{\mcitedefaultseppunct}\relax
\EndOfBibitem
\bibitem[Shinde \latin{et~al.}(2015)Shinde, Sami, and
  Lee]{shinde2015electrocatalytic}
Shinde,~S.; Sami,~A.; Lee,~J.-H. Electrocatalytic hydrogen evolution using
  graphitic carbon nitride coupled with nanoporous graphene co-doped by S and
  Se. \emph{Journal of Materials Chemistry A} \textbf{2015}, \emph{3},
  12810--12819\relax
\mciteBstWouldAddEndPuncttrue
\mciteSetBstMidEndSepPunct{\mcitedefaultmidpunct}
{\mcitedefaultendpunct}{\mcitedefaultseppunct}\relax
\EndOfBibitem
\bibitem[Shinde \latin{et~al.}(2015)Shinde, Sami, and Lee]{shinde2015nitrogen}
Shinde,~S.~S.; Sami,~A.; Lee,~J.-H. Nitrogen-and phosphorus-doped nanoporous
  graphene/graphitic carbon nitride hybrids as efficient electrocatalysts for
  hydrogen evolution. \emph{ChemCatChem} \textbf{2015}, \emph{7},
  3873--3880\relax
\mciteBstWouldAddEndPuncttrue
\mciteSetBstMidEndSepPunct{\mcitedefaultmidpunct}
{\mcitedefaultendpunct}{\mcitedefaultseppunct}\relax
\EndOfBibitem
\end{mcitethebibliography}

\end{document}


\title{Supporting Information: First-Principles Study of  Penta-CN$_{2}$
Quantum Dots for Efficient Hydrogen Evolution Reaction }
\author{Rupali Jindal, Rachana Yogi, and Alok Shukla}
\affiliation{Department of Physics, Indian Institute of Technology Bombay, Powai,
Mumbai 400076, India}
\email{rupalijindalpu@gmail.com, yogirachana04@gmail.com, shukla@iitb.ac.in}

\maketitle
\begin{figure}[H]
\centering{}\includegraphics[scale=0.4]{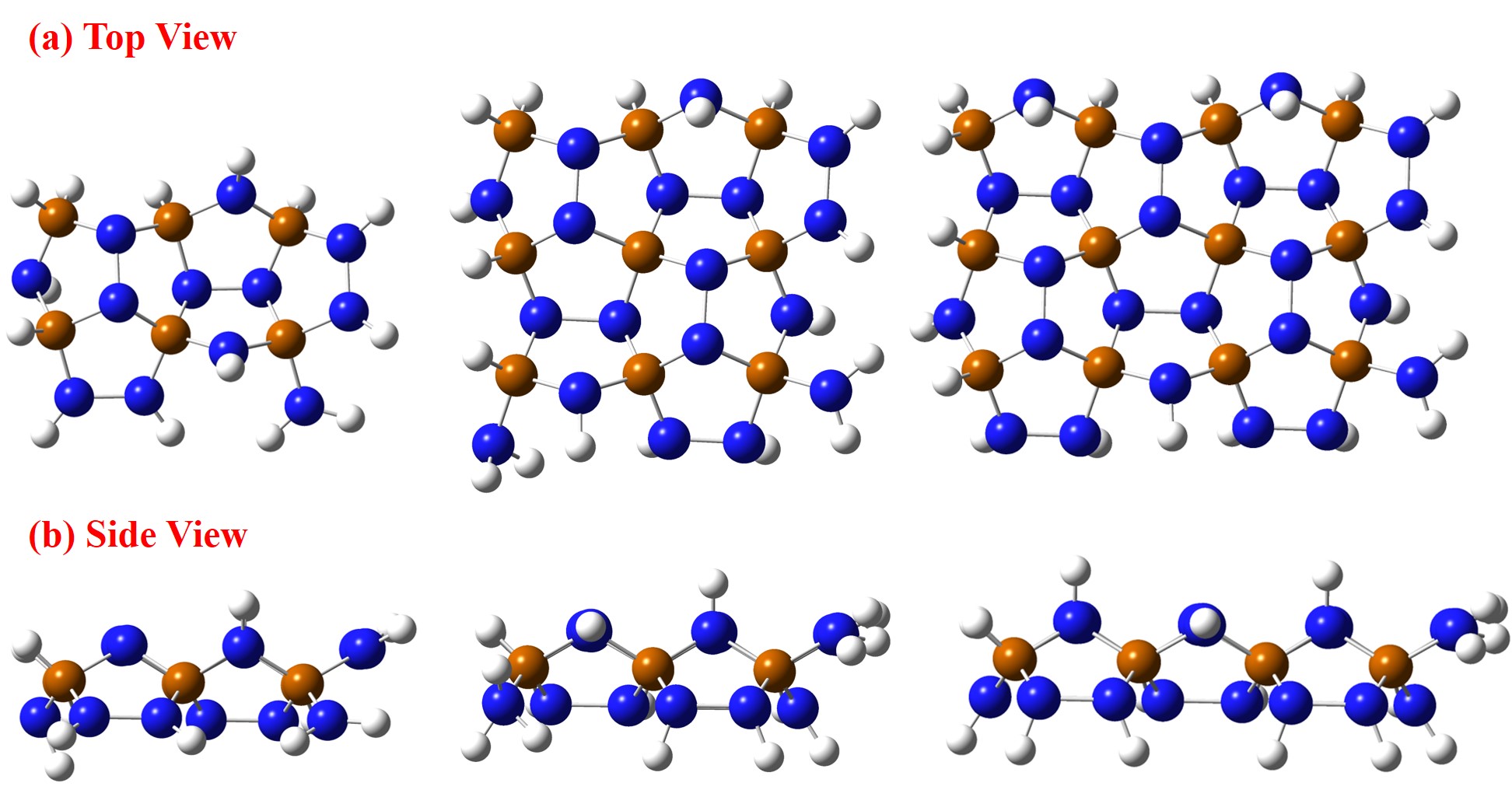}\caption{\label{fig:initial CN2-penta}Schematics of initial structures of
$3\times3$, $3\times4$, $4\times4$ $\mathrm{\mathrm{penta-CN_{2}}}$
QDs with: (a) Top view and (b) side view, where brown, blue, and grey
spheres represent carbon, nitrogen, and hydrogen atoms, respectively.}
\end{figure}
\begin{figure}[H]
\centering{}\includegraphics[scale=0.45]{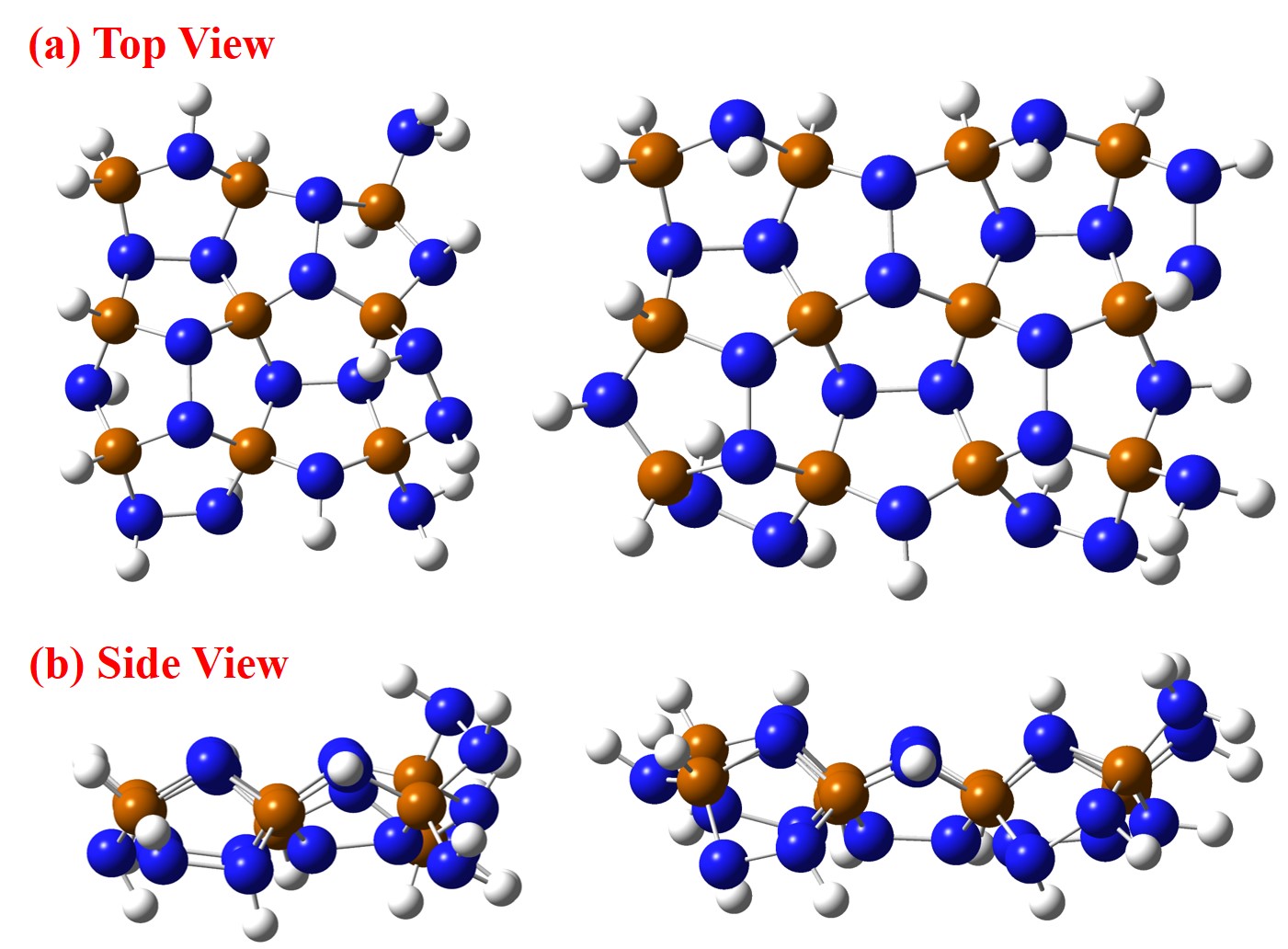}\caption{\label{fig:optimized CN2-penta-1}The optimized structures of $3\times4$,
$4\times4$ $\mathrm{\mathrm{penta-CN_{2}}}$ QDs with (a) Top view
and (b) Side view, where brown, blue, and grey represents carbon,
nitrogen, and hydrogen respectively.}
\end{figure}
\begin{figure}[H]
\centering{}\includegraphics[scale=0.52]{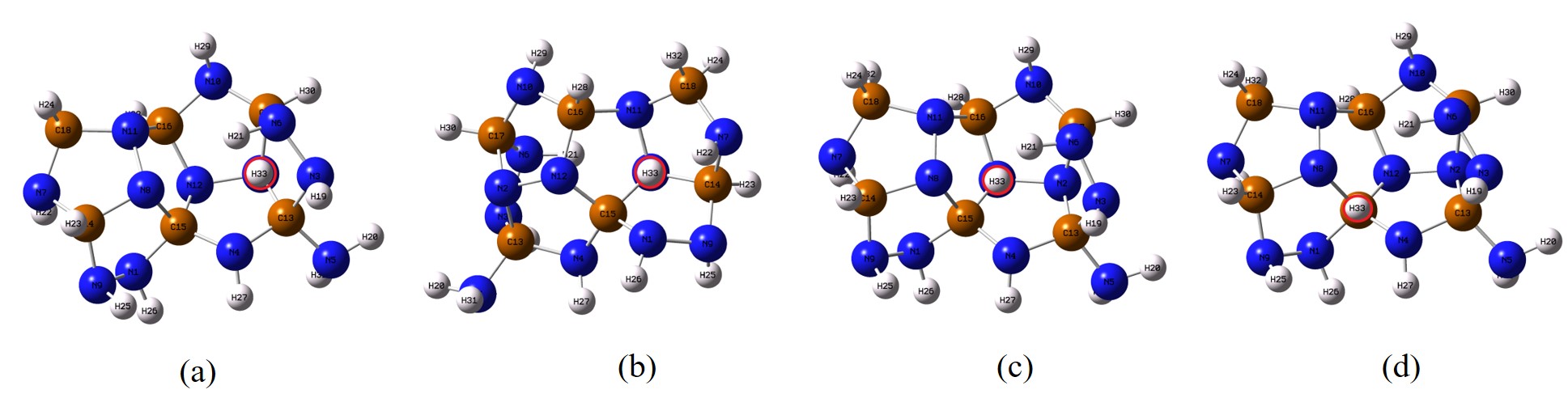}\caption{\label{fig:all-Initial-structure}Initial structures of $\mathrm{\mathrm{3\times3\:penta-CN_{2}}}$
QD with H adsorbed on the top of (a) N2 (b) N8 (c) N12 (d) C15}
\end{figure}
\begin{figure}[H]
\centering{}\includegraphics[scale=0.6]{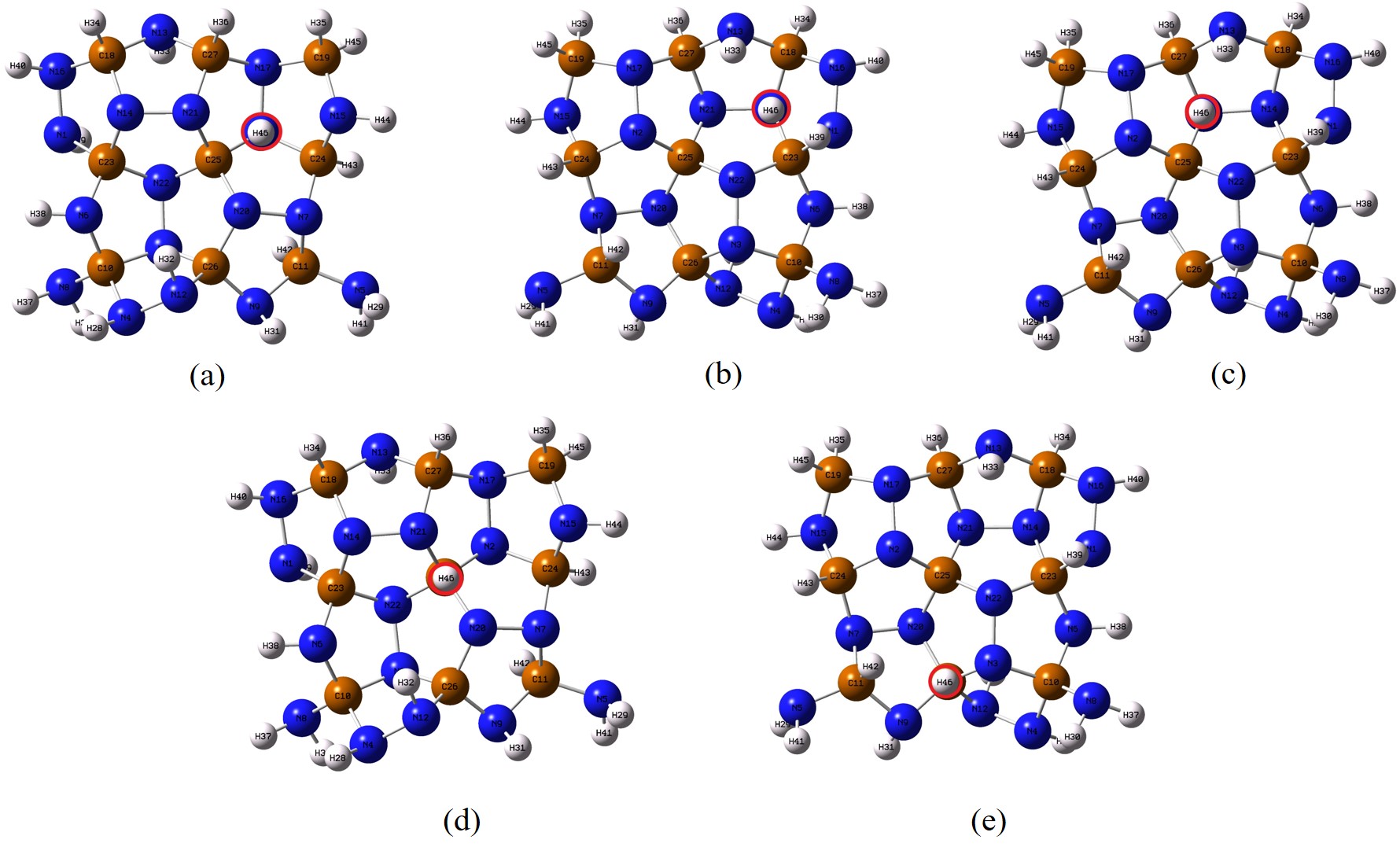}\caption{\label{fig:all-Initial-structure-3x4}Initial structures of $\mathrm{\mathrm{3\times4\:\,penta-CN_{2}}}$
QD with H placed on the top of (a) N2 (b) N14 (c) N21 (d) C25 (e)
C26.}
\end{figure}
\begin{figure}[H]
\centering{}\includegraphics[scale=0.55]{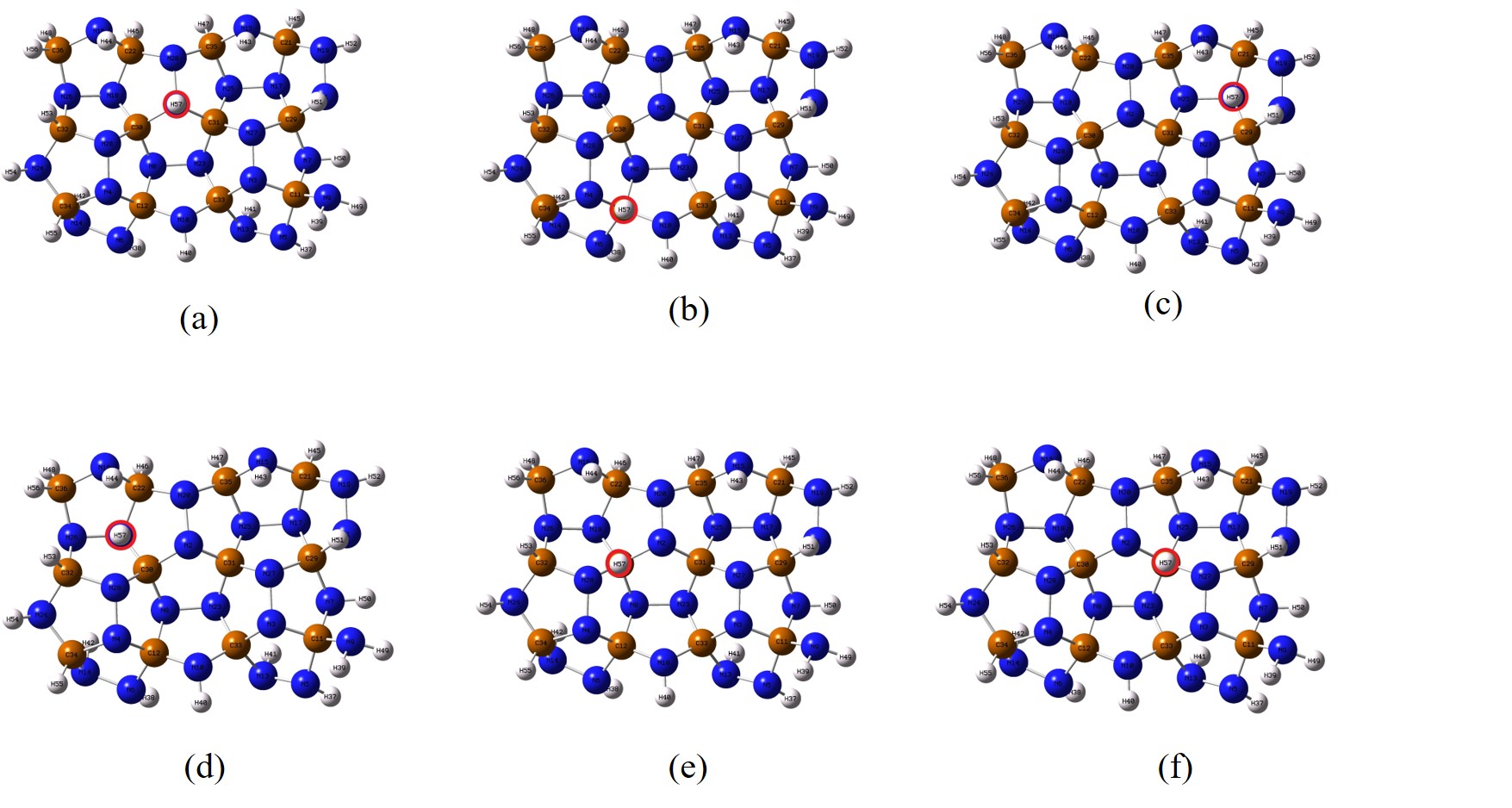}\caption{\label{fig:all-Initial-structure-4x4}Initial structures of $\mathrm{\mathrm{4\times4\:\,penta-CN_{2}}}$
QD with H placed on the top of (a) N2 (b) C12 (c) N17 (d) N18 (e)
C30 (f) C31.}
\end{figure}
\begin{figure}[H]
\centering{}\includegraphics[scale=0.52]{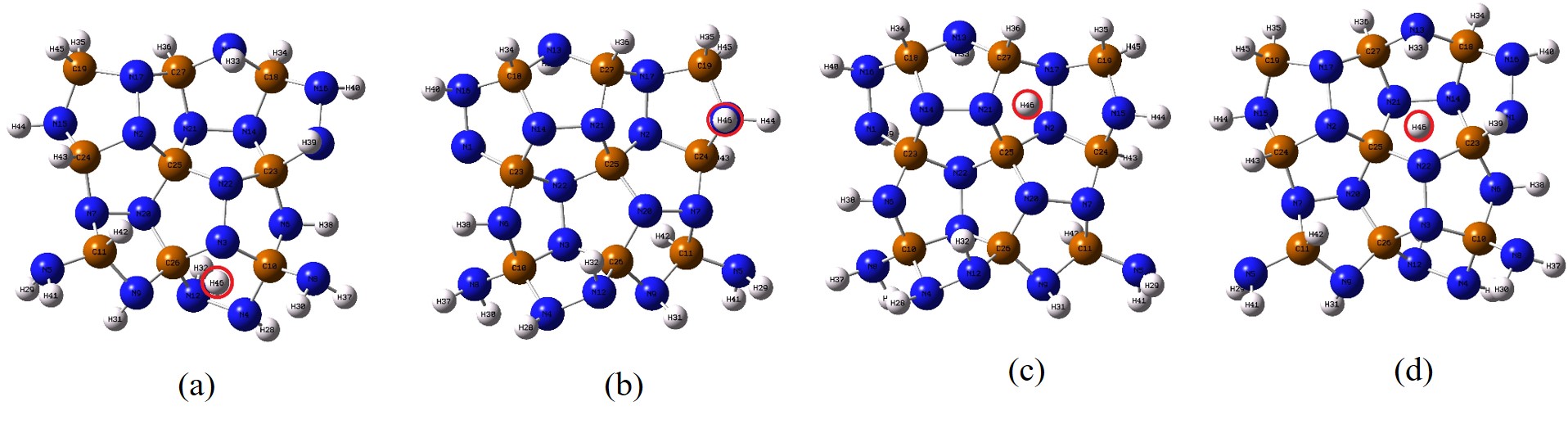} \caption{\label{fig:all-Optimized-structure-3x4}Optimized structures of $\mathrm{\mathrm{3\times4\:\,penta-CN_{2}}}$
QD with H-adsorbed positions (a) N3 (b) N15 (c) N21 (d) N22.}
\end{figure}
\begin{figure}[H]
\centering{}\includegraphics[scale=0.55]{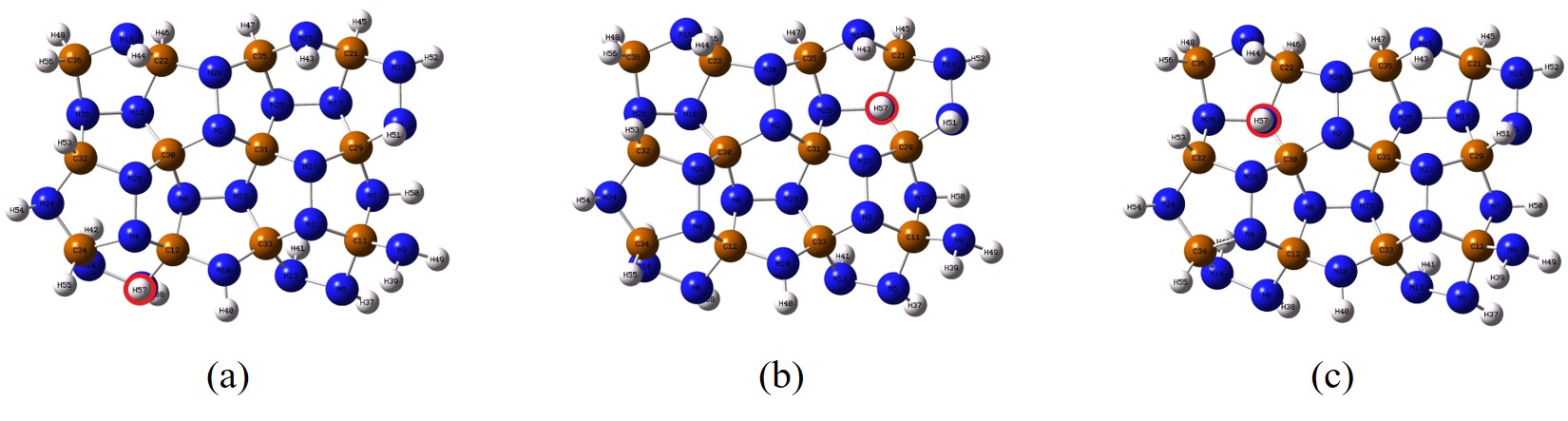} \caption{\label{fig:all-Optimized-structure-4x4}Optimized structures of $\mathrm{\mathrm{4\times4\:\,penta-CN_{2}}}$
QD with H-adsorbed positions (a) N4 (c) N27 (d) N28.}
\end{figure}

\textcolor{black}{In the first step, the geometries of the penta-CN$_{2}$
quantum dots (QDs) of increasing sizes, $3\times3$, $3\times4$,
$4\times4$ are optimized with purpose of their possible use as catalysts
in the hydrogen evolution reaction. The initial structures of the
QDs are illustrated in Fig. \ref{fig:initial CN2-penta}, and the
final optimized structures of $3\times4$ and $4\times4$ are shown
in Fig. \ref{fig:optimized CN2-penta-1}, while the optimized structure
of }$3\times3$\textcolor{black}{{} is displayed in the main manuscript.
In the next step, the H atom is placed on the top of the optimized
structures of $3\times3$, $3\times4$ and $4\times4$ penta-CN$_{2}$
QDs (see Figs. \ref{fig:all-Initial-structure}, \ref{fig:all-Initial-structure-3x4}
and \ref{fig:all-Initial-structure-4x4}). After optimization, the
H-adsorbed structures of }$3\times4$\textcolor{black}{{} and }$4\times4$\textcolor{black}{{}
QDs are presented in Fig. \ref{fig:all-Optimized-structure-3x4} and
Fig. \ref{fig:all-Optimized-structure-4x4}, while those for $3\times3$
are shown in the main manuscript. For a more detailed understanding
of the geometries, we have listed the individual bond lengths and
bond angles of the optimized }$3\times3$\textcolor{black}{{} penta-CN$_{2}$
QD in Tables \ref{tab:bond_lengths} and \ref{tab:bond_angles}, respectively.
Tables \ref{tab:Vertical_distances-3x4} and \ref{tab:Vertical_distances-4x4-1}
present the information regarding the initial and final structures
of the }$3\times4$\textcolor{black}{{} and }$4\times4$\textcolor{black}{{}
penta-CN$_{2}$ QDs with H on top, along with their respective optimized
distances (i.e., distance of H from the closest QD atom). The electronic
properties depicted by the global reactivity descriptors are presented
in Table \ref{tab:GRD-3x4} and \ref{tab:GRD-4x4}. The essential
descriptors utilized to evaluate the HER catalytic activity are displayed
in Table \ref{tab:HER-descriptors-3x4} and Table \ref{tab:HER-descriptors-4x4}.
The atomic coordinates of all structures can be provided upon request.}

\begin{table}[H]
\centering{}%
\begin{tabular}{cccccccccccccccc}
\toprule 
Bond lengths ($\text{Å}$) &  &  & Pure &  &  & N2 &  &  & N8 &  &  & N12 &  &  & C15\tabularnewline
\midrule
\midrule 
R1 & N1-N9 &  & 1.4431 &  &  & 1.4435 &  &  & 1.4387 &  &  & 1.4428 &  &  & 1.4429\tabularnewline
\midrule 
R2 & N1-C15 &  & 1.4618 &  &  & 1.4604 &  &  & 1.4628 &  &  & 1.4625 &  &  & 1.4625\tabularnewline
\midrule 
R3 & N1-H26 &  & 1.0147 &  &  & 1.0148 &  &  & 1.0143 &  &  & 1.0147 &  &  & 1.0147\tabularnewline
\midrule 
R4 & N2-N12 &  & 1.4793 &  &  & 1.4808 &  &  & 1.4778 &  &  & 1.4789 &  &  & 1.4789\tabularnewline
\midrule 
R5 & N2-C13 &  & 1.5015 &  &  & 1.5007 &  &  & 1.5011 &  &  & 1.5013 &  &  & 1.5012\tabularnewline
\midrule 
R6 & N2-C17 &  & 1.4931 &  &  & 1.4929 &  &  & 1.4935 &  &  & 1.4933 &  &  & 1.4933\tabularnewline
\midrule 
R7 & N3-N6 &  & 1.4437 &  &  & 1.4434 &  &  & 1.4433 &  &  & 1.443 &  &  & 1.4429\tabularnewline
\midrule 
R8 & N3-C13 &  & 1.4779 &  &  & 1.4772 &  &  & 1.4776 &  &  & 1.4786 &  &  & 1.4787\tabularnewline
\midrule 
R9 & N3-H19 &  & 1.0164 &  &  & 1.0164 &  &  & 1.0164 &  &  & 1.0166 &  &  & 1.0166\tabularnewline
\midrule 
R10 & N4-C13 &  & 1.4614 &  &  & 1.4635 &  &  & 1.4615 &  &  & 1.4601 &  &  & 1.46\tabularnewline
\midrule 
R11 & N4-C15 &  & 1.4432 &  &  & 1.4448 &  &  & 1.4424 &  &  & 1.4375 &  &  & 1.4375\tabularnewline
\midrule 
R12 & N4-H27 &  & 1.0131 &  &  & 1.013 &  &  & 1.013 &  &  & 1.0124 &  &  & 1.0124\tabularnewline
\midrule 
R13 & N5-C13 &  & 1.4391 &  &  & 1.4386 &  &  & 1.4387 &  &  & 1.4384 &  &  & 1.4384\tabularnewline
\midrule 
R14 & N5-H20 &  & 1.0163 &  &  & 1.0162 &  &  & 1.0163 &  &  & 1.0162 &  &  & 1.0162\tabularnewline
\midrule 
R15 & N5-H31 &  & 1.0172 &  &  & 1.0171 &  &  & 1.0172 &  &  & 1.0171 &  &  & 1.0171\tabularnewline
\midrule 
R16 & N6-C17 &  & 1.4745 &  &  & 1.4742 &  &  & 1.4745 &  &  & 1.4727 &  &  & 1.4726\tabularnewline
\midrule 
R17 & N6-H21 &  & 1.0294 &  &  & 1.0289 &  &  & 1.0295 &  &  & 1.0291 &  &  & 1.0291\tabularnewline
\midrule 
R18 & N7-C14 &  & 1.4545 &  &  & 1.4541 &  &  & 1.4531 &  &  & 1.4528 &  &  & 1.4529\tabularnewline
\midrule 
R19 & N7-C18 &  & 1.4535 &  &  & 1.4534 &  &  & 1.4534 &  &  & 1.454 &  &  & 1.454\tabularnewline
\midrule 
R20 & N7-H22 &  & 1.0195 &  &  & 1.0194 &  &  & 1.0204 &  &  & 1.0195 &  &  & 1.0195\tabularnewline
\midrule 
R21 & N8-N11 &  & 1.4697 &  &  & 1.4684 &  &  & 1.4698 &  &  & 1.4598 &  &  & 1.4599\tabularnewline
\midrule 
R22 & N8-C14 &  & 1.5227 &  &  & 1.524 &  &  & 1.5222 &  &  & 1.5232 &  &  & 1.5231\tabularnewline
\midrule 
R23 & N8-C15 &  & 1.5091 &  &  & 1.5088 &  &  & 1.5071 &  &  & 1.5129 &  &  & 1.5128\tabularnewline
\midrule 
R24 & N9-C14 &  & 1.4501 &  &  & 1.4492 &  &  & 1.4509 &  &  & 1.448 &  &  & 1.448\tabularnewline
\midrule 
R25 & N9-H25 &  & 1.0262 &  &  & 1.0269 &  &  & 1.0259 &  &  & 1.0261 &  &  & 1.026\tabularnewline
\midrule 
R26 & N10-C16 &  & 1.4623 &  &  & 1.4616 &  &  & 1.4624 &  &  & 1.4605 &  &  & 1.4604\tabularnewline
\midrule 
R27 & N10-C17 &  & 1.4359 &  &  & 1.4361 &  &  & 1.4355 &  &  & 1.4366 &  &  & 1.4365\tabularnewline
\midrule 
R28 & N10-H29 &  & 1.0154 &  &  & 1.0154 &  &  & 1.0154 &  &  & 1.0153 &  &  & 1.0153\tabularnewline
\midrule 
R29 & N11-C16 &  & 1.4769 &  &  & 1.4774 &  &  & 1.4767 &  &  & 1.4797 &  &  & 1.4797\tabularnewline
\midrule 
R30 & N11-C18 &  & 1.4878 &  &  & 1.4877 &  &  & 1.4879 &  &  & 1.4892 &  &  & 1.4892\tabularnewline
\midrule 
R31 & N12-C15 &  & 1.4657 &  &  & 1.4653 &  &  & 1.465 &  &  & 1.4654 &  &  & 1.4654\tabularnewline
\midrule 
R32 & N12-C16 &  & 1.473 &  &  & 1.4735 &  &  & 1.4729 &  &  & 1.4716 &  &  & 1.4715\tabularnewline
\midrule 
R33 & C14-H23 &  & 1.0929 &  &  & 1.0926 &  &  & 1.0932 &  &  & 1.0924 &  &  & 1.0924\tabularnewline
\midrule 
R34 & C16-H28 &  & 1.0968 &  &  & 1.0967 &  &  & 1.0964 &  &  & 1.0969 &  &  & 1.0969\tabularnewline
\midrule 
R35 & C17-H30 &  & 1.093 &  &  & 1.0929 &  &  & 1.0929 &  &  & 1.093 &  &  & 1.093\tabularnewline
\midrule 
R36 & C18-H24 &  & 1.0958 &  &  & 1.0957 &  &  & 1.0958 &  &  & 1.0954 &  &  & 1.0954\tabularnewline
\midrule 
R37 & C18-H32 &  & 1.0935 &  &  & 1.0935 &  &  & 1.0933 &  &  & 1.0932 &  &  & 1.0932\tabularnewline
\midrule 
 &  &  &  &  &  &  &  &  &  &  &  &  &  &  & \tabularnewline
\bottomrule
\end{tabular}\caption{\label{tab:bond_lengths}The individual bond lengths of $\mathrm{3\times3\:penta-CN_{2}}$
QD after optimization}
\end{table}

\begin{table}
\centering{}%
\begin{tabular}{ccccccccccccccccc}
\toprule 
Angles ( in degrees) &  &  & Pure &  &  & N2 &  &  & N8 &  &  & N12 &  &  & C15 & \tabularnewline
\midrule
\midrule 
A1 & N9-N1-C15 &  & 107.3151 &  &  & 107.2866 &  &  & 107.8466 &  &  & 107.6708 &  &  & 107.6686 & \tabularnewline
\midrule 
A2 & N9-N1-H26 &  & 111.3891 &  &  & 111.3844 &  &  & 111.9234 &  &  & 111.4356 &  &  & 111.4251 & \tabularnewline
\midrule 
A3 & C15-N1-H26 &  & 115.8695 &  &  & 115.9461 &  &  & 116.4809 &  &  & 115.7411 &  &  & 115.734 & \tabularnewline
\midrule 
A4 & N12-N2-C13 &  & 107.0589 &  &  & 107.1508 &  &  & 107.0961 &  &  & 107.0497 &  &  & 107.0492 & \tabularnewline
\midrule 
A5 & N12-N2-C17 &  & 105.107 &  &  & 105.075 &  &  & 105.1275 &  &  & 105.135 &  &  & 105.1334 & \tabularnewline
\midrule 
A6 & C13-N2-C17 &  & 104.674 &  &  & 104.7098 &  &  & 104.6704 &  &  & 104.7 &  &  & 104.6986 & \tabularnewline
\midrule 
A7 & N6-N3-C13 &  & 106.4701 &  &  & 106.7059 &  &  & 106.4097 &  &  & 106.5427 &  &  & 106.5339 & \tabularnewline
\midrule 
A8 & N6-N3-C19 &  & 109.5942 &  &  & 109.6496 &  &  & 109.6438 &  &  & 109.4104 &  &  & 109.4081 & \tabularnewline
\midrule 
A9 & C13-N3-C19 &  & 110.7627 &  &  & 110.8627 &  &  & 110.8114 &  &  & 110.7231 &  &  & 110.7158 & \tabularnewline
\midrule 
A10 & C13-N4-C15 &  & 111.0479 &  &  & 111.0118 &  &  & 111.1865 &  &  & 111.4698 &  &  & 111.4801 & \tabularnewline
\midrule 
A11 & C13-N4-H27 &  & 110.6196 &  &  & 110.7105 &  &  & 110.7938 &  &  & 111.3918 &  &  & 111.3962 & \tabularnewline
\midrule 
A12 & C15-N4-H27 &  & 114.2598 &  &  & 114.3101 &  &  & 114.5393 &  &  & 115.0549 &  &  & 115.0671 & \tabularnewline
\midrule 
A13 & C13-N5-H20 &  & 107.8688 &  &  & 107.8539 &  &  & 107.9168 &  &  & 107.943 &  &  & 107.9398 & \tabularnewline
\midrule 
A14 & C13-N5-H31 &  & 109.2075 &  &  & 109.2667 &  &  & 109.2689 &  &  & 109.3239 &  &  & 109.3204 & \tabularnewline
\midrule 
A15 & H20-N5-H31 &  & 107.3409 &  &  & 107.4252 &  &  & 107.372 &  &  & 107.4857 &  &  & 107.4842 & \tabularnewline
\midrule 
A16 & N3-N6-C17 &  & 98.273 &  &  & 98.4414 &  &  & 98.2493 &  &  & 98.4877 &  &  & 98.4892 & \tabularnewline
\midrule 
A17 & N3-N6-C21 &  & 108.8434 &  &  & 108.9229 &  &  & 108.8913 &  &  & 108.9822 &  &  & 108.986 & \tabularnewline
\midrule 
A18 & C17-N6-C21 &  & 107.4994 &  &  & 107.6301 &  &  & 107.5429 &  &  & 107.9662 &  &  & 107.9594 & \tabularnewline
\midrule 
A19 & C14-N7-C18 &  & 103.5784 &  &  & 103.6396 &  &  & 103.6871 &  &  & 103.4494 &  &  & 103.4375 & \tabularnewline
\midrule 
A20 & C14-N7-H22 &  & 106.0783 &  &  & 106.1087 &  &  & 106.7688 &  &  & 106.3083 &  &  & 106.3103 & \tabularnewline
\midrule 
A21 & C18-N7-H22 &  & 109.45 &  &  & 109.4593 &  &  & 109.5944 &  &  & 109.5594 &  &  & 109.5594 & \tabularnewline
\midrule 
A22 & N11-N8-C14 &  & 105.799 &  &  & 105.8231 &  &  & 105.8098 &  &  & 106.2463 &  &  & 106.245 & \tabularnewline
\midrule 
A23 & N11-N8-C15 &  & 106.6079 &  &  & 106.5632 &  &  & 106.6727 &  &  & 107.0082 &  &  & 107.013 & \tabularnewline
\midrule 
A24 & C14-N8-C15 &  & 104.4648 &  &  & 104.4456 &  &  & 104.7317 &  &  & 104.9314 &  &  & 104.9358 & \tabularnewline
\midrule 
A25 & N1-N9-C14 &  & 99.7354 &  &  & 99.7668 &  &  & 100.1191 &  &  & 99.9463 &  &  & 99.9477 & \tabularnewline
\midrule 
A26 & N1-N9-H25 &  & 109.9986 &  &  & 109.9475 &  &  & 109.937 &  &  & 110.1462 &  &  & 110.1448 & \tabularnewline
\midrule 
A27 & C14-N9-H25 &  & 106.559 &  &  & 106.41 &  &  & 106.7926 &  &  & 106.5064 &  &  & 106.5038 & \tabularnewline
\midrule 
A28 & C16-N10-C17 &  & 107.9727 &  &  & 107.859 &  &  & 108.0857 &  &  & 108.0897 &  &  & 108.0968 & \tabularnewline
\midrule 
A29 & C16-N10-H29 &  & 111.0273 &  &  & 111.0791 &  &  & 111.1073 &  &  & 111.3224 &  &  & 111.3354 & \tabularnewline
\midrule 
A30 & C17-N10-H29 &  & 112.8501 &  &  & 112.8583 &  &  & 112.9174 &  &  & 112.9974 &  &  & 113.006 & \tabularnewline
\midrule 
A31 & N8-N11-C16 &  & 107.9912 &  &  & 108.0693 &  &  & 108.0392 &  &  & 108.1378 &  &  & 108.1359 & \tabularnewline
\midrule 
A32 & N8-N11-C18 &  & 105.5326 &  &  & 105.5635 &  &  & 105.5324 &  &  & 105.3299 &  &  & 105.324 & \tabularnewline
\midrule 
A33 & C16-N11-C18 &  & 118.0448 &  &  & 118.0833 &  &  & 118.0251 &  &  & 117.984 &  &  & 117.9695 & \tabularnewline
\midrule 
A34 & N2-N12-C15 &  & 108.4675 &  &  & 108.4473 &  &  & 108.5631 &  &  & 108.5444 &  &  & 108.5504 & \tabularnewline
\midrule 
A35 & N2-N12-C16 &  & 107.6872 &  &  & 107.6148 &  &  & 107.7551 &  &  & 107.7389 &  &  & 107.7422 & \tabularnewline
\midrule 
A36 & C15-N12-C16 &  & 108.0305 &  &  & 107.984 &  &  & 108.1003 &  &  & 108.2652 &  &  & 108.2702 & \tabularnewline
\midrule 
A37 & N2-C13-N3 &  & 102.1647 &  &  & 102.3579 &  &  & 102.131 &  &  & 102.1978 &  &  & 102.1973 & \tabularnewline
\midrule 
A38 & N2-C13-N4 &  & 105.46 &  &  & 105.2994 &  &  & 105.3755 &  &  & 105.2447 &  &  & 105.2444 & \tabularnewline
\midrule 
A39 & N2-C13-N5 &  & 116.4825 &  &  & 116.4311 &  &  & 116.6076 &  &  & 116.6061 &  &  & 116.6072 & \tabularnewline
\midrule 
A40 & N3-C13-N4 &  & 115.3218 &  &  & 115.4699 &  &  & 115.2546 &  &  & 115.473 &  &  & 115.4657 & \tabularnewline
\midrule 
A41 & N3-C13-N5 &  & 108.1011 &  &  & 108.1907 &  &  & 108.1323 &  &  & 108.047 &  &  & 108.048 & \tabularnewline
\midrule 
A42 & N4-C13-N5 &  & 109.4226 &  &  & 109.2293 &  &  & 109.4471 &  &  & 109.4008 &  &  & 109.406 & \tabularnewline
\midrule 
A43 & N7-C14-N8 &  & 107.4015 &  &  & 107.3283 &  &  & 107.4964 &  &  & 107.0001 &  &  & 107.0038 & \tabularnewline
\midrule 
A44 & N7-C14-N9 &  & 111.6942 &  &  & 111.7538 &  &  & 111.9378 &  &  & 112.2106 &  &  & 112.2146 & \tabularnewline
\midrule 
 &  &  &  &  &  &  &  &  &  &  &  &  &  &  &  & \tabularnewline
\bottomrule
\end{tabular}
\end{table}

\begin{table}
\centering{}%
\begin{tabular}{ccccccccccccccccc}
\toprule 
Angles ( in degrees) &  &  & Pure &  &  & N2 &  &  & N8 &  &  & N12 &  &  & C15 & \tabularnewline
\midrule
\midrule 
A45 & N7-C14-H23 &  & 111.7304 &  &  & 111.8621 &  &  & 111.6958 &  &  & 111.9675 &  &  & 111.9563 & \tabularnewline
\midrule 
A46 & N8-C14-N9 &  & 105.0399 &  &  & 104.9884 &  &  & 105.0772 &  &  & 104.8091 &  &  & 104.8091 & \tabularnewline
\midrule 
A47 & N8-C14-H23 &  & 109.8711 &  &  & 109.752 &  &  & 109.7761 &  &  & 109.5069 &  &  & 109.5139 & \tabularnewline
\midrule 
A48 & N9-C14-H23 &  & 110.8218 &  &  & 110.8536 &  &  & 110.5849 &  &  & 110.9896 &  &  & 110.9874 & \tabularnewline
\midrule 
A49 & N1-C15-N4 &  & 114.9632 &  &  & 114.8785 &  &  & 114.6755 &  &  & 115.3807 &  &  & 115.3791 & \tabularnewline
\midrule 
A50 & N1-C15-N8 &  & 102.5254 &  &  & 102.5477 &  &  & 102.5055 &  &  & 101.9448 &  &  & 101.9478 & \tabularnewline
\midrule 
A51 & N1-C15-N12 &  & 112.6947 &  &  & 112.8939 &  &  & 112.8189 &  &  & 112.686 &  &  & 112.6839 & \tabularnewline
\midrule 
A52 & N4-C15-N8 &  & 113.006 &  &  & 113.0442 &  &  & 113.2245 &  &  & 113.559 &  &  & 113.5677 & \tabularnewline
\midrule 
A53 & N4-C15-N12 &  & 106.1316 &  &  & 105.9598 &  &  & 106.0781 &  &  & 106.0341 &  &  & 106.0266 & \tabularnewline
\midrule 
A54 & N8-C15-N12 &  & 107.4264 &  &  & 107.4448 &  &  & 107.4766 &  &  & 107.1059 &  &  & 107.1061 & \tabularnewline
\midrule 
A55 & N10-C16-N11 &  & 113.8106 &  &  & 113.8728 &  &  & 113.8331 &  &  & 114.048 &  &  & 114.0521 & \tabularnewline
\midrule 
A56 & N10-C16-N12 &  & 104.6702 &  &  & 104.5823 &  &  & 104.6532 &  &  & 104.7229 &  &  & 104.7233 & \tabularnewline
\midrule 
A57 & N10-C16-H28 &  & 109.1837 &  &  & 109.202 &  &  & 109.2705 &  &  & 109.2434 &  &  & 109.2435 & \tabularnewline
\midrule 
A58 & N11-C16-N12 &  & 108.2206 &  &  & 108.1511 &  &  & 108.1444 &  &  & 108.0694 &  &  & 108.0674 & \tabularnewline
\midrule 
A59 & N11-C16-H28 &  & 108.9526 &  &  & 109.0065 &  &  & 109.0221 &  &  & 108.718 &  &  & 108.7147 & \tabularnewline
\midrule 
A60 & N12-C16-H28 &  & 112.0059 &  &  & 112.0239 &  &  & 111.9103 &  &  & 112.0571 &  &  & 112.0581 & \tabularnewline
\midrule 
A61 & N2-C17-N6 &  & 106.6964 &  &  & 106.7239 &  &  & 106.6527 &  &  & 106.8014 &  &  & 106.7986 & \tabularnewline
\midrule 
A62 & N2-C17-N10 &  & 103.2926 &  &  & 103.1777 &  &  & 103.3076 &  &  & 103.282 &  &  & 103.28 & \tabularnewline
\midrule 
A63 & N2-C17-H30 &  & 109.012 &  &  & 109.0208 &  &  & 108.9818 &  &  & 108.9349 &  &  & 108.9347 & \tabularnewline
\midrule 
A64 & N6-C17-N10 &  & 117.2672 &  &  & 117.2436 &  &  & 117.3335 &  &  & 117.2583 &  &  & 117.2591 & \tabularnewline
\midrule 
A65 & N6-C17-H30 &  & 107.7527 &  &  & 107.8596 &  &  & 107.7097 &  &  & 107.8261 &  &  & 107.8283 & \tabularnewline
\midrule 
A66 & N10-C17-H30 &  & 112.3733 &  &  & 112.3616 &  &  & 112.4009 &  &  & 112.2895 &  &  & 112.2909 & \tabularnewline
\midrule 
A67 & N7-C18-N11 &  & 108.4447 &  &  & 108.394 &  &  & 108.4569 &  &  & 108.2965 &  &  & 108.2983 & \tabularnewline
\midrule 
A68 & N7-C18-H24 &  & 109.4028 &  &  & 109.4354 &  &  & 109.4549 &  &  & 109.4472 &  &  & 109.4448 & \tabularnewline
\midrule 
A69 & N7-C18-H32 &  & 111.9436 &  &  & 111.9458 &  &  & 111.8768 &  &  & 112.0095 &  &  & 112.0117 & \tabularnewline
\midrule 
A70 & N11-C18-H24 &  & 107.7177 &  &  & 107.7355 &  &  & 107.6846 &  &  & 107.7317 &  &  & 107.73 & \tabularnewline
\midrule 
A71 & N11-C18-H32 &  & 110.8518 &  &  & 110.8379 &  &  & 110.8419 &  &  & 110.759 &  &  & 110.758 & \tabularnewline
\midrule 
A72 & H24-C18-H32 &  & 108.3845 &  &  & 108.3981 &  &  & 108.4317 &  &  & 108.5012 &  &  & 108.5021 & \tabularnewline
\midrule 
 &  &  &  &  &  &  &  &  &  &  &  &  &  &  &  & \tabularnewline
\bottomrule
\end{tabular}\caption{\label{tab:bond_angles}The individual bond angles of $\mathrm{3\times3\:penta-CN_{2}}$
QD after optimization}
\end{table}

\subsection{\textcolor{black}{Other Electronic Properties}}

\textcolor{black}{The Ionization Potential (IP) values among different
configurations represented by N4 (5.671), N6 (5.729), N11 (5.728),
C14 (5.67), C15 (5.688), exhibit minimal variation both within each
other and in comparison to the pristine penta-CN$_{2}$ QD.}

\textcolor{black}{}
\begin{table}[H]
\centering{}\textcolor{black}{}%
\begin{tabular}{cccc}
\toprule 
\textcolor{black}{}%
\begin{tabular}{c}
\textcolor{black}{Initial Structure}\tabularnewline
\end{tabular} & \textcolor{black}{}%
\begin{tabular}{c}
\textcolor{black}{Final Structure}\tabularnewline
\end{tabular} & \textcolor{black}{}%
\begin{tabular}{c}
\textcolor{black}{Initial Distance ($\text{Å}$)}\tabularnewline
\end{tabular} & \textcolor{black}{}%
\begin{tabular}{c}
\textcolor{black}{Final Distance ($\text{Å}$)}\tabularnewline
\end{tabular}\tabularnewline
\midrule
\midrule 
\textcolor{black}{}%
\begin{tabular}{c}
\textcolor{black}{$\mathrm{N2}$}\tabularnewline
\end{tabular} & \textcolor{black}{}%
\begin{tabular}{c}
\textcolor{black}{$\mathrm{N15}$}\tabularnewline
\end{tabular} & \textcolor{black}{}%
\begin{tabular}{c}
\textcolor{black}{2}\tabularnewline
\end{tabular} & \textcolor{black}{}%
\begin{tabular}{c}
\textcolor{black}{2.44}\tabularnewline
\end{tabular}\tabularnewline
\textcolor{black}{}%
\begin{tabular}{c}
\textcolor{black}{$\mathrm{N14}$}\tabularnewline
\end{tabular} & \textcolor{black}{}%
\begin{tabular}{c}
\textcolor{black}{N22}\tabularnewline
\end{tabular} & \textcolor{black}{}%
\begin{tabular}{c}
\textcolor{black}{2}\tabularnewline
\end{tabular} & \textcolor{black}{}%
\begin{tabular}{c}
\textcolor{black}{2.41}\tabularnewline
\end{tabular}\tabularnewline
\textcolor{black}{}%
\begin{tabular}{c}
\textcolor{black}{$\mathrm{N21}$}\tabularnewline
\end{tabular} & \textcolor{black}{}%
\begin{tabular}{c}
\textcolor{black}{N22}\tabularnewline
\end{tabular} & \textcolor{black}{}%
\begin{tabular}{c}
\textcolor{black}{2}\tabularnewline
\end{tabular} & \textcolor{black}{}%
\begin{tabular}{c}
\textcolor{black}{2.41}\tabularnewline
\end{tabular}\tabularnewline
\textcolor{black}{}%
\begin{tabular}{c}
\textcolor{black}{$\mathrm{C25}$}\tabularnewline
\end{tabular} & \textcolor{black}{}%
\begin{tabular}{c}
\textcolor{black}{N21}\tabularnewline
\end{tabular} & \textcolor{black}{}%
\begin{tabular}{c}
\textcolor{black}{2}\tabularnewline
\end{tabular} & \textcolor{black}{}%
\begin{tabular}{c}
\textcolor{black}{2.54}\tabularnewline
\end{tabular}\tabularnewline
\textcolor{black}{}%
\begin{tabular}{c}
\textcolor{black}{$\mathrm{C26}$}\tabularnewline
\end{tabular} & \textcolor{black}{}%
\begin{tabular}{c}
\textcolor{black}{N3}\tabularnewline
\end{tabular} & \textcolor{black}{}%
\begin{tabular}{c}
\textcolor{black}{2}\tabularnewline
\end{tabular} & \textcolor{black}{}%
\begin{tabular}{c}
\textcolor{black}{2.58}\tabularnewline
\end{tabular}\tabularnewline
\bottomrule
\end{tabular}\textcolor{black}{\caption{\label{tab:Vertical_distances-3x4}The initial and final adsorption
sites of the H atoms along with the distances from the closest atom
of the $3\times4$ penta-CN$_{2}$ QD.}
}
\end{table}

\textcolor{black}{}
\begin{table}[H]
\centering{}\textcolor{black}{}%
\begin{tabular}{cccc}
\toprule 
\textcolor{black}{}%
\begin{tabular}{c}
\textcolor{black}{Initial Structure}\tabularnewline
\end{tabular} & \textcolor{black}{}%
\begin{tabular}{c}
\textcolor{black}{Final Structure}\tabularnewline
\end{tabular} & \textcolor{black}{}%
\begin{tabular}{c}
\textcolor{black}{Initial Distance ($\text{Å}$)}\tabularnewline
\end{tabular} & \textcolor{black}{}%
\begin{tabular}{c}
\textcolor{black}{Final Distance ($\text{Å}$)}\tabularnewline
\end{tabular}\tabularnewline
\midrule
\midrule 
\textcolor{black}{}%
\begin{tabular}{c}
\textcolor{black}{$\mathrm{N2}$}\tabularnewline
\end{tabular} & \textcolor{black}{}%
\begin{tabular}{c}
\textcolor{black}{$\mathrm{N28}$}\tabularnewline
\end{tabular} & \textcolor{black}{}%
\begin{tabular}{c}
\textcolor{black}{2}\tabularnewline
\end{tabular} & \textcolor{black}{}%
\begin{tabular}{c}
\textcolor{black}{2.35}\tabularnewline
\end{tabular}\tabularnewline
\textcolor{black}{}%
\begin{tabular}{c}
\textcolor{black}{$\mathrm{C12}$}\tabularnewline
\end{tabular} & \textcolor{black}{}%
\begin{tabular}{c}
\textcolor{black}{$\mathrm{N4}$}\tabularnewline
\end{tabular} & \textcolor{black}{}%
\begin{tabular}{c}
\textcolor{black}{2}\tabularnewline
\end{tabular} & \textcolor{black}{}%
\begin{tabular}{c}
\textcolor{black}{2.40}\tabularnewline
\end{tabular}\tabularnewline
\textcolor{black}{}%
\begin{tabular}{c}
\textcolor{black}{$\mathrm{N17}$}\tabularnewline
\end{tabular} & \textcolor{black}{}%
\begin{tabular}{c}
\textcolor{black}{$\mathrm{N27}$}\tabularnewline
\end{tabular} & \textcolor{black}{}%
\begin{tabular}{c}
\textcolor{black}{2}\tabularnewline
\end{tabular} & \textcolor{black}{}%
\begin{tabular}{c}
\textcolor{black}{2.51}\tabularnewline
\end{tabular}\tabularnewline
\textcolor{black}{}%
\begin{tabular}{c}
\textcolor{black}{$\mathrm{N18}$}\tabularnewline
\end{tabular} & \textcolor{black}{}%
\begin{tabular}{c}
\textcolor{black}{$\mathrm{N28}$}\tabularnewline
\end{tabular} & \textcolor{black}{}%
\begin{tabular}{c}
\textcolor{black}{2}\tabularnewline
\end{tabular} & \textcolor{black}{}%
\begin{tabular}{c}
\textcolor{black}{2.35}\tabularnewline
\end{tabular}\tabularnewline
\textcolor{black}{}%
\begin{tabular}{c}
\textcolor{black}{$\mathrm{C30}$}\tabularnewline
\end{tabular} & \textcolor{black}{}%
\begin{tabular}{c}
\textcolor{black}{$\mathrm{N28}$}\tabularnewline
\end{tabular} & \textcolor{black}{}%
\begin{tabular}{c}
\textcolor{black}{2}\tabularnewline
\end{tabular} & \textcolor{black}{}%
\begin{tabular}{c}
\textcolor{black}{2.35}\tabularnewline
\end{tabular}\tabularnewline
\textcolor{black}{}%
\begin{tabular}{c}
\textcolor{black}{$\mathrm{C31}$}\tabularnewline
\end{tabular} & \textcolor{black}{}%
\begin{tabular}{c}
\textcolor{black}{$\mathrm{N27}$}\tabularnewline
\end{tabular} & \textcolor{black}{}%
\begin{tabular}{c}
\textcolor{black}{2}\tabularnewline
\end{tabular} & \textcolor{black}{}%
\begin{tabular}{c}
\textcolor{black}{2.51}\tabularnewline
\end{tabular}\tabularnewline
\bottomrule
\end{tabular}\textcolor{black}{\caption{\label{tab:Vertical_distances-4x4-1}The initial and final adsorption
sites of the H atoms along with the distances from the closest atom
of the $4\times4$ penta-CN$_{2}$ QD.}
}
\end{table}

\textcolor{black}{}
\begin{table}[H]
\centering{}\textcolor{black}{}%
\begin{tabular}{ccccccccccccccccccc}
\toprule 
\addlinespace[1mm]
\textcolor{black}{$3\times4$ QD} & \textcolor{black}{$E_{HOMO}$} & \textcolor{black}{$E_{LUMO}$} &  &  & \textcolor{black}{$E_{g}$} &  &  & \textcolor{black}{$S$} &  &  & \textcolor{black}{$\mu$} &  &  & \textcolor{black}{$\omega$} &  &  & \textcolor{black}{$IP$} & \textcolor{black}{$EA$}\tabularnewline\addlinespace[1mm]
\midrule
\addlinespace[1mm]
\addlinespace[1mm]
\textcolor{black}{Pure} & \textcolor{black}{-5.669} & \textcolor{black}{1.495} &  &  & \textcolor{black}{7.164} &  &  & \textcolor{black}{0.279} &  &  & \textcolor{black}{-2.087} &  &  & \textcolor{black}{0.608} &  &  & \textcolor{black}{5.669} & \textcolor{black}{-1.495}\tabularnewline\addlinespace[1mm]
\addlinespace[1mm]
\addlinespace[1mm]
\textcolor{black}{$\mathrm{N3}$} & \textcolor{black}{-5.701} & \textcolor{black}{-1.309} &  &  & \textcolor{black}{4.392} &  &  & \textcolor{black}{0.455} &  &  & \textcolor{black}{-3.505} &  &  & \textcolor{black}{2.798} &  &  & \textcolor{black}{5.701} & \textcolor{black}{1.309}\tabularnewline\addlinespace[1mm]
\addlinespace[1mm]
\addlinespace[1mm]
\textcolor{black}{$\mathrm{N15}$} & \textcolor{black}{-5.688} & \textcolor{black}{-1.155} &  &  & \textcolor{black}{4.533} &  &  & \textcolor{black}{0.441} &  &  & \textcolor{black}{-3.421} &  &  & \textcolor{black}{2.583} &  &  & \textcolor{black}{5.688} & \textcolor{black}{1.155}\tabularnewline\addlinespace[1mm]
\addlinespace[1mm]
\addlinespace[1mm]
\textcolor{black}{$\mathrm{N21}$} & \textcolor{black}{-5.721} & \textcolor{black}{-0.982} &  &  & \textcolor{black}{4.740} &  &  & \textcolor{black}{0.422} &  &  & \textcolor{black}{-3.352} &  &  & \textcolor{black}{2.370} &  &  & \textcolor{black}{5.721} & \textcolor{black}{0.982}\tabularnewline\addlinespace[1mm]
\addlinespace[1mm]
\addlinespace[1mm]
\textcolor{black}{$\mathrm{N22}$} & \textcolor{black}{-5.702} & \textcolor{black}{-1.839} &  &  & \textcolor{black}{3.862} &  &  & \textcolor{black}{0.518} &  &  & \textcolor{black}{-3.771} &  &  & \textcolor{black}{3.681} &  &  & \textcolor{black}{5.702} & \textcolor{black}{1.839}\tabularnewline\addlinespace[1mm]
\addlinespace[1mm]
\addlinespace[1mm]
 &  &  &  &  &  &  &  &  &  &  &  &  &  &  &  &  &  & \tabularnewline\addlinespace[1mm]
\bottomrule
\addlinespace[1mm]
\end{tabular}\textcolor{black}{\caption{\label{tab:GRD-3x4}Global reactivity descriptors, namely, HOMO energy
($E_{HOMO}$), LUMO energy ($E_{LUMO}$), H-L gap ($E_{g}$), chemical
softness ($S$), chemical potential ($\mu$), electrophilicity index
($\omega$), ionization potential ($IP$), and electron affinity ($EA$)
for the pristine $3\times4$ penta-CN$_{2}$ QD, and those with H
adsorbed on the sites N3, N15, N21, and N22. Above, $S$ is in the
units of eV$^{-1}$, while all other quantities are in eV units.}
}
\end{table}
\textcolor{black}{}
\begin{table}[H]
\centering{}\textcolor{black}{}%
\begin{tabular}{ccccccccccccccccc}
\toprule 
\addlinespace[1mm]
\textcolor{black}{$4\times4$ QD} & \textcolor{black}{$E_{HOMO}$} & \textcolor{black}{$E_{LUMO}$} &  &  & \textcolor{black}{$E_{g}$} &  &  & \textcolor{black}{$S$} &  &  & \textcolor{black}{$\mu$} &  &  & \textcolor{black}{$\omega$} & \textcolor{black}{$IP$} & \textcolor{black}{$EA$}\tabularnewline\addlinespace[1mm]
\midrule
\addlinespace[1mm]
\addlinespace[1mm]
\textcolor{black}{Pure} & \textcolor{black}{-5.790} & \textcolor{black}{1.130} &  &  & \textcolor{black}{6.920} &  &  & \textcolor{black}{0.289} &  &  & \textcolor{black}{-2.330} &  &  & \textcolor{black}{0.785} & \textcolor{black}{5.790} & \textcolor{black}{-1.130}\tabularnewline\addlinespace[1mm]
\addlinespace[1mm]
\addlinespace[1mm]
\textcolor{black}{$\mathrm{N4}$} & \textcolor{black}{-5.889} & \textcolor{black}{-1.154} &  &  & \textcolor{black}{4.736} &  &  & \textcolor{black}{0.422} &  &  & \textcolor{black}{-3.522} &  &  & \textcolor{black}{2.619} & \textcolor{black}{5.889} & \textcolor{black}{1.154}\tabularnewline\addlinespace[1mm]
\addlinespace[1mm]
\addlinespace[1mm]
\textcolor{black}{$\mathrm{N27}$} & \textcolor{black}{-5.807} & \textcolor{black}{-1.969} &  &  & \textcolor{black}{3.838} &  &  & \textcolor{black}{0.521} &  &  & \textcolor{black}{-3.888} &  &  & \textcolor{black}{3.938} & \textcolor{black}{5.807} & \textcolor{black}{1.969}\tabularnewline\addlinespace[1mm]
\addlinespace[1mm]
\addlinespace[1mm]
\textcolor{black}{$\mathrm{N28}$} & \textcolor{black}{-5.893} & \textcolor{black}{-1.859} &  &  & \textcolor{black}{4.034} &  &  & \textcolor{black}{0.496} &  &  & \textcolor{black}{-3.876} &  &  & \textcolor{black}{3.725} & \textcolor{black}{5.893} & \textcolor{black}{1.859}\tabularnewline\addlinespace[1mm]
\addlinespace[1mm]
\addlinespace[1mm]
 &  &  &  &  &  &  &  &  &  &  &  &  &  &  &  & \tabularnewline\addlinespace[1mm]
\bottomrule
\addlinespace[1mm]
\end{tabular}\textcolor{black}{\caption{\label{tab:GRD-4x4}Global reactivity descriptors, namely, HOMO energy
($E_{HOMO}$), LUMO energy ($E_{LUMO}$), H-L gap ($E_{g}$), chemical
softness ($S$), chemical potential ($\mu$), electrophilicity index
($\omega$), ionization potential ($IP$), and electron affinity ($EA$)
for the pristine $4\times4$ penta-CN$_{2}$ QD, and those with H
adsorbed on the sites N4, N27, and N28. Above, $S$ is in the units
of eV$^{-1}$, while all other quantities are in eV units.}
}
\end{table}

\textcolor{black}{}
\begin{table}[H]
\centering{}\textcolor{black}{}%
\begin{tabular}{cccccc}
\toprule 
\textcolor{black}{H-adsorbed configurations} & \textcolor{black}{Functional} & \textcolor{black}{Basis-Set} & \textcolor{black}{$\Delta E_{ads}$(eV)} & \textcolor{black}{$\Delta G$(eV)} & \textcolor{black}{$\varphi$(mV)}\tabularnewline
\midrule
\midrule 
\textcolor{black}{$\mathrm{N3}$} & \multirow{4}{*}{\textcolor{black}{B3LYP}} & \multirow{4}{*}{\textcolor{black}{6-31g(d,p)}} & \textcolor{black}{-0.029} & \textcolor{black}{0.211} & \textcolor{black}{211}\tabularnewline
\cmidrule{1-1} \cmidrule{4-6} \cmidrule{5-6} \cmidrule{6-6} 
\textcolor{black}{$\mathrm{N15}$} &  &  & \textcolor{black}{-0.032} & \textcolor{black}{0.208} & \textcolor{black}{208}\tabularnewline
\cmidrule{1-1} \cmidrule{4-6} \cmidrule{5-6} \cmidrule{6-6} 
\textcolor{black}{$\mathrm{N21}$} &  &  & \textcolor{black}{-0.028} & \textcolor{black}{0.212} & \textcolor{black}{212}\tabularnewline
\cmidrule{1-1} \cmidrule{4-6} \cmidrule{5-6} \cmidrule{6-6} 
\textcolor{black}{$\mathrm{N22}$} &  &  & \textcolor{black}{-0.044} & \textcolor{black}{0.196} & \textcolor{black}{196}\tabularnewline
\bottomrule
\end{tabular}\textcolor{black}{\caption{\label{tab:HER-descriptors-3x4}Values of adsorbtion energy, Gibbs
free energy, over potential, and the exchange current density for
all the adsorption sites for the $3\times4$ penta-CN$_{2}$ QD.}
}
\end{table}

\textcolor{black}{}
\begin{table}[H]
\centering{}\textcolor{black}{}%
\begin{tabular}{cccccc}
\toprule 
\textcolor{black}{H-adsorbed configurations} & \textcolor{black}{Functional} & \textcolor{black}{Basis-Set} & \textcolor{black}{$\Delta E_{ads}$(eV)} & \textcolor{black}{$\Delta G$(eV)} & \textcolor{black}{$\varphi$(mV)}\tabularnewline
\midrule
\midrule 
\textcolor{black}{$\mathrm{N4}$} & \multirow{3}{*}{} & \multirow{3}{*}{} & \textcolor{black}{-0.035} & \textcolor{black}{0.205} & \textcolor{black}{205}\tabularnewline
\cmidrule{1-1} \cmidrule{4-6} \cmidrule{5-6} \cmidrule{6-6} 
\textcolor{black}{$\mathrm{N27}$} &  &  & \textcolor{black}{-0.030} & \textcolor{black}{0.210} & \textcolor{black}{210}\tabularnewline
\cmidrule{1-1} \cmidrule{4-6} \cmidrule{5-6} \cmidrule{6-6} 
\textcolor{black}{$\mathrm{N28}$} &  &  & \textcolor{black}{-0.049} & \textcolor{black}{0.191} & \textcolor{black}{191}\tabularnewline
\cmidrule{1-1} \cmidrule{4-6} \cmidrule{5-6} \cmidrule{6-6} 
\end{tabular}\textcolor{black}{\caption{\label{tab:HER-descriptors-4x4}Values of adsorbtion energy, Gibbs
free energy, over potential, and the exchange current density for
all the adsorption sites for the $4\times4$ penta-CN$_{2}$ QD.}
}
\end{table}